\documentclass[iop]{emulateapj}
\usepackage{natbib,graphicx,color}

\slugcomment{Accepted to ApJ}

\shorttitle{CO survey of disks}
\shortauthors{Brown et al.}

\def\grapprox{$_>\atop{^\sim}$}

\begin{document}

\title{VLT-CRIRES survey of rovibrational CO emission from protoplanetary disks}

\author{{J.M. Brown\altaffilmark{1,2}, K.M. Pontoppidan \altaffilmark{3}, E.F. van
    Dishoeck\altaffilmark{2,4}, G.J. Herczeg\altaffilmark{5}, G.A. Blake\altaffilmark{6},
    A. Smette \altaffilmark{7}}}
\altaffiltext{1}{Harvard-Smithsonian Center for Astrophysics, 60 Garden St., MS 78, Cambridge, MA 02138; joannabrown@cfa.harvard.edu}
\altaffiltext{2}{Max-Planck-Institut f{\"u}r extraterrestrische Physik, Postfach 1312,  85741 Garching, Germany }
\altaffiltext{3}{Space Telescope Science Institute, 3700 San Martin Drive, Baltimore, MD 21218}
\altaffiltext{4}{Leiden Observatory, Leiden University, P.O. Box 9513, NL-2300 RA Leiden, The Netherlands}
\altaffiltext{5}{The Kavli Institute  for Astronomy and Astrophysics, Peking University, Yi He Yuan Lu 5, Hai Dian Qu, Beijing 100871, P. R. China} 
\altaffiltext{6}{Division of Geological \& Planetary Sciences, California
Institute of Technology, Pasadena, CA 91125}
\altaffiltext{7}{ESO, Alonso de Cordova 3107, Casilla 19001, Vitacura, Chile}

\begin{abstract}
  We present a large, comprehensive survey of rovibrational CO line
  emission at 4.7 $\mu$m from 69 protoplanetary disks, obtained with
  CRIRES on the ESO Very Large Telescope at the highest available
  spectral resolving power ($R$=95,000, $\Delta
  v$=$3.2\rm\,km\,s^{-1}$). The CO fundamental band ($\Delta v$=1) is
  a well-known tracer of warm gas in the inner, planet-forming regions
  of gas-rich disks around young stars, with the lines formed in the
  super-heated surfaces of the disks at radii of 0.1--10 AU.
  Consistent with earlier studies, the presence of 100-1000 K CO is
  found to be ubiquitous around young stars which still retain
  disks. Our high spectral resolution data provide new insight into
  the kinematics of the inner disk gas. The observed line profiles are
  complex and reveal several different components.  Pure double-peaked
  Keplerian profiles are surprisingly uncommon in our sample, beyond
  the frequency expected based on disk inclination. The majority of
  the profiles are consistent with emission from a disk plus a slow
  (few km s$^{-1}$) molecular disk wind.  This is evidenced by
  analysis of different classes as well as an overall tendency for
  line profiles to have excess emission on their blue side.  The data
  support the notion that thermal molecular winds are common for young
  disks. Thanks to the high spectral resolution, narrow absorption
  lines and weak emission lines from isotopologues and from
  vibrationally excited levels are readily detected. In general,
  $^{13}$CO lines trace cooler gas than the bulk $^{12}$CO emission
  and may arise from further out in the disk, as indicated by narrower
  line profiles. A high fraction of the sources show vibrationally
  excited emission ($\sim$50\%) which is correlated with accretion
  luminosity, consistent with ultra-violet (UV) fluorescent
  excitation.  Disks around early-type Herbig AeBe stars have narrower
  line profiles, on average, than their lower-mass late-type
  counterparts, due to their increased luminosity. Evolutionary
  changes in CO are also seen. Removal of the protostellar envelope
  between class I and II results in the disappearance of the strong
  absorption lines and CO ice feature characteristic of class I
  spectra. However, CO emission from class I and II objects are
  similar in detection frequency, excitation and line shape,
  indicating that inner disk characteristics are established early.

\end{abstract}


\section{Introduction}

The inner regions of gas-rich protoplanetary disks ($R\lesssim
10\,$AU) are thought to be the birth places of most giant planets
\citep{armitage10, kley12}. The chemical and physical processes sculpting
these environments in the first few million years of the life of a
star are critical for determining many properties of mature planetary
systems. These include the formation of rocky ``oligarchs'' -- the
building blocks of terrestrial planets \citep{nagasawa07} -- and 
comets and water-rich asteroids important for the delivery of water
and organics to planetary surfaces \citep{raymond04}. The interaction
between the gas-rich inner disk and protoplanets has the power to
re-arrange the orbital structure of the entire planetary system
\citep{armitage11}, allowing for significant modification of radial
chemical abundance structures; for instance, the chemical boundary
defined by the snow-line may not always be predictive for the
compositions of planets in mature systems.

A key diagnostic of the structure of planet-forming regions is the
fundamental ($\Delta v=1$) rovibrational band of CO at 4.7\,$\mu$m.
It is particularly sensitive to gas temperatures of 100--1000\,K,
corresponding to radii of 0.1--10\,AU in typical protoplanetary disks
around Solar-mass pre-main sequence stars. Because of the high opacity
of dust at 5\,$\mu$m and the low temperatures of disk midplanes beyond
$\sim 1$\,AU, the CO fundamental band typically traces the disk at
high altitude, specifically the so-called ``warm molecular layer''
\citep{aikawa02,gorti08,woitke09}. In comparison with the total disk
surface densities in the planet-forming region of $\Sigma =
10-1000\,\rm g\,cm^{-2}$, the CO fundamental band traces roughly
$N_{\rm H}\sim 10^{21}-10^{23}\,\rm cm^{-2}$, corresponding to
$10^{-3}-10^{-1}\,\rm g\,cm^{-2}$, where $N_{\rm H}$ is the total
column of hydrogen nuclei. 

While the warm molecular layer represents a small fraction of the
total vertical disk column, its structure is intimately linked to key
properties of the bulk gas in protoplanetary disks, relevant to their
evolution and ability to form planets. Through high temperatures and
interactions with ionizing stellar radiation, the molecular layer acts
as a chemical factory, producing water and complex organics
\citep{markwick02,glassgold09,woods09,walsh12}. It also forms a
boundary between the deep, neutral and inactive disk midplane and the
uppermost ionized layers. As such, the physical and kinematical
structure of the molecular layer traces thermal and photo-evaporative
flows from the disk surface, controlling outward radial mixing and
mass loss \citep{hollenbach94,owen10}. It is in the molecular layer
that the stellar magnetic field can couple to the disk and drive
turbulence and accretion flows \citep{gammie96,perez11}. Finally, the
molecular layer responds to dynamical perturbations caused by the
presence of giant protoplanets, leading to potentially observable
effects \citep{regaly10}.

Rovibrational CO emission is present throughout a wide range of
protoplanetary disks from still embedded protostars
\citep{pontoppidan03} to transitional disks with inner dust holes
\citep{salyk09}, and from low mass T Tauri stars \citep{najita03} to
higher mass Herbig stars \citep{brittain07}. The CO lines are formed
by a combination of collisional excitation, infrared (vibrational)
pumping and UV (electronic) fluorescence. The relative importance of
these excitation processes depends on location within the disk and the
shape and strength of the radiation field from the central star
\citep{blake04}.

The advent of the CRyogenic high-resolution InfraRed Echelle
Spectrograph (CRIRES) instrument on the Very Large Telescope (VLT) has
opened the possibility to observe the CO fundamental bands at higher
spectral resolving power ($R\approx$95000) and higher spatial
resolution than before, thus providing new insight into this critical
planet-forming region of the disk. We present here the results of a
large VLT-CRIRES programme of 69 disks around low- and
intermediate-mass stars and 22 embedded young stellar objects
\citep{pontoppidan11_mess}.  Previous papers have used subsets of our
CRIRES database to address a variety of questions. \citet{bast11}
investigate the presence and origin of a class of broad single-peaked
CO rovibrational line profiles indicating non-Keplerian gas
motions. Using spectro-astrometry, \citet{pontoppidan08,pontoppidan11}
constrain the structure of the gas emission and velocity fields on
milli-arcsecond scales in a smaller sample of disks. Some disks show
CO emission consistent with simple Keplerian models, whereas other
disks, especially those in the \citet{bast11} sample, show a
spectro-astrometry pattern consistent with a slow molecular disk wind.
For transitional disks with a large inner dust hole or gap,
\citet{pontoppidan08} pinpointed the CO rovibrational emission to
originate from inside the dust gap whereas \citet{brown12} resolved
the CO emission near the outer wall of the hole around the Herbig star
Oph IRS 48. \citet{herczeg11} characterized the progenitors to
protoplanetary disks, when the sources are still embedded in
protostellar envelopes.  \citet{thi10} found evidence for episodic
outflow activity with winds up to 100 km s$^{-1}$ in broad
blue-shifted lines toward one object, whereas \citet{herczeg11} found
the same phenomenon for a handful of other embedded sources.

Finally, the high spectral resolution of CRIRES also boosts the line
to continuum ratio and thus allows the detection of weak lines from
minor species.  \citet{smith09} accurately measured the isotopologue
ratios of the four major CO species -- $^{12}$CO, $^{13}$CO,
  C$^{18}$O and C$^{17}$O -- in the circumstellar environment of two
  young stars.  The aim was to search for mass-independent oxygen
  isotope fractionation relative to the interstellar medium in order
  to understand the Solar System oxygen anomaly. \citet{mandell12}
  searched for near-infrared lines of small organic molecules (HCN,
  C$_2$H$_2$, CH$_4$) in a few sources. While the these studies
  discuss the detailed properties of specific sub-samples, this paper
  provides an overview of the basic properties of CO rovibrational
  emission from protoplanetary disks around young Solar-type stars at
  the highest available spectral resolution.

Using the entire CRIRES data set, we aim to address the following
  questions related to the structure and evolution of planet-forming
  regions of protoplanetary disks: 1) Where is the CO gas located?
  What fraction of sources show emission from a radially flowing
  surface such as a slow disk wind? 2) What are typical temperatures
  probed by CO and its isotopologues and how do these values compare
  with current thermo-chemical models of the inner disk? How are the
  CO fundamental bands excited, and what consequences does the
  excitation mechanism have for the lines as a tracer of inner disk
  surfaces? 3) Do CO line profiles depend on stellar spectral type,
  and is this a tracer of the prevalence of UV fluorescent excitation
  over thermal excitation?  4) What evolutionary changes occur in the
  gas?

\section{Observations and Sample Characteristics}
\label{obs}

\subsection{The CRIRES survey}
In this paper, we analyze high resolution spectra of the 4.7 micron CO
v=1--0 fundamental emission band using the CRIRES on the VLT of the
European Southern Observatory (ESO) \citep{kaeufl04}. CRIRES operates
at high resolution ($R=95,000$, $\Delta v=3.2\rm \,km\, s^{-1}$) using
a $0\farcs 2$ slit. It is fed by a Multi-Application Curvature
Adaptive Optics (MACAO) system, which allows correction of atmospheric
turbulence and can provide diffraction limited images at the focal
plane, therefore improving the overall instrument sensitivity.

The sample consists of 91 young stars -- 69 protoplanetary disks
around young stars and 22 embedded protostars that are still
surrounded by massive remnant envelopes. The data were obtained as
part of an {\it ESO Large Programme}\footnote{This work is based on
observations collected at the European Southern Observatory Very Large
Telescope under programme ID 179.C-0151.} to study infrared molecular
emission from solar-type protostars and protoplanetary disks
\citep{pontoppidan11_mess}\footnote{Fully processed spectra are
available on {\tt http://www.stsci.edu/$\sim$pontoppi}}.  We discuss
here primarily the disk sample, but compare our results with those of
the embedded protostars presented in \citet{herczeg11} to search for
evolutionary trends between embedded and classical protoplanetary
disks.

\subsection{Sample Selection}

\begin{figure}
\includegraphics[angle=90,width=8cm]{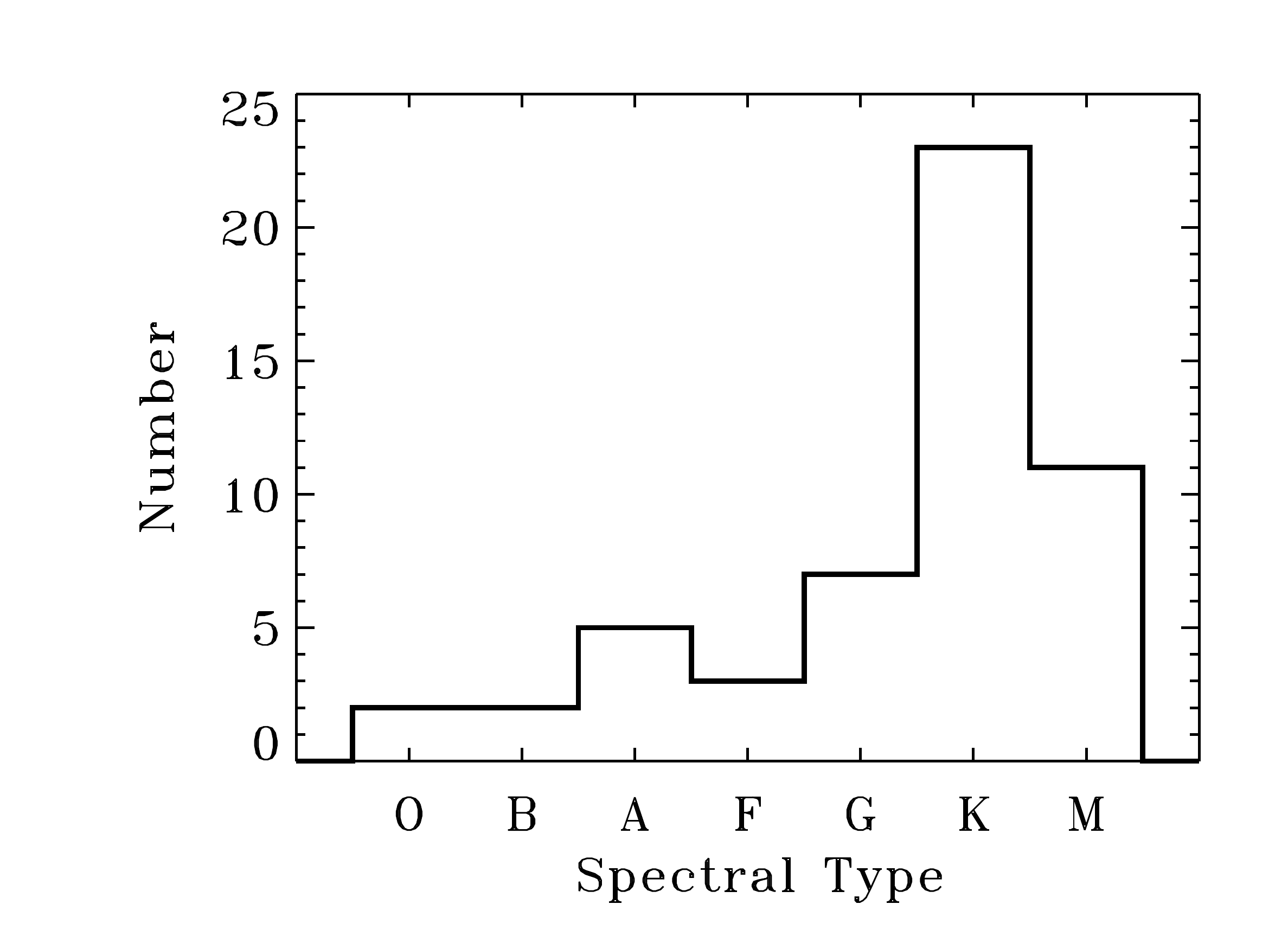} 
\caption{A histogram of the spectral types of the stars in our disk
  sample. The sample is dominated by K type stars. When a range of
  spectral types were reported in the literature the mean value was
  taken. \label{fig:spt}}
\end{figure}

The disk sample spans a range of physical properties including spectral
type, stellar mass and luminosity, and inclination. Such a large
sample of CO emission profiles provides an
opportunity for a broad examination of the gas distribution in a wide
variety of circumstellar environments. The sample was chosen to
include protoplanetary disks around solar-mass ($0.5-2.0\,M_{\odot}$)
young stars (see Figure \ref{fig:spt}). A range of evolutionary states
were sampled from massive gas-rich disks that are optically thick to
visible and UV radiation to transition disks with inner dust holes or
gaps. The sources are located in the nearby star forming regions
visible from Paranal Observatory, including Taurus, Ophiuchus,
Serpens, Corona Australis and Chamaeleon. All these regions have
undergone recent active star formation and have ages of approximately
1--5 Myr \citep{greene95,armitage03,oliveira09}. Protoplanetary disks
were selected based on brightness ($\gtrsim 100\,$mJy at $4.7\,\mu$m),
as well as the existence of prior datasets characterizing the
structure of the disks and supporting the presence of significant gas
reservoirs. The prior datasets defining the sample include Spitzer
spectroscopy from the cores to disks (c2d) Spitzer Legacy survey
\citep{evans03}, the Keck-NIRSPEC 3--5\,$\mu$m protoplanetary disk
survey (e.g. \citealt{blake04,salyk11}) and the VLT-ISAAC protostellar
survey \citep{vanDishoeck03}. We limited the survey to disks around
Solar-type stars ($0.1\,M_{\odot}\lesssim M_* \lesssim
2.0\,M_{\odot}$), excluding most Herbig AeBe stars -- the focus of
complementary CRIRES surveys \citep[e.g.,][]{vanderplas09}.

\subsection{Observing procedure and data reduction} 

The high spectral resolution and high dynamic range of CRIRES spectra,
a factor of four improvement in spectral resolution over most previous
CO surveys, fully resolves the individual line profiles so that
velocity information can be used to locate the gas within the
disk. The resolution is particularly needed in the cases of disks with
low inclinations where the lines are intrinsically narrow and in the
cases where multiple components, particularly absorption features,
contribute to the line profile.  Because of the adaptive optics
system, line emission can be spatially resolved down to angular scales
of $\sim 0\farcs 1$.

The observations were taken between April 2007 and March 2010.
Table \ref{table:obs} lists the targets observed and the dates and
wavelength settings of the observations. Multiple spectral settings
were taken to cover a range of rotational $J$ lines.  Each spectral
setting is observed with four different detectors, leading to four
discontinuous wavelength regions. The wavelength listed is the center
of the third detector and the range is $-$70 to +39 nm on either side
of the listed wavelength, with $\sim$6 nm gaps between the
detectors. The spectra cover the CO $P$ branch lines with $\Delta
J=-1$ and a few low energy lines from the $R$ branch ($\Delta
J=1$). $^{12}$CO v=1--0 lines are the most prominent. Gas phase lines
from higher vibrational states and rarer CO isotopologues such as
$^{13}$CO, C$^{18}$O and C$^{17}$O are also included in the wavelength
settings.

The spectra were obtained using an ABBA 10" nodding pattern permitting
a first order correction of the infrared background by pair
subtraction. Integration times were generally 8--16 minutes per
spectral setting.  The reduction of the spectra is described in detail
in \cite{pontoppidan08,pontoppidan11}.  Wavelength calibration used
the prevalent telluric features in the standard star
spectra. Atmospheric features were removed by dividing the source
spectrum by the standard star spectrum. Standard stars were observed close in
time and elevation to minimize atmospheric differences. Remaining
small discrepancies from airmass differences and subpixel variations
in wavelength solution were corrected by scaling or shifting the
standard spectrum to minimize telluric residuals. Remaining strong
telluric features were blanked from the analyzed spectra. The
observations were preferentially scheduled for periods when the
Earth's velocity around the sun created large shifts of the source
lines relative to the telluric features. When possible, complete
wavelength coverage was obtained using the Earth's orbital motion
between separated observation dates to Doppler shift the telluric
features relative to the source lines. The spectra from two or more
dates were then combined, barring large differences in line profiles.

The accuracy of the absolute flux calibration is $\sim 30\%$,
estimated by comparing raw counts of consecutive nod pairs. This flux
calibration is limited by differences in Strehl ratio between the
target disk and the standard star, as well as by pointing jitter. We
therefore scaled the spectra to known M band fluxes (see Table C1),
with many from the Spitzer IRAC 2 band. Note, however, that some
protoplanetary disks are known to be variable on short time scales at
the 10--60\% level \citep{flaherty12} and discrepancies could exist
based on the different aperture sizes. In cases where no Spitzer
photometry was available a linear relationship between the maximum
counts received and the flux was determined based on the sources with
known fluxes. Determining the flux using this method led to a median
error in derived fluxes of $\sim$20\% for sources with previously
known fluxes. However, errors were as large as a factor of 2 in
exceptional cases. The determination was more accurate with larger
numbers of exposures.

\begin{figure}[]
\includegraphics[angle=180,width=8cm]{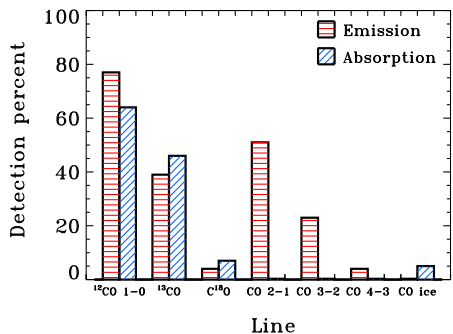} 
\vspace{-0.7cm}
\caption{Percentage of disk sources with different lines detected in
  emission (red, horizontally hashed) and absorption (blue, diagonally
  hashed). In general, rarer isotopologues and higher energy lines are
  less frequently detected. The CO isotopologues are more commonly
  seen in absorption while the vibrationally excited transitions are
  seen exclusively in emission. Table \ref{table:feat} lists common
  lines detected for individual sources.
  \label{fig:detstats}}
\end{figure}

\subsection{Detection rates}

CO gas phase lines are common in our sample, both in emission and in
absorption, as shown graphically in Fig.~\ref{fig:detstats}. Lines are
commonly a combination of emission and absorption resulting from two
distinct gas components such as disk emission with foreground
absorption. Overall, emission features are seen from 53/69 (77\%) of
the disks. Absorption is seen in 44/69 (64 \%) of the spectra. CO
v=2--1 lines are seen from 36/69 (52\%) of the sample, always in
emission. $^{13}$CO emission lines are seen in 27 (39\%) of the
sources. However, $^{13}$CO is more commonly seen in absorption
(46\%), often at cold temperatures indicative of foreground material.
C$^{18}$O is seen in emission from 4 sources (6\%) and in absorption
from 7 (10\%).

\subsection{Multiplicity}

In cases where a binary companion was known, we oriented the slit to
observe both stars. There are 8 disk systems where both components are
well separated and bright enough for analysis: AS 205, DoAr 24 E, S
CrA, SR 24, SR 9, Sz 68, T Tau, VV CrA. Six additional disk systems
had detectable companions but the M-band secondary was too faint for
any further analysis: Haro 1-4, HD 144432S, HD 144965, SR 9, SX Cha,
VW Cha. In some cases, the two components are blended in our data. HBC
680 consists of a pair of similiar luminosity stars with a separation
of 0\farcs22 \citep{koehler08}. The northern component of SR 24 is a
blend of the B and C components with a separation on 0\farcs081
\citep{correia06}.

The spectra from the two components of a binary can look very
different. AS 205, DoAr 24 E, T Tau and VV CrA all have one component
seen in emission and the other in absorption. In the case of T Tau,
the northern component is viewed face on while the southern component
is seen through its disk \citep{walter03}. In other cases, one star
may be seen through the disk of the other \citep{smith09}.

\subsection{Extended emission}
\label{extent}

Good AO correction allows spatially extended sources to be seen
directly in the CRIRES sample without using
spectroastrometry. \citet{brown12} present a very extended ring in IRS
48, whereas \citet{herczeg11} examine extended emission in two
embedded sources. Extended emission is not common in our sample,
however.  Only four additional sources have directly imaged extended
emission: EC 82, LLN 19, R CrA and T CrA (see Appendix). Limits for
the remaining point-like objects are generally around 3 to 4 AU (see
also \citealt{bast11}).

\subsection{Temporal variability}
\label{time}
 
 Young stars are variable on relatively short timescales in a variety
of diagnostics from optical fluxes, accretion line diagnostics and
even at longer mid-IR fluxes \citep{flaherty12,muzerolle09}. 
Two or more epochs in the same setting were obtained for 11
sources. We approach the search for variability in two ways: firstly,
by simply comparing the line fluxes from the two observations and
searching for significant systematic differences (taking into account
our large error bars), and secondly, by closely examining the line
profiles for differences in shape.
 
Overall, we find on timescales of up to two years a perhaps surprising
lack of variability in most of the line profiles.  A clear exception
is EX Lup, which was undergoing a large outburst during the period of
our programme (see \citealt{goto11} for further details). RU Lup shows a
systematic decrease in line flux by 60\% between April 2007 and April
2008. The decrease could be due to either an increase in the continuum
level or a decrease in the CO flux.  Another example is a
disappearance of the VV CrA outflow absorption from 2007 April to 2007
August. Further details on variability will be available in Pontoppidan et al. (2013, in
prep).

\section{Classification of CO line profiles}
\label{profiles}

\begin{figure}
\includegraphics[angle=180,width=9cm]{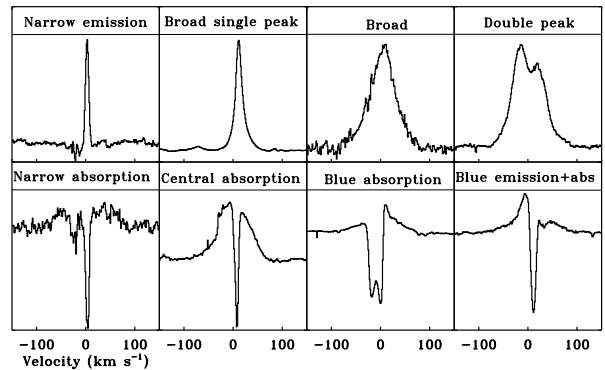} 
\vspace{-0.7cm}
\caption{Gallery of the 8 morphological line profile
types discussed in Section \ref{profiles}. The profiles are calculated
by averaging observed $^{12}$CO $v=1-0$ lines between R(10) and P(32), avoiding
blended lines. \label{fig:profiles}}
\end{figure}

\begin{figure*}
\includegraphics[angle=0,width=20cm]{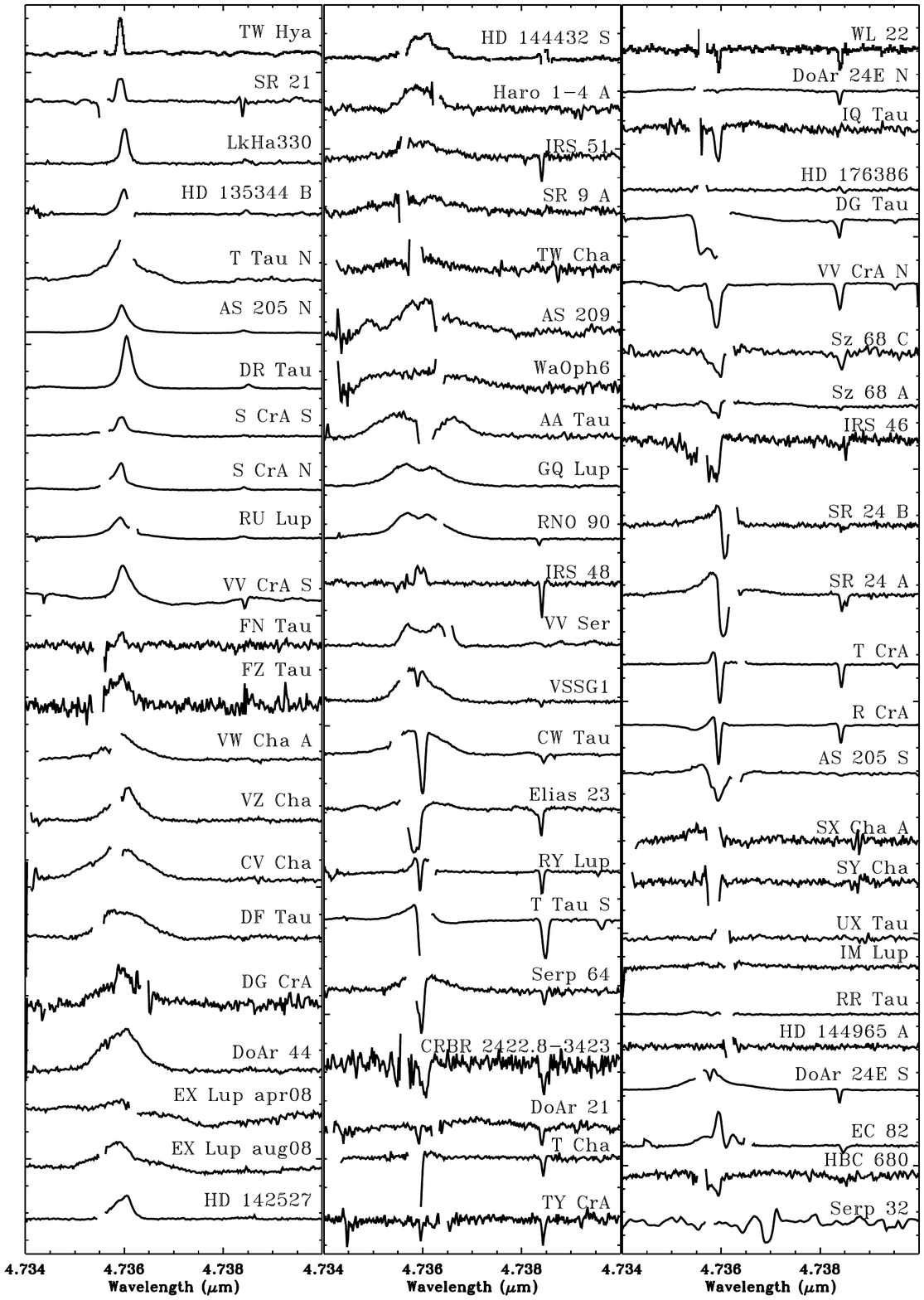} 
\vspace{-2cm}
\caption{A gallery of the P(8) $^{12}$CO line at 4.7359 $\mu$m with
the $^{13}$CO R(3) line seen in several spectra at 4.7383 $\mu$m where
available. Sources are ordered by similarity of line profile, roughly
following the classification scheme in Table \ref{table:cat}. RW Aur,
SO 411 and WX Cha were not covered at these wavelengths and are
missing from the plot. \label{fig:gallery}}
\end{figure*}

Figure \ref{fig:profiles} shows representative examples of the
different CO line profiles of the disk sources. The profiles are often shaped by both
emission and absorption components.  In order to estimate the relative
importance of the various physical components contributing to the
observed CO lines, we divide the sources into groups based on the line
profiles. The classification is summarized in Table
\ref{table:cat}. We discuss physical interpretations of these
different categories in Section \ref{discussion}. Figure
\ref{fig:gallery} shows a close up of the P(8) line at 4.7359 $\mu$m
for all sources. In most cases we do not find any significant
dependence of the emission line shape with rotational quantum number
within the observed range (typically R(1)--P(32)).  S CrA B and DG CrA
are exceptions with a colder narrow component and a broader warm
component in the $^{12}$CO profile \citep[see also][]{bast11}.

\subsection{Emission profiles.} 

\begin{figure*}[ht!]
\vspace{-2cm}
\includegraphics[angle=180,width=17cm]{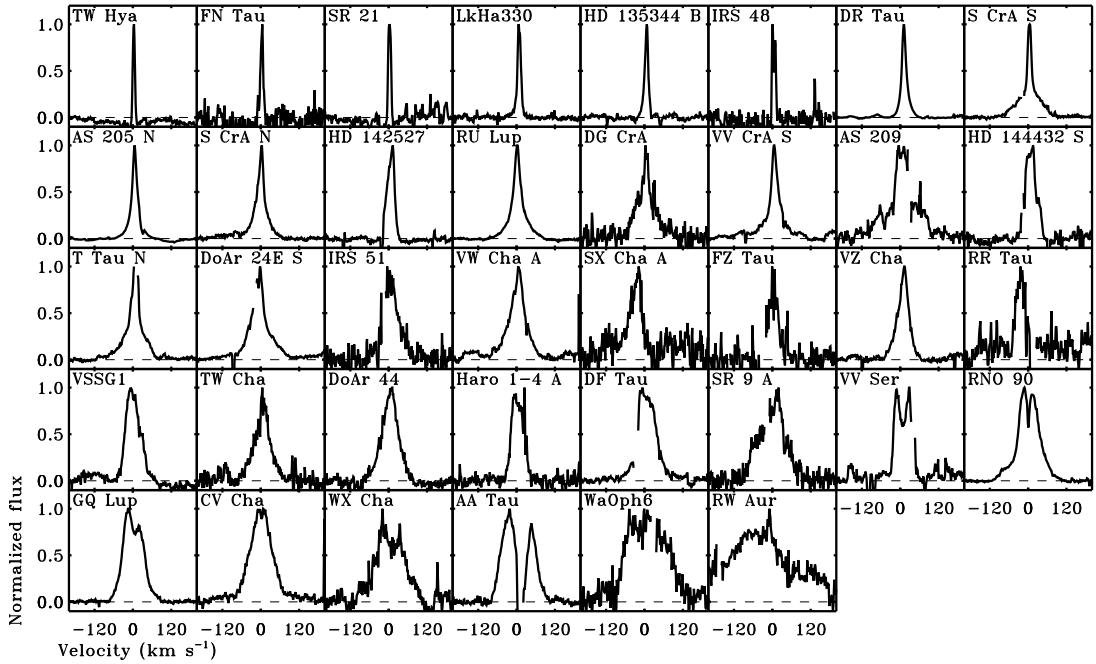}
\vspace{-1.5cm}
\caption{Normalized $^{12}$CO line profiles of the sources with any
emission lines free of absorption, arranged by increasing
FWHM. \label{fig:miniprofiles}}
\end{figure*}

Sources with clean emission lines can be classified based on line
width (Fig.~\ref{fig:miniprofiles}), with four distinct categories:
(1) narrow, (2) broad single peaked, (3) broad, and (4) double
peaked. The narrow category sources have line widths at base of less
that 35 km s$^{-1}$. Disks in this category are known to be face-on
and/or transitional in nature (TW Hya, LkH$\alpha$ 330, HD 135344 B,
IRS 48), with CO likely arising from further out in the disk
\citep{salyk11}. Lines which are slightly broader (FWHM $\sim$ 13--40
km s$^{-1}$) are generally part of the single-peaked line profile
sample of \citet{bast11}. These lines are characterized by a single
central peak with broad wings and have large line-to-continuum
values. The broad category objects retain a single peak but the
contrast between the line width at the top and bottom (the $p$ value
of \citealt{bast11}) decreases. The last group shows a double peaked
structure.

\begin{figure}
\includegraphics[angle=180,width=8cm]{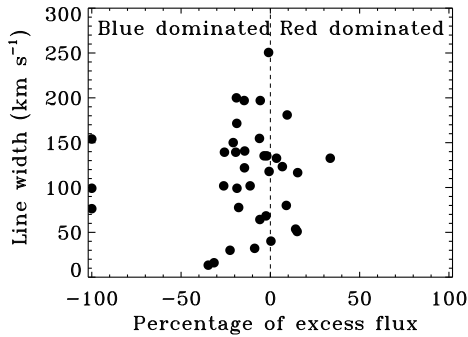} 
\vspace{-0.7cm}
\caption{The percentage of excess blue/red flux compared to the total
  flux contrasted with the line width.  The sources overall tend to
  have excess emission on the blue side of the line. This trend
  affects lines of all widths indicating that it is unlikely to be a
  geometric effect.  \label{fig:asym}}
\end{figure}

Many of the CO emission lines in our sample are slightly
asymmetric. To evaluate the magnitude of any asymmetries, the
difference between the integrated flux in the blue and red sides of
the lines was calculated for the stacked line profiles (see Figure
\ref{fig:asym}). Where possible, the line center was determined from
stellar radial velocity measurements. However, such measurements do not exist
for many of the stars in the sample. In these cases, the line center
was estimated from fitting and checked for large deviations from the
known molecular cloud velocities. The lines show an overall tendency
for excess emission on the blue side, regardless of line profile
category. This trend affects lines of all widths indicating that it is
unlikely to be a geometric effect. Telluric features introduce some
uncertainty into the integrated fluxes but no systematic correlation
with the telluric feature placement is seen. The most dramatic cases
of this shift are those in the category which show strong emission
only on the blue side of the line while absorption dominates the red
side of the line. However, for the majority of the sample, the
magnitude of the effect is generally small, $\sim$10\% compared to the
total flux.

\subsection{Absorption profiles.} 
\begin{figure}
\includegraphics[angle=180,width=9cm]{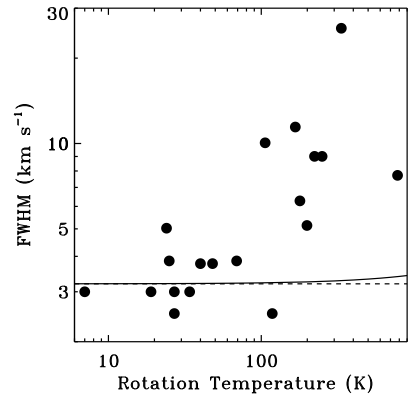} 
\caption{Comparison of FWHM with rotation temperature for absorption
components. The resolution limit is marked by the dashed
line. Unresolved lines generally arise from much colder gas than
broader absorption lines. The solid line marks the width expected from
thermal broadening convolved with the instrumental resolution.
\label{fig:fwhmt}}
\end{figure}

\begin{figure}
\includegraphics[angle=90,width=9cm]{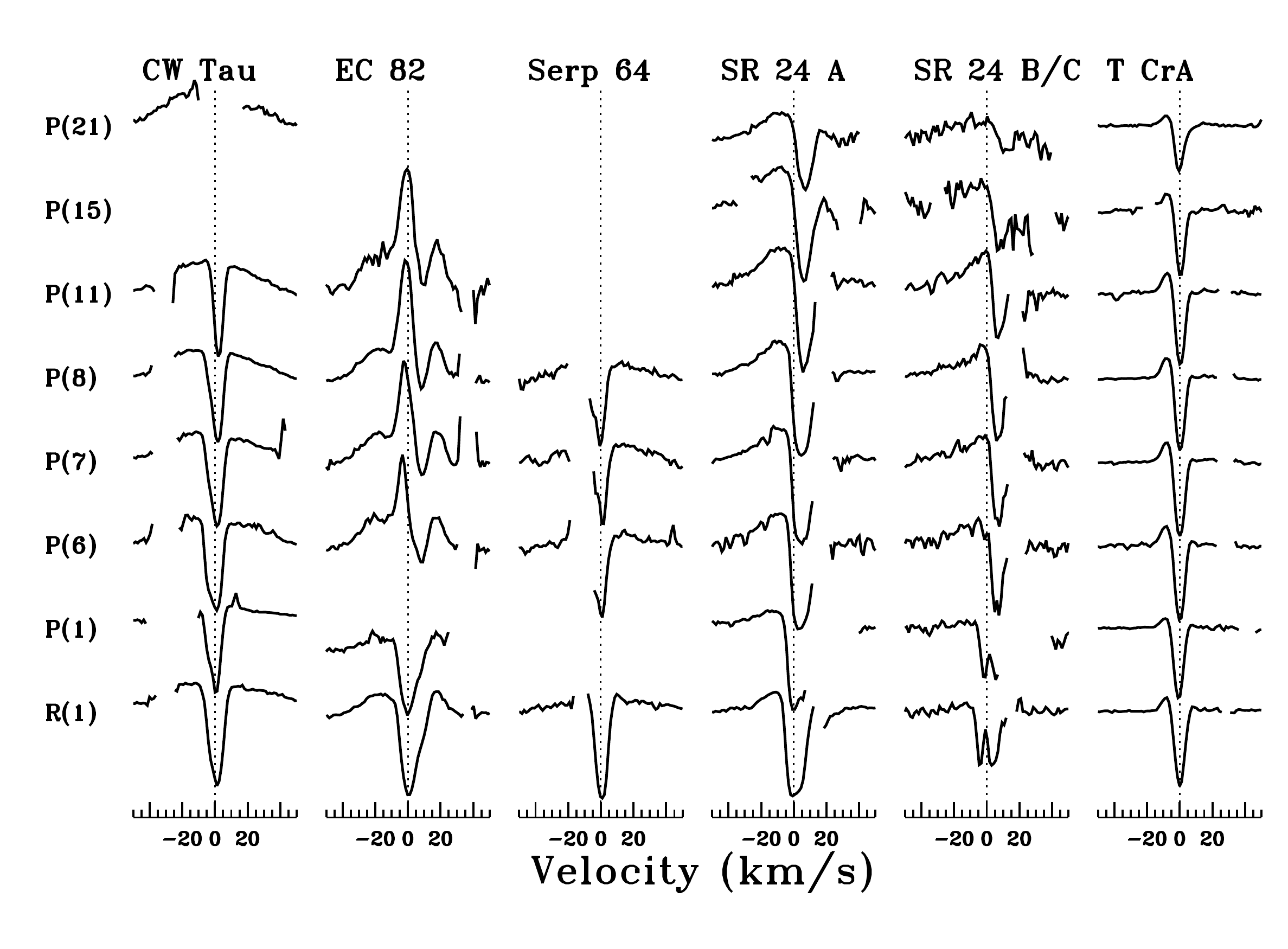} 
\caption{Sources with broad absorption lines. Most also have emission
lines as well, generally stronger on the blue side of the line
(e.g. SR 24, T CrA). The absorption line centers shift redwards at
higher $J$ for CW Tau, SR 24 A, and SR 24 B/C.  \label{fig:windj2}}
\end{figure}

Absorption lines are common throughout the sample, with a detection
rate of 67\% (46 sources). Line widths range from the resolution limit
of $\sim$3 km s$^{-1}$ to tens of km s$^{-1}$.  The energy levels with
detectable absorption also vary from only the lowest $J$ levels to
throughout the observed range, reflecting underlying temperature
differences. In general, hotter gas produces broader lines, at a level
much stronger than expected from thermal broadening
(Fig. \ref{fig:fwhmt}). The lines occur both with and without
emission. We divide the lines based on profile shape and temperature
into four categories: (5) unresolved absorption, (6) broad central
absorption, (7) blue absorption, and (8) absorption with blue
emission. The narrow absorption lines in category 5 are the most
commonly seen absorption lines in our disk sample at 59\% of all
sources that show absorption (27/46). For 21 of these sources, listed
as FA (for foreground absorption) in Table \ref{table:feat}, the lines
are strongest in the lowest $J$ levels and disappear for higher energy
transitions ($J>$8). The remaining 6 sources in this category have
hotter CO absorption and are listed in Table \ref{table:cat}. Warm
central absorption (category 6) is less common and almost always
occurs with resolved line profiles.  The \grapprox 100~K absorption is
generally symmetric and centered at the stellar velocity
(Figure~\ref{fig:windj2}). These warm absorption lines usually have
FWHM of $\sim$ 10 km s$^{-1}$ and occur together with broader emission
lines. Absorption feature velocities generally agree well with stellar
radial velocities (see Table \ref{table:radvel}). However, CW Tau and
SR 24 A and B/C show a shift in the absorption line centers redwards
for higher energy lines.

\begin{figure}
\includegraphics[angle=180,width=10cm]{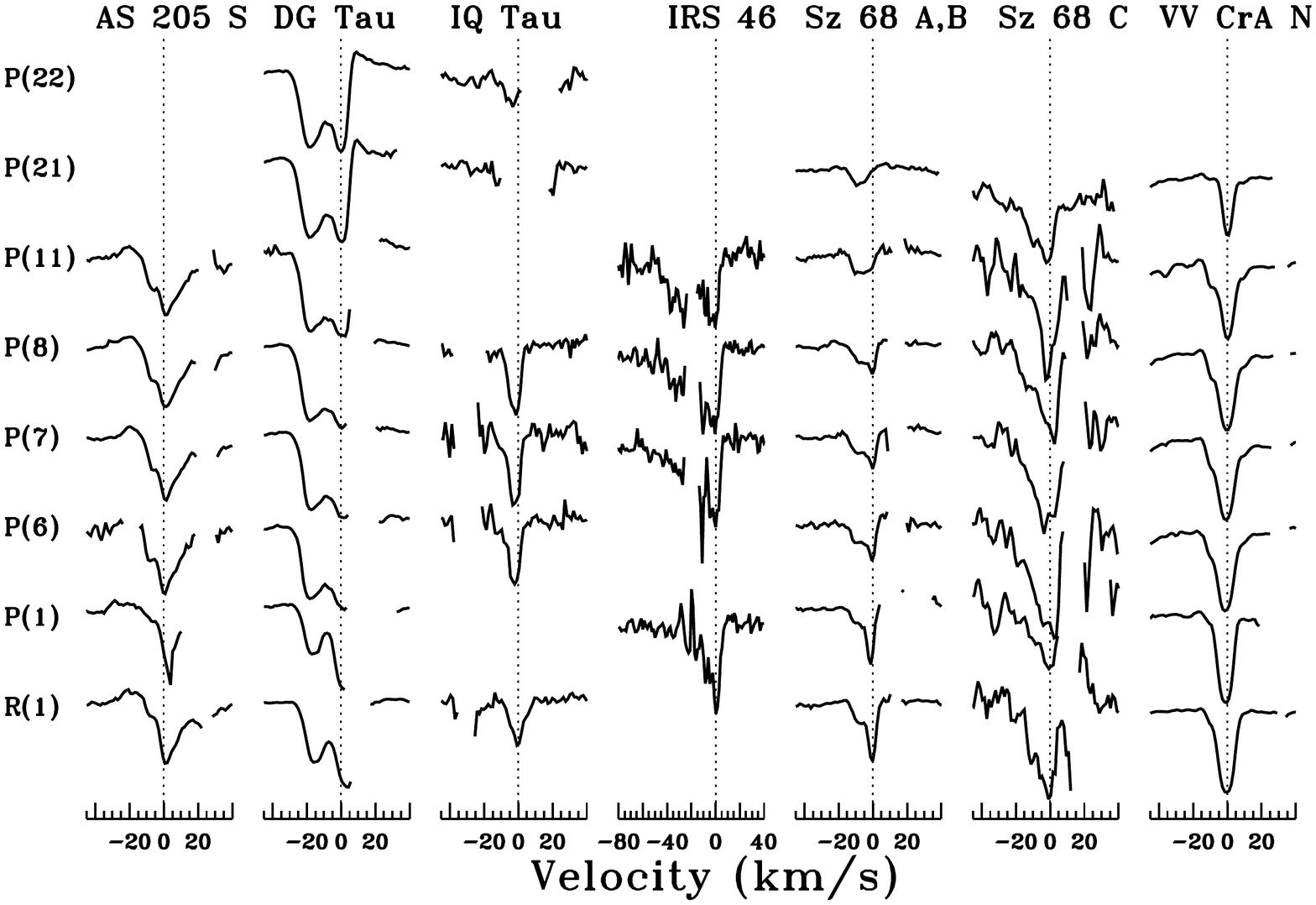} 
\caption{Sources with blue absorption wings up to high $J$, attributed
to outflows. Most show maximum outflow velocities of $<$30 km s$^{-1}$
although IRS 46 reaches velocities of $\sim$60 km
s$^{-1}$. The VV CrA N data are from 2007 April, by 2007 August the
blue absorption shoulder is no longer visible. \label{fig:windj}}
\end{figure}

Six sources belong to category 7 with clearly resolved blue
wings in the absorption lines (Fig.~\ref{fig:windj}). These absorption
features are usually seen throughout the covered $J$ range. The
maximum velocities are less than 50 km s$^{-1}$, except for IRS 46
where the gas reaches $\sim$60 km s$^{-1}$. The final category
(category 8) have lines that are
characterized by emission on the blue side and absorption on the red
(Figure~\ref{fig:windj2}).

\subsection{Absence of CO lines.} Only six of our sources show no indication
of CO lines in either emission or absorption. If CO is present, the
lines are either weak or narrow enough to be lost in the telluric
features ($<$20 km/s). In all cases, the continuum is detected
(generally S/N of $\sim$50), indicating the presence of circumstellar
dust in late-type stars. The CO non-detection from UX Tau A is
particularly surprising as \citet{salyk09} show a clear detection
within this band with NIRSPEC data. However, the telluric feature is
poorly placed covering -5 to 20 km s$^{-1}$, which includes much of
the line center. Also, the lack of CO from the IM Lup disk, which has
been detected and imaged in CO millimeter emission out to several
hundred AU is puzzling \citep{vankempen07,panic09}. HD 176386 is a B9
star \citep{torres06} while HD 144965 A has a spectral type of B3 so
both disks may simply have no CO close to the star due to high
photodissociation rates. In general, very few sources have no CO
visible in the spectrum.

\section{CO excitation}
\label{temp}

\subsection{Rotational temperatures}

The high resolving power of CRIRES allows for the detection of
low-contrast emission lines, leading to an enhanced potential for
measuring the rotational temperatures and accurate column densities of
the weak optically thin isotopologues. In this section, we take
advantage of this to compare the rotational temperatures of $^{12}$CO,
$^{13}$CO $v=1-0$ and $^{12}$CO $v=2-1$ transitions.

A linear fit to the continuum was determined for each line, using
uncontaminated continuum within 300 km s$^{-1}$ of the line center and
50 km s$^{-1}$ greater than the line width. For accurate measurements
of integrated line fluxes, we constructed template line profiles by
stacking isolated lines free of strong telluric absorption. Absorption
components were removed using a Gaussian profile. The integrated line
fluxes were determined by scaling the template to each line (or a
superposition of templates in case of line blends) using
Levenberg-Marquardt least-squares minimization. This method proved
advantageous to direct integration of the line flux: while direct
integration works well for strong lines, it is problematic for
multi-component line shapes, weak lines and lines affected by telluric
absorption. Each line fit was examined by eye and those that clearly
did not recover a meaningful line flux due to strong telluric
residuals or line blending were removed from further analysis.

The fluxes are put into rotational diagrams to determine rotational
temperatures and column densities. Based on an
assumed Boltzmann distribution,
the observed flux $F^{'}_{J}$ is such that:
\begin{equation}
\frac{4 \pi F_J^{'}}{hc \nu_J \Omega g_{J} A_{Ji}} = \frac{N_{\rm tot}}{Q(T_{\rm rot})} e^{-E_J/kT_{\rm rot}},
\end{equation}
where $\Omega$ is the emitting area, $A_{Ji}$ is the Einstein A
coefficient of each level $i$ to which the $J$ level can decay,
$g_{J}$ is the statistical weight of each level (i.e., 2$J$+1),
$N_{\rm tot}$ is the column density, $Q(T_{\rm rot})$ is the partition
function, $E_J$ is the energy of the transition, and $T_{\rm rot}$ is
the rotation temperature. For optically thin, isothermal gas in
Thermodynamic Equilibrium (TE), rotation diagrams produce a straight
line with a slope of inverse temperature and an intercept proportional
to the total mass of gas, but are insensitive to emitting area. In the
optically thick limit, temperature and emitting area are degenerate,
while the rotation diagrams are insensitive to the total column
density. Thus, curved rotation diagrams can be a sign of high line
optical depths varying with $J$, although multiple temperature
components can also cause curvature. When both optically thick and
thin tracers are used, it may be possible to constrain all three
parameters, $N$, $T_{\rm rot}$ and $\Omega$, if both isotopologues
arise from the same region, but, as shown below, this is not the case.

\begin{figure*}
\begin{minipage}{0.5\linewidth}
\hspace{0.4cm}
\includegraphics[angle=90,scale=0.34]{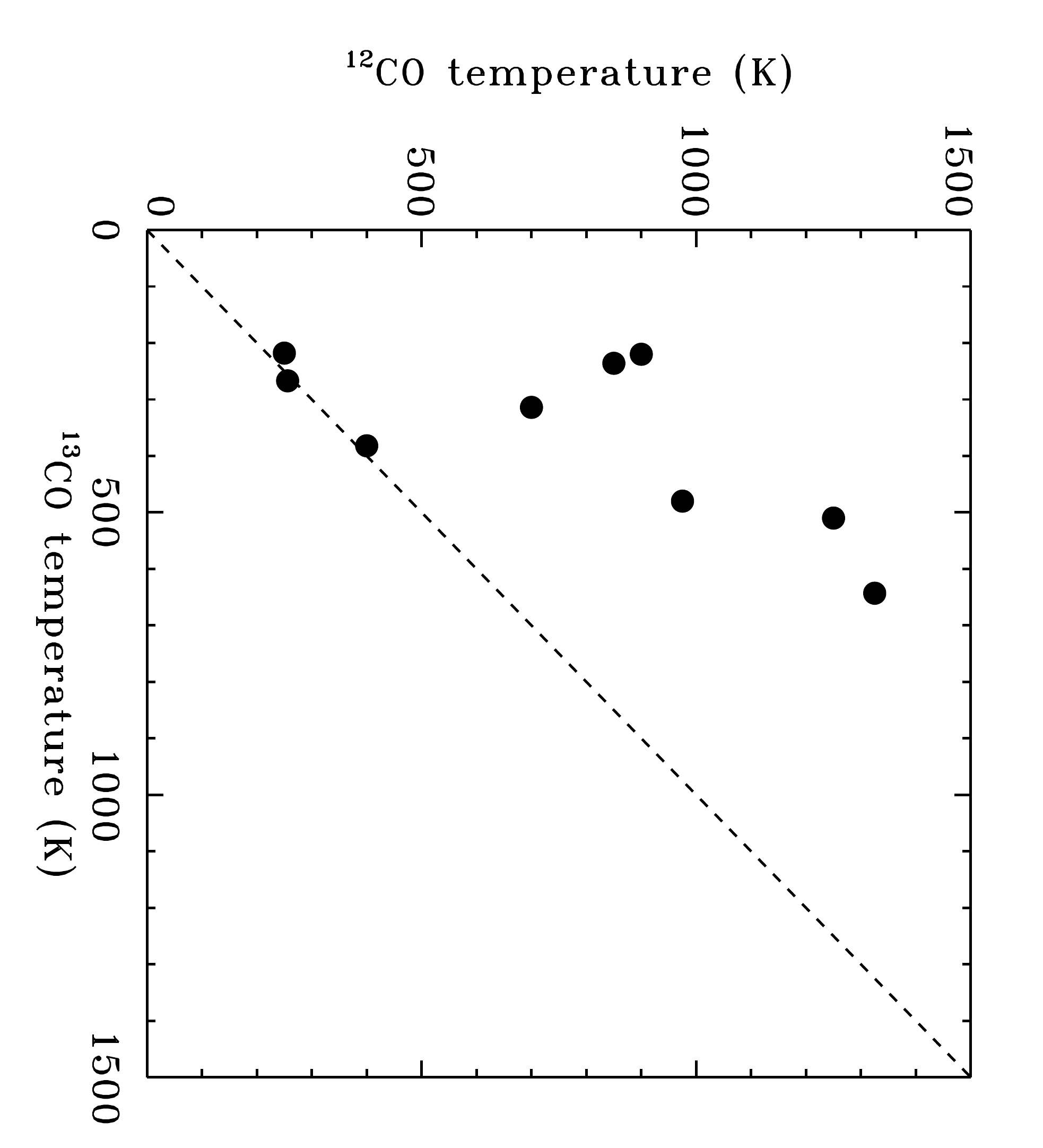} 
\end{minipage}
\begin{minipage}{0.5\linewidth}
\hspace{-1cm}
\includegraphics[angle=180,scale=0.35]{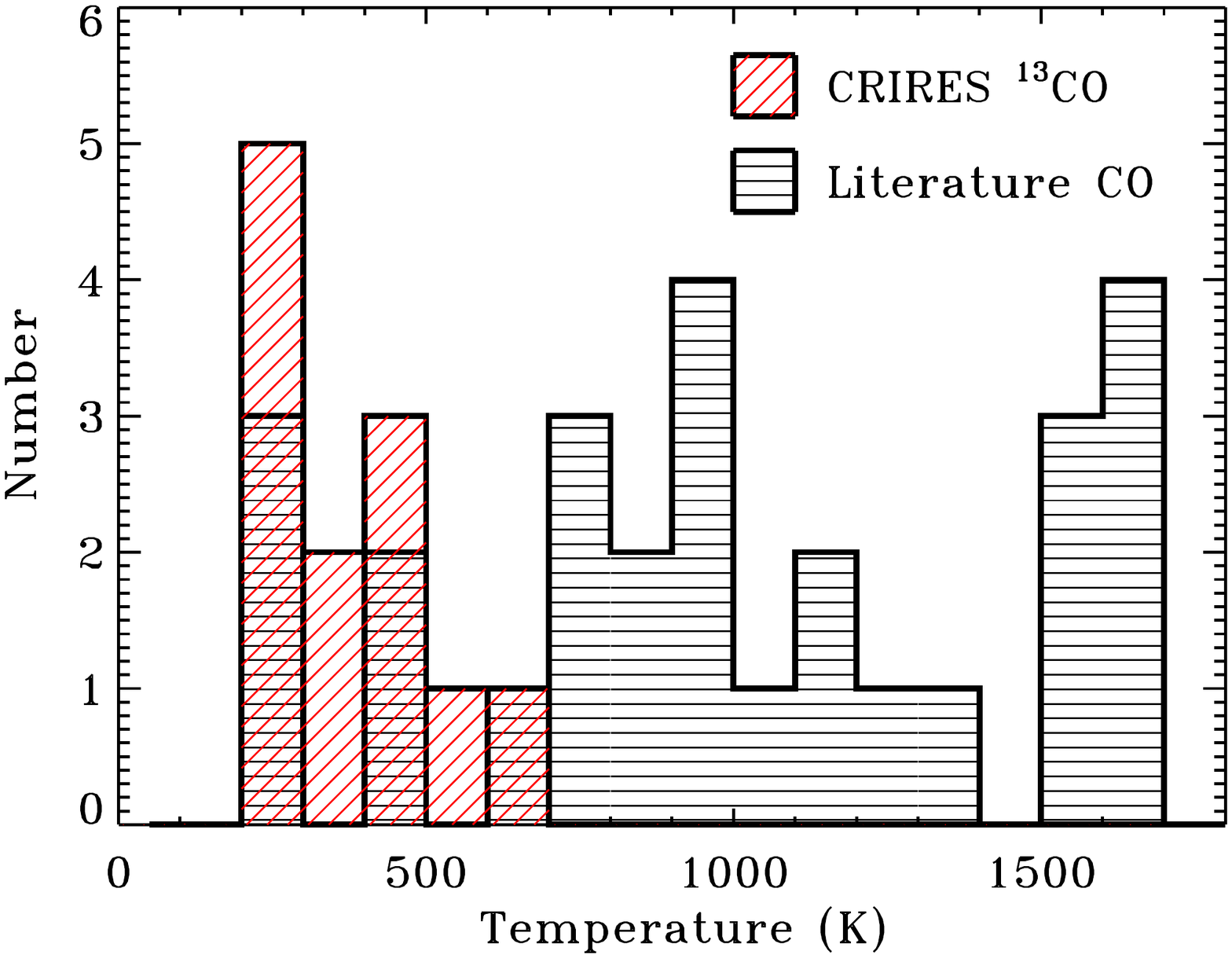} 
\end{minipage}
\caption{(left) $^{13}$CO temperatures derived from optically thin
  fits of disk emission sources compared to temperatures derived from
  fitting the $^{12}$CO lines.  (right) $^{13}$CO temperatures derived
  from optically thin fits compared to CO temperatures reported in the
  literature \citep{salyk09,salyk11}. The three literature sources
  with $^{12}$CO temperatures in the 200--300 K range are all sources
  for which $^{13}$CO was previously detected. \label{fig:tco13}}
\end{figure*}

\subsubsection{$^{13}$CO}
 \label{13corot}

$^{13}$CO is detected in emission from 27 sources and suitable line
fits (minumum of good fits to five clean lines) could be determined
for 12 sources (see Table \ref{table:rotpar}). The sources consist
almost exclusively of a mixture of transitional disks and the broad
single peaked sources of \citet{bast11}. The lines are relatively weak
compared to the $^{12}$CO v=1-0 and v=2-1 lines with fluxes down by
factors of $\sim$10 and $\sim$4, respectively.

\begin{figure}
\hspace{-0.5cm}
\includegraphics[angle=180,scale=0.35]{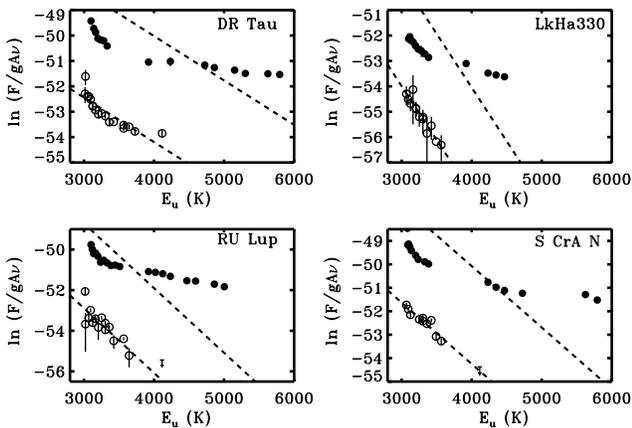} 
\vspace{-.7cm}
\caption{Rotational diagrams of $^{12}$CO (filled circles) and
$^{13}$CO (open circles). A linear fit was made to the $^{13}$CO,
lower dashed line, and was scaled up by an isotope ratio of 65 to
compare to the expected $^{12}$CO if optically thin. The high $J$
lines are clearly significantly underpredicted and require a hotter
gas component to fit. The lower $J$ lines are likely underpredicted
due to optical depth effects. \label{fig:rotplotexample}}
\end{figure}

The $^{13}$CO fluxes are distributed along a straight line within the
errors in the rotational diagrams -- consistent with single
temperature optically thin emission (see Fig. \ref{fig:rotplotexample}
for examples). Possible exceptions include AS 205 N and DR Tau where
some curvature may be apparent in the highest and lowest $J$
levels. For the case of DR Tau, determination of the
$^{13}$CO/C$^{18}$O flux ratios indicates optical depths of the
$^{13}$CO lines of only 0.3$\pm$0.2 \citep{bast11}. The emitting radius at
which the $^{13}$CO is optically thin is listed in Table
\ref{table:rotpar}, assuming a circular slab model with uniform
temperature and mass.

Optically thin linear fits uniformly reveal temperatures between 200
and 600 K with the majority in the 200--400 K range
(Fig. \ref{fig:tco13}), consistent with the analysis of \citet{bast11}
for a subset of the sources. This is significantly lower than has been
previously reported for CO fundamental band temperatures, generally
determined from optically thick $^{12}$CO, which may trace different
gas. Literature values are included in Fig.~\ref{fig:tco13}. In the
optically thin case, the range of energies covered should have no
effect on the linear fit. For the sources with detectable $^{13}$CO,
lines are detected generally up to P/R(15) (E$_u$=3656 K/3826 K) and
up to R(23) (E$_u$=4722 K) in AS 205 N.

The fact that the $^{13}$CO lines appear optically thin makes it possible
to directly correlate the y-intercept and total $^{13}$CO gas
mass. Table \ref{table:rotpar} includes the inferred CO column
densities assuming an emitting area of 100 AU$^2$ ($\sim$5 AU radius)
and $^{12}$CO/$^{13}$CO=65. The exact area for which the $^{13}$CO
emission becomes optically thin depends on the specific sources and
also on the assumed local line profile, taken here to have a total
broadening of 2 km s$^{-1}$, the sound speed of H$_2$ at 1000 K, in
agreement with \citet{herczeg11} and \citet{salyk11}. Larger
broadening parameters result in more optically thin gas while smaller
broadening parameters have the opposite effect.

For a few sources, the inferred CO column densities are close to
10$^{20}$ cm$^{-2}$ (SR 21, LkH$\alpha$ 330) within the assumed 5 AU
radius, which would imply values of $N_H>10^{24}$ cm$^{-2}$ for a
standard conversion factor of CO/H$_2$=10$^{-4}$. These column
desities are very large compared to the surface column densities of
$N_H$ $\sim$ 10$^{21}$-10$^{23}$ cm$^{-2}$ that are expected down to
the layer where the 5 $\mu$m continuum becomes optically thick (e.g.,
\citealt{aikawa02,gorti08,woitke09}). Another way to look at this
problem is to derive total gas masses from the $^{13}$CO data. Values
computed under the optically thin assumption (but independent from any
assumed emitting area) are included in Table \ref{table:rotpar}. For
some sources, the masses are $>$10$^{-4}$ M$_\odot$, implying that a
significant fraction of the disk mass would be contained in just the
surface layers of the inner few AU of the disk, under these
assumptions. One possible solution to this conundrum of the large
inferred column densities and masses is that the dust grains are
settled to the midplane so that the gas/dust ratio is much larger than
the standard value, which allows us to look deeper into the disk at 5
$\mu$m. A correlation between mid-IR SED slopes, a tracer of settling, and CO
equivalent widths indicates that dust settling may increase the
observable CO (see \S \ref{trans}). Another possibility is that UV or
IR radiative excitation and resonant scattering contributes to the
$^{13}$CO lines fluxes (see \S \ref{12corot}), in which case the
inferred column densities and masses are upper limits.

The results from optically thin fits of the $^{13}$CO gas, detected
here with a much larger frequency than in previous studies, imply a
different location for at least some of the CO than indicated from
studies of $^{12}$CO. The low temperatures and large emitting areas
suggest that much of the $^{13}$CO arises from either larger disk radii
or deeper into the disk.

\subsubsection{$^{12}$CO}
\label{12corot}

The $^{12}$CO lines throughout our sample show curvature in the
rotation diagram which has been attributed to optical depth effects
\citep{blake04}. They look qualitatively similar to previously published
rotation diagrams from disks \citep{salyk09,najita03}. In cases where
the sources had been previously observed at lower spectral resolution
(e.g. NIRSPEC, \citealt{salyk09}), benchmark tests ensured that our
CRIRES rotation diagrams quantitatively agree with previous
observations.

\begin{figure}
\hspace{-1.5cm}
\includegraphics[angle=180,scale=0.38]{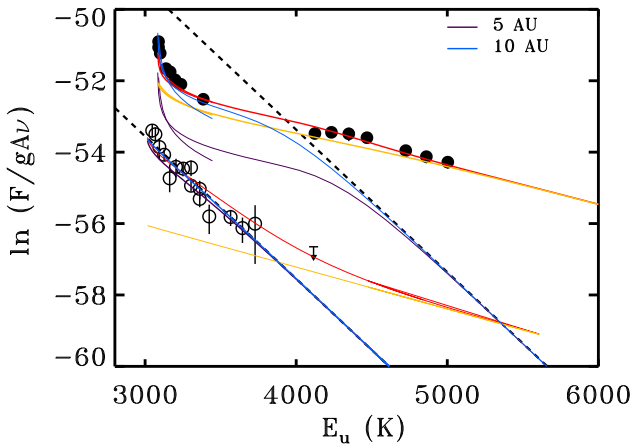} 
\vspace{-1.cm}
\caption{The rotational diagram for $^{12}$CO (filled circles) and
  $^{13}$CO (open circles) from HD 135344 B. The blue and purple lines
  are isothermal slab models based on the rotational temperature and
  mass determined from the optically thin fits to the $^{13}$CO (250
  K, 8 $10^{16}$ cm$^{-2}$) for two emitting radii. The yellow line is a
  higher temperature model (850 K, $N$($^{12}$CO)=10$^{17.5}$
  cm$^{-2}$ column density) to fit the high $J$ $^{12}$CO and the
  red line is this model combined with the 5 AU radius $^{13}$CO
  model. \label{fig:hd135344example}}
\end{figure}

In theory, optically thin $^{13}$CO can constrain the degeneracies in
modeling the optically thick $^{12}$CO emission, assuming that the
emission arises from the same gas. However, the low $^{13}$CO
temperatures are incapable of explaining the observed $^{12}$CO high
$J$ lines in most cases, indicating the presence of a warm component
seen in the $^{12}$CO (Fig.~\ref{fig:rotplotexample}). We examine HD
135344 B in detail to determine whether a two temperature model is
capable of fitting this behavior (Fig.~\ref{fig:hd135344example}). We
use a simple slab model which calculates the optical depth in each
transition (see e.g., \citealt{salyk09, brown08t}). The $^{13}$CO data
are best fitted with $T_{\rm rot}$=250 K, $N$($^{13}$CO)= $8\times
10^{16}$ cm$^{-2}$ and an emitting region with a radius of 5 AU to
ensure that the $^{13}$CO emission is optically thin. Using such a
large radius and multiplying the column density by 65, the low
$^{13}$CO temperature is incapable of explaining the high $J$
P(20--27) $^{12}$CO lines which are clearly detected (formal errors
are smaller than the points) (Fig.~\ref{fig:rotplotexample}). A
smaller, hotter (850 K) region of $^{12}$CO emission needs to be added
to fit the $^{12}$CO data. We conclude that a two temperature model
(black line) is capable of fitting the data as the expected $^{13}$CO
fluxes from the warm $^{12}$CO component are below our detection
threshholds if close to optically thin.

Previous studies have also had difficulty reconciling the $^{12}$CO
and $^{13}$CO line fluxes for some sources. \citet{salyk09} found a
similarly low temperature of 250--350 K for $^{13}$CO in SR 21. This
is compatible with the $^{12}$CO lines due to a lack of detected high
$J$ lines for that source. However, the required mass of 6x10$^{25}$ g
of CO (gas mass of 3 x 10$^{-4}$ M$_\odot$) and emitting area of 78
AU$^2$ are large, although a gas location at radii as large as 7 AU is
confirmed by spectro-astrometry \citep{pontoppidan11}. \citet{blake04}
were unable to explain the high relative fluxes from $^{13}$CO in AB
Aurigae and \citet{brittain09} found anomalously high $^{13}$CO line
fluxes from the disk around A-type star HD 100546, with an apparent
ratio of $^{12}$CO/$^{13}$CO of 4 rather than the interstellar ratio
of $\sim$65. In both cases, non-thermal processes were invoked to
explain the discrepancy. \citet{blake04} find that resonant scattering
of a large fraction of the IR continuum photons can reproduce the AB Aur
fluxes. \citet{brittain09}, on the other hand, suggest that UV pumping
is the primary excitation mechanism and the overabundance of $^{13}$CO
is the result of additional sources of opacity in the disk.

IR resonant scattering (IR pumping) of $^{13}$CO primarily enhances
the low energy $J$ lines, reflecting the thermal distribution of the
lower vibrational level population in the excited state. Optical depth
effects may reduce the strength of this component in the $^{12}$CO
lines. Excitation via IR pumping is expected to be particularly
effective for disk wind sources and transitional disks, explaining why
we see $^{13}$CO mostly from these sources. Winds lift material above
the disk where the CO molecules are more exposed to radiation. In
transitional disks, the lower 5 $\mu$m continuum optical depth may
result in less shielding from the dust and thus enhanced
absorption. Resonant scattering is primarily dependant on the received
flux and would imply column densities several orders of magnitude
smaller to reproduce our observed line fluxes.

\subsection{Vibrational excitation}
\label{vibration}

Vibrational lines from CO v=2--1, v=3--2 and v=4--3 are seen from a
subset of sources. CO v=2--1 lines are the
strongest and thus most commonly detected.  These higher energy lines
are likely pumped by UV flux as the upper level temperatures of
$\sim$7000 K are too high for collisional excitation by the
500--1000 K gas seen in the $^{12}$CO lines. 

A vibrational flux ratio was calculated for each source based on the
v=2--1 flux divided by the v=1--0 flux for the equivalent rotational
line (Table \ref{table:vib}. UV pumping does not induce large changes in population
distribution in the different rotational states since the selection
rule only allows quantum number changes of $\Delta J$ $\pm$1. Thus, the
shape of the v=2-1 rotational diagram is similar to the v=1-0 lines
and the offset between the two reflects the difference in upper state
population. Measuring the vibrational flux ratios of equivalent $J$
rotation lines thus traces the population difference. Typically five
flux ratios were found and the median was taken for the final value to
remove outliers. The vibrational flux ratios range from 0.05 to 0.5
and upper limits were calculated for high S/N spectra where the CO
v=2-1 lines were not detected.

\begin{figure}
\hspace{-1cm}
\includegraphics[angle=180,scale=0.37]{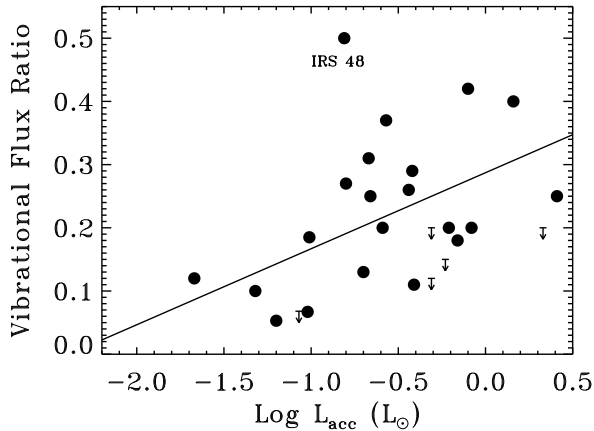} 
\vspace{-.7cm}
\caption{Flux ratio in vibrationally excited levels derived from the
  2--1/1--0 flux ratios versus accretion luminosity determined from
  simultaneous Pf$\beta$ emission. The accretion
  luminosity in the UV drives UV fluorescent pumping which is
  reflected in increased CO v=2-1 line fluxes. IRS 48 is an outlier in
  the plot likely due to its higher spectral type of A0 resulting in
  photospheric UV flux rather than accretion luminosity pumping the
  v=2--1 lines, and is therefore excluded from the linear
  fit. \label{fig:vibpfb}}
\end{figure}

The observed vibrational flux ratio generally increases with accretion
luminosity (Fig. \ref{fig:vibpfb}). The accretion luminosity is
determined based on simultaneous measurements of the Pf$\beta$ line
which is covered serendipitously in the CRIRES spectrum using the
conversion to accretion luminosity as given in \citet{salyk12}. IRS 48
is an outlier in the plot, probably due to its higher spectral type of
A0 resulting in a photospheric UV flux rather than accretion
luminosity pumping the v=2--1 lines and the 30 AU gas location
\citep{brown12} reducing collisional excitation, and is therefore
excluded from the linear fit.

We modeled the effects of a stellar UV field on the CO emission in a
UV excitation model similar to \citet{brittain07}. UV fluorescent
pumping of the CO vibrational lines in the fundamental band involves
excitation by UV photons from the $X^1\Sigma ^+$ ground electronic
state to the $A^1 \Pi$ excited electronic state. The decay from the
excited electronic state to the ground state can increase the
vibrational energy of the molecule by populating higher vibrational
levels in the ground electronic state than would be expected from
purely thermal excitation. We assume that the system is in a steady
state and balance pumping into the $A^1 \Pi$ state from the incident
UV radiation field with spontaneous emission out of the $A^1 \Pi$
state. Spontaneous emission within the $X^1\Sigma ^+$ ground state
vibrational levels is also included. We expanded the vibrational
states included in the model to 35 $X^1\Sigma ^+$ ground levels and 25
$A^1 \Pi$ levels; rotational levels are not explicitly taken into
account.  For this purpose, we computed the oscillator strengths and
Einstein-$A$ coefficients between individual A--X vibrational levels
using the Rydberg-Klein-Rees (RKR) program of
\citet{leroy04}. Accurate potential curves based on the spectroscopic
data of \citet{lefloch91} for the X state and \citet{field72} for the A
state were used together with the A--X transition dipole moment of
\citet{gilijamse07}.

One of the main reasons for the expansion to higher vibrational states
was to probe the effects of Lyman $\alpha$ emission (1216 \AA), which
dominates the far ultraviolet emission from classical T Tauri stars
\citep{schindhelm12} and can pump the v=14--0 line of CO
\citep{france11}. Much of the observed Lyman $\alpha$ is absorbed by
the ISM before reaching Earth so the flux seen by the disk must be
reconstructed from fluoresced H$_2$ lines \citep{herczeg02}. However, even with
reconstructed Ly$\alpha$ line profiles, the effects of Lyman $\alpha$
are only noticeable at high vibrational levels but are marginal at the
v=2--1 level that is seen in our sample. Some of this result can be
explained by the fact that the oscillator strength for the 14--0 band
is more than three orders of magnitude smaller than those for the
1--0, 2--0 and 3--0 bands which dominate the UV pumping. The other
factor is that the downward relaxation in the X band is much faster
than the UV excitation, spreading the excess flux through all the
lower energy vibrational levels.

\begin{figure}
\hspace{-1cm}
\includegraphics[angle=180,scale=0.37]{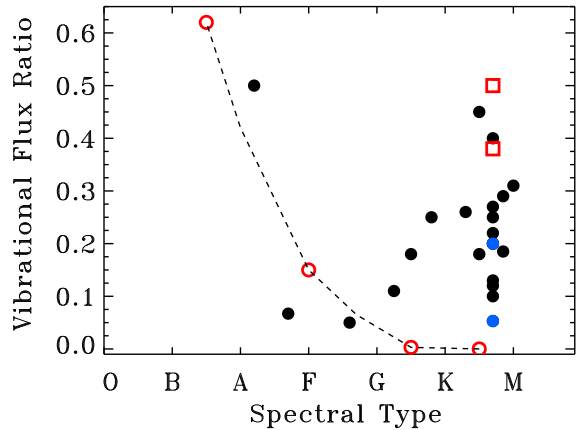} 
\vspace{-.7cm}
\caption{Vibrational flux ratio (v=2-1/v=1-0) versus spectral type
  compared to the pure fluorescence model for vibrational emission
  (dashed line). Filled black circles are the observed values, open
  circles are model results for blackbodies at 4000, 6000, 8000 and
  10000 K and squares are model results using the observed UV spectra
  of DR Tau and TW Hya while the blue dots are the observed
  values. The observed increased UV flux over blackbody is clearly
  essential at late spectral types. \label{fig:vibsp}}
\end{figure}

Fig.~\ref{fig:vibsp} presents our model results for the vibrational
population fraction of the v=2/v=1 vibrational levels for a pure UV
fluorescent pumping model using different blackbody radiation
fields. The vibrational population fraction decreases with stellar
effective temperature due to the smaller amount of UV available to
pump the higher vibrational levels of the A state. The observed
vibrational flux ratios are included in Fig.~\ref{fig:vibsp} and
generally show much greater values than would be expected from simple
blackbodies, especially for the later spectral types.  This is in
agreement with the correlation found with accretion luminosity in
Fig.~\ref{fig:vibpfb}. For the cases of TW Hya and DR Tau, both
strongly accreting objects, we have also run models using the observed
stellar spectra including the enhanced UV emission
\citep{herczeg02,yang12}, resulting in vibrational flux ratios due to
pure UV pumping of $\sim$0.4--0.5. The observed flux ratios for these
sources are 0.05 and 0.2, respectively, indicated with blue dots in
Figure \ref{fig:vibsp}. The difference between models and observations
is likely due to excess population in the v=1 levels from IR pumping
or collisional excition, which results in a lower vibration flux
ratio. More detailed modeling and observations of higher energy
vibrational lines are needed to disentangle these two excitation
processes (e.g. \citealt{brittain07}, \citealt{brown12},
\citealt{bast11}). However, the ubiquity and strength of v=2--1 lines
in the sample appear consistent with UV fluorescent pumping primarily
from accretion luminosity.

\section{Discussion}
\label{discussion}

\subsection{Where is the CO gas located?}

The presence of multiple line profile components and temperatures
suggests that separate, physically unrelated gas reservoirs are
contributing to the spectra. Such reservoirs could include 1)
superheated gas from the inner ($\lesssim 1\,$AU) disk surface
\citep{najita03,blake04}, 2) emission/absorption from an extended disk
wind or other outflow activity \citep{pontoppidan11,herczeg11}, 3)
absorption from the outer disk in an inclined system \citep{rettig06},
4) absorption from another disk around a binary companion
\citep{smith09}, 5) absorption from remnant envelope
\citep{boogert02b,herczeg11}, and 6) absorption from cold foreground
molecular gas, entirely unrelated to the young star
\citep{boogert02a}.

{\it Inner disk:} The surface layers of disks have long been thought to be
one of the primary producers of fundamental CO emission
(e.g. \citealt{najita03}). The gas in the disk atmosphere has regions
with the 100-1000 K temperature range needed to thermally excite the v=1
level of CO and produce the abundant strong CO v=1-0 lines. The line
widths at the base are consistent with expected Keplerian rotation in
the inner regions of disks (\citealt{salyk11}, \S \ref{rin}).

Disks with a pure Keplerian velocity field (no radial motions and $V_{\rm
phi}\propto R^{-1/2}$) produce a characteristic line profile based on
inclination. A double peak arises from inclined disks due to the
radial decrease in temperature and resulting decrease in low-velocity
CO emission from the cold outer disk causing a central dip.  The
double peaked line profiles (category 4) are good examples of this
classic Keplerian profile and can be seen in the cases of GQ Lup and
RNO 90. The central dip disappears at low inclinations due to the lack
of velocity contrast between the inner and outer disk resulting in
narrow single peaked lines (category 1), such as TW Hya and HD 135344
B. In these cases, the two peaks can still be separated using
spectro-astrometry \citep{pontoppidan08}.

The list of sources with clear classical Keplerian line profiles
consistent with these models is given in Table \ref{table:cat} and is
surprisingly small. As can be seen in Figure \ref{fig:miniprofiles},
the majority of our line profiles are single peaked: our percentage of
double peaked sources is only 10\% (7/69) with 6 additional sources
(category 6) which may be self-absorbed highly-inclined disks. This
lack of double peaked profiles has been noted in even the earliest CO
surveys \citep{najita03}. However, the constraints on Keplerian models
are greater with the improved spectral and spatial resolution and much
large sample. If the inclinations of the full CRIRES sample are
randomly distributed and all line profiles are taken to be Keplerian,
these low statistics would imply an inclination of $>$70$^\circ$
before a double peaked line profile would be resolved. For a 1
M$_\odot$ star at the CRIRES spectral resolution, this would imply
some CO emission from $\sim$10 AU or greater, which is not seen in the
2-D spectral traces. For less massive stars, limits on the outer
extent are less extreme (e.g. 0.5 M$_\odot$ corresponds to $\sim$5
AU). However, no trend is seen with spectral type and uniformly low
masses are excluded by the range of measured stellar types. We
conclude that the warm surface gas in the inner disks in the majority
of our sources has kinematics that are inconsistent with only
Keplerian rotation.

{\it Disk winds:} Many of the sources with the strongest line to continuum
ratios are single peaked with broad wings (category 2), including AS
205 A, DR Tau and RU Lup. As discussed in \citet{bast11}, the
combination of a narrow single peak and broad wings is incompatible
with pure Keplerian profiles without high temperatures in the outer
disk. The lack of spatial extent rules out this model.  Indeed,
\citet{pontoppidan11} use spectroastrometry to show that the spatial
distribution of the emission is far more compact and asymmetric than
can be explained by such a Keplerian disk. They propose a slow moving
disk wind as the source of the central emission, but with the broad
line wings still due to Keplerian rotation imparted from the launching
region of the disk. The line profiles appear symmetric and centered
close to the source radial velocity (within $\sim$ 5 km s$^{-1}$, see
Table \ref{table:radvel}) constraining the physical characteristics of
the wind. Sources with these profiles show higher vibrational CO lines
indicating fluorescent excitation by the UV radiation produced by the
accretion shocks.

The largest category of emission lines is category 3 consisting of the
broad profiles ($>$ 200 km s$^{-1}$). These were not included in the
\citet{bast11} sample due to lower contrast between the line width at
10\% compared to 90\% of peak height and/or lower line to continuum
ratios. However, based on the prevalence of these single peaked lines,
they may have the same origin in a wind.

{\it Outer disk:} Category 6 (emission with broad central absorption)
sources have strong absorption lines above P(10) (4.7545 $\mu$m,
$E_{\rm upper}$=3330 K) and likely arise from the surface layers of the
outer disk. These absorption lines commonly occur in conjunction with
disk emission lines although the emission can be difficult to
categorize due to the absorption in the line center. Examples of warm
CO absorption occur in CW Tau, SR 24, and T Tau S.  Temperatures range
from 100 to 300 K (\S~\ref{temp}). At near edge-on inclinations, the
outer disk atmosphere could absorb CO as has been suggested in the
case of T Tau S \citep{rettig06}.  In the case of binaries, it is
possible that the absorption can arise from looking at the star
through the companion's disk as may be the case in the VV CrA system
where VV CrA N would be seen through the VV CrA S disk
\citep{smith09}.

{\it Outflows:} Molecular outflows and winds moving towards us in our
line of sight result in absorption lines with broad blue-shifted wings
seen in category 7 sources. The gas producing the lines is warm (up to
1000~K) likely originating from currently shocked gas rather than
entrained outflow gas. The lines are blue shifted due to the disk
blocking the red-shifted component and its position behind the
illuminating star. The wings are not always smooth indicating clumps
in the wind as in the case of DG Tau (shown in Figure
\ref{fig:profiles}). LLN 19 has episodic outflows \citep{thi10}. VV
CrA N also has outflow components which disappear between April and
August 2007. The profiles are similar to those found for the embedded
sources in this CRIRES sample highlighted in \citet{herczeg11}.

The category 8 sources with blue emission and red absorption are
probably from a combination of outflow in emission and residual
envelope in absorption.  Both the emission and absorption have similar
warm temperatures of a few hundred K. These lines are the most
asymmetric of the categories. The warm absorption is close to the
cloud velocity and the emission could then be from an outflow. In the
case of T CrA, which is known to drive a jet \citep{wang04}, the
emission component is spatially extended definitively indicating an
origin in an outflow.

{\it Foreground molecular clouds:} Many of the CRIRES sources are
towards high extinction regions of nearby star forming clouds. CO gas
should be present in the foreground molecular cloud with low turbulent
broadening and cold temperatures ($\sim$10-100 K). We attribute the
Category 5 narrow unresolved absorption lines which appear only at low
$J$ levels to this foreground material. This combination of cold
temperatures and low densities is confirmed by rotational diagrams
(see \S~\ref{temp}). These lines are very common throughout the sample
with a detection rate of 38\%. The high spectral resolution of CRIRES
is a strong advantage for complex line profiles because the foreground
lines can be readily separated from emission components and other
dips. The binary pairs, such as DoAr 24E also show the clumpiness of
foreground material with even cold absorption components varying
between binary pairs on spatial scales of a few 100 AU.

\subsection{Temperature structure of the inner disk region}

\label{therm}

Typical temperatures of the $^{13}$CO lines appear lower ($<$500 K)
than has generally been derived from likely optically thick $^{12}$CO
lines (500-1000 K). $^{13}$CO is close to being optically thin and
thus may be able to probe deeper into the disk's vertical structure
than $^{12}$CO. The lower temperatures indicate that the $^{13}$CO
lines are dominated by emission from either further out or deeper
vertically in the disk. As discussed in \S \ref{13corot} and
\ref{12corot}, both UV and IR pumping may contribute to the line
fluxes, but these processes should not change the inferred rotational
temperatures significantly.

\begin{figure}[]
\includegraphics[angle=180,scale=0.37]{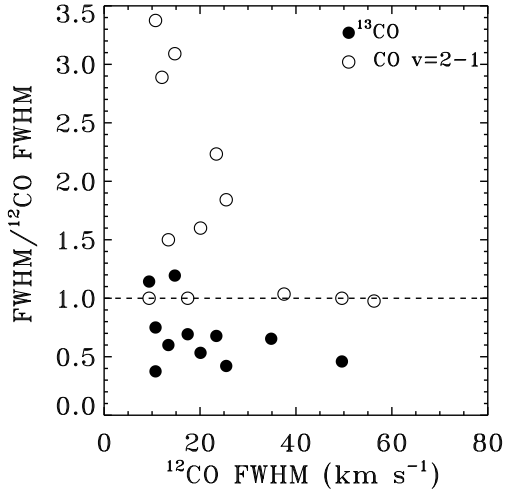}
\caption{$^{13}$CO FWHM/$^{12}$CO FWHM (dots) and $^{12}$CO v=2-1
  FWHM/$^{12}$CO v=1--0 FWHM (open circles) versus FWHM of the v=1-0
  lines. The $^{12}$CO v=2-1 line widths range from similar to the
  v=1--0 lines to much broader. The $^{13}$CO line widths are
  generally down by a factor of $\sim$2. The two sources with similar
  FWHM are SR 21, where all emission lines appear to arise from the
  same region, and S CrA S, which has two distinct components in the
  $^{12}$CO with the cooler $^{12}$CO component having the same line
  shape as the $^{13}$CO. \label{fig:fwhmiso}}
\end{figure}

Line widths systematically differ between $^{12}$CO v=1--0, v=2--1 and
$^{13}$CO, indicating that the lines do not arise from identical
locations in the disk.  For each disk, the CO v=2--1 line profiles are
generally broader than v=1--0 while the $^{13}$CO lines are generally
narrower than $^{12}$CO v=1-0 (Fig.~\ref{fig:fwhmiso}). This trend was noted
by \citet{bast11} in their broad single peaked sources but appears
more generally throughout the sample. The change in average width
between the $^{12}$CO and $^{13}$CO lines corresponds to a change in
radius by a factor of $\sim$4, assuming Keplerian rotation. If the
entire temperature difference between the isotopes is assumed to come
from a radial difference, this would imply a temperature profile $T
\propto R^p$ where p is between -0.5 and -1.5.

We can also compare our temperature and thermal column density
measurements with current chemical models. Many thermo-chemical models
of inner disk regions exist and they typically have gas kinetic
temperatures of 1000 K or larger in the layer where CO becomes
abundant (e.g., \citealt{gorti08, gorti11, glassgold09, woitke09,
najita11, walsh12, bruderer12}). Although there is a large spread in
the model results, these temperatures are broadly consistent with
those measured for $^{12}$CO.

\citet{woitke09}, their Figure 14, provide gas temperatures weighted
by the CO abundance as functions of CO column density for different
disk radii. Their model indicates that the low $J$ CO rovibrational
lines become optically thick at 1 AU for a column density of 10$^{15}$
cm$^{-2}$, at which point the CO weighted temperature is 600 K. Rarer
isotopes, such as $^{13}$CO, are sensitive to larger column densities
before becoming optically thick and thus probe lower temperatures. Our
CRIRES $^{13}$CO measurements follow the predicted relation between
column density and gas temperature in their Figure 14 but lie above
the 1 AU contour by a factor of $\sim$3x10$^3$ in column density. This
may be further indication that the $^{13}$CO lines are not just thermally
excited and thus the column densities are overestimated.

\citet{gorti11}, their Figure 8, model the CO rovibrational lines in
the transitional TW Hya disk and find that inner disk masses between
10$^{-4}$ to 10$^{-6}$ M$_\odot$ fit the rovibrational lines well to
within a factor of 2. Most of the model emission is dominated
by the warm surface layers rather than deeper material. This amount of
inner disk gas is similar to those derived from our $^{13}$CO data
(Table \ref{table:rotpar}).

\begin{figure}[]
\hspace{-1cm}
\includegraphics[angle=180,scale=0.36]{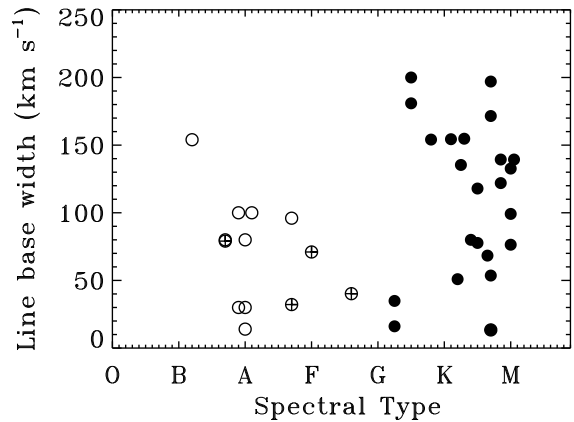} 
\caption{A plot of the line width at base vs spectral type. This
  figure shows that earlier type stars tend to have narrower line
  widths. Our CRIRES sample (filled circles for late-type stars, crossed
  circles for early type)  consists mainly of later type stars.
  Additional Herbig stars from \citet{brittain07,brittain09} and
  \citet{vanderplas09} are included as open
  circles. \label{fig:herbigwidth}}
\end{figure}

\subsection{Effects of stellar mass on CO}

The fundamental band of CO from Herbig AeBe stars has been studied extensively
\citep{najita03,blake04,brittain07,vanderplas09}. One of the most obvious
differences is the prevalence of high vibration lines in Herbig stars
with transitions up to v=8--7 being seen \citep{brittain09}. In
comparison, the T Tauri stars in our sample show vibrationally excited
emission generally only to v=2--1 with a significant fraction showing
no vibrationally excited lines at all. This is probably due to the
weaker UV fields emitted from T Tauri stars as is clearly predicted
from models (Figure~\ref{fig:vibsp}). 

\begin{figure*}[]
\vspace{-9cm}
\hspace{-3.5cm}
\includegraphics[angle=180,scale=0.9]{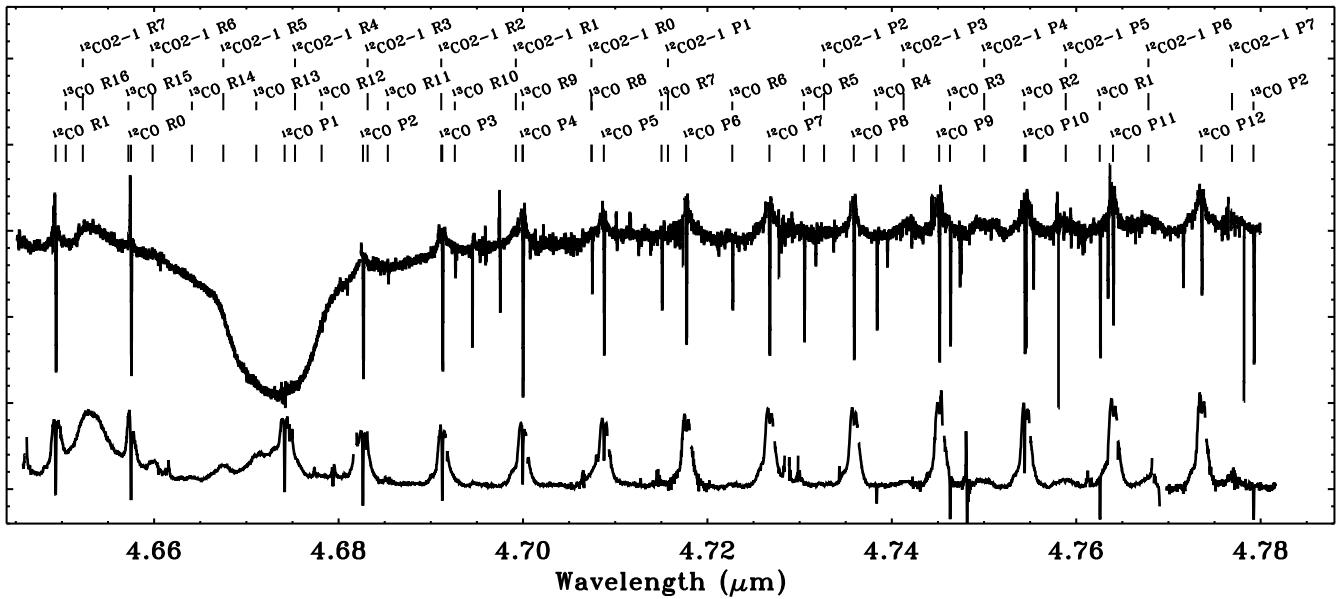} 
\vspace{-2cm}
\caption{A comparison of the spectra from the class I source IRS 43 (top)
  and the class II disk RNO 90 (bottom). The deep ice feature at 4.67 $\mu$m is
  only present in the class I source. RNO 90 has absorption lines only
  in the lowest $J$ $^{12}$CO lines from cold foreground gas. Most of
  the differences between the two spectra can be attributed to the
  disappearance of the envelope. \label{fig:evolfig}}
\end{figure*}

A comparison of the widths of the line bases (see Appendix) reveals
that Herbig stars tend to have narrower lines (Figure
\ref{fig:herbigwidth}). This indicates that the inner edge of the gas
is further out in the disk compared to T Tauri stars.  As our sample
consists mainly of later type stars, additional Herbig stars from
\citet{brittain07,brittain09} and \citet{vanderplas09} are
included. Our few Herbig stars show similar line widths indicating
that this is unlikely to be due to any differences in methodology. A
Kolmogorov-Smirnov test places the probability at 99\% that the
samples have different underlying distributions.

For a given temperature, the radial location will be further out in a
Herbig disk than in a T Tauri disk due to the larger luminosities of
Herbig stars \citep{salyk11}. As the CO lines are sensitive to only a
certain range of gas temperatures, the lines would be expected to
arise from further out in Herbig disks. While this explains the
general trend that we see, it does not explain the prevalence of
Herbig stars with large ($>$ 10 AU) gas holes (IRS 48 in our sample,
\citealt{brown12}; see also \citealt{goto06}, \citealt{vanderplas09},
\citealt{brittain09}). Some additional effect, e.g. photodissociation
of CO, is needed to explain these holes.

\subsection{Evolutionary changes in CO} 
\label{trans}

Complexity in line profiles generally decreases at each evolutionary
stage. Transition disks thus show the most straightforward profiles
and embedded class I sources the most kinematically complex.

\begin{figure}[]
\hspace{-1cm}
\includegraphics[angle=180,scale=0.37]{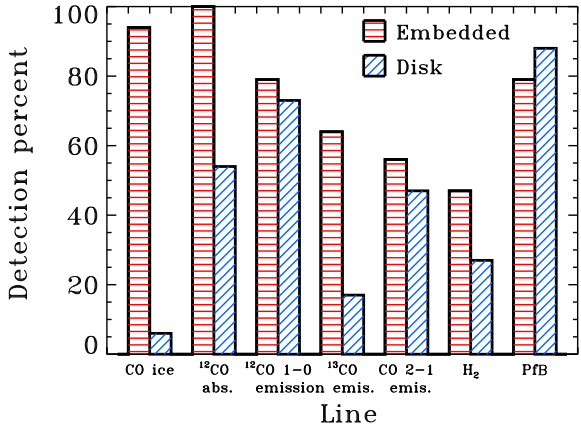} 
\caption{A comparison of the detection percentages of different lines
  in the embedded (class I) and disk (class II) samples.
  \label{fig:detclasses}}
\end{figure}

{\it Class I to Class II:} The 18 class I embedded objects discussed
by \cite{herczeg11} show some systematic differences from the 69 disk
sources. The first is that the CO ice band seen prominently in class I
sources \citep{pontoppidan03} disappears from the spectra of the disks
due to the disappearance of the cold envelope
(Fig.~\ref{fig:evolfig},~\ref{fig:detclasses}). There are only a few
exceptions of class II disks with ice bands, with the ice most likely
arising in dense foreground gas. Disk winds/outflows as reflected in
blue absorption wings are seen in both classes. However, these sources
are less common in the disk sample and, when seen, the maximum
velocities are smaller (generally $<$20 km s$^{-1}$ as opposed to
50-100 km s$^{-1}$).

The number of sources with gas phase absorption lines decreases from
100\% to $\sim$60\% as sources evolve from the embedded class I to the
class II phase (Fig.~\ref{fig:detclasses}). A comparison of the
absorption line temperatures reveals that the class II sources can
show very low temperature ($<$50 K) components
(Fig.~\ref{fig:fwhmt}). However, both class I and class II sources can
have warm absorption lines. The inner regions of class I envelopes
probably result in higher average temperatures along the line of sight
while class II sources may have absorption lines arising solely from
cold foreground material. The few class II sources with hundred K
absorption lines likely have an inclination angle such that absorption
arises through the surface layers in the outer disk.

The emission line characteristics of the class I sources are similar
to those of the class II sources (see \citealt{herczeg11} for further
details). The class I profiles are complicated by absorption from the
envelope and outflow but the underlying emission seems similar to the
class II disk emission with the addition of more common blueshifted
components. In particular, the class I emission lines have similar
widths, temperatures and lack of spatial extent to the class II
sources. 8/16 of class I sources show CO v=2--1 emission lines
(although 3 show no $^{12}$CO in emission) which is similar to the
47\% of class II sources which show CO v=2--1 emission. Unlike the
class II sources, the class I sources show no double peaked line
profiles at all, but are always single-peaked indicative of a slow
disk wind. Overall, the inner few AU of class I and II disks are very
similar indicating that disk characteristics are established early.

{\it Full to transitional disks:} Transitional disks with inner gaps
and holes provide an opportunity to study inner disk gas at a time
when the dust is disappearing. The transitional disks generally fall
into the narrow line profile category, partly coincident with the
known close to face-on inclinations of many of the disks and partly
due to the generally narrower line widths of transitional disks found
by \citet{salyk11}. The lower dust continuum should increase the
effectiveness of radiative excitation in the inner regions. This
appears to be the case with all transitional disks showing UV-pumped
CO v=2-1 emission and most showing the rarer $^{13}$CO emission lines
which may be IR pumped.

\begin{figure}[]
\hspace{-1cm}
\includegraphics[angle=180,scale=0.38]{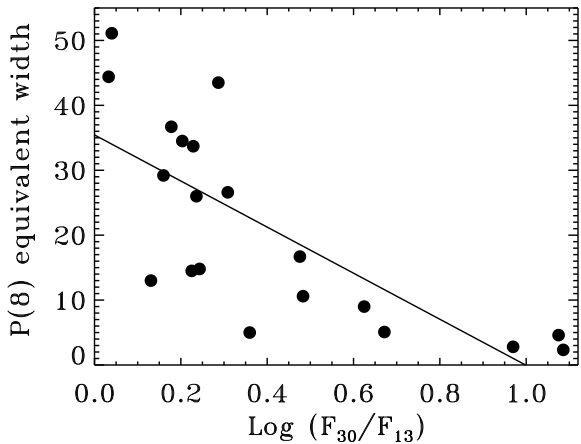} 
\caption{A trend is seen for declining SEDs, as seen in the 13-30 $\mu$m
  flux ratios,  to have larger CO P(8) equivalent widths. The flux
  ratios are taken from \citet{brown07, furlan09}, and the Spitzer
  archive. The correlation has a Pearson's correlation coefficient of
  -0.75 indicating a strong correlation. \label{fig:1330p8}}
\end{figure}

SED slopes provide clues to disk structure and evolutionary states. We
focus on the ratio between 13 and 30 $\micron$ (F$_{\nu {\rm 30 \mu
    m}}$/F$_{\nu {\rm 13 \mu m}}$) as defined in \citet{brown07}.  Low
numbers likely indicate strong settling while extremely high numbers
mark transitional disks with inner gaps and holes (see
Figure~\ref{fig:1330p8}). A trend is seen with the lowest $n_{13-30}$
sources having the strongest P(8) equivalent widths and the highest
$n_{13-30}$ sources having weaker P(8) equivalent widths. This
correlation is not reflected in the fluxes, indicating that the cause
lies in the interplay between dust continuum and gas lines. This may be
an optical depth effect where more CO resides above the dust $\tau =
1$ surface in the settled disks leading to strong CO lines. 

\section{Conclusions}
\label{conclusions}

CO fundamental lines are commonly seen from the circumstellar
environments of young stars and provide information on molecular gas
in the planet-forming region in disks that ALMA will not be able to
probe. Our extensive collection of high spectral resolution CRIRES
observations reveal complex environments with signatures of disks,
envelopes, foreground molecular clouds, winds and outflows in diverse
combinations.

Young disks display a variety of complex line profiles with simple
    double-peaked Keplerian disk profiles being surprisingly
    rare. Most line profiles have wings consistent with Keplerian
    rotation in the inner disk close to the dust sublimation
    radius. However, the majority of the emission lines have excess
    flux at the line center, which in a disk model requires emission
    from large radii. In most cases, there is no evidence of spatial
    extent in the CO lines, ruling out standard Keplerian disk models. This
    trend, plus a tendency for slight excess emission on the blue side
    of the lines, leads us to conclude that slow disks winds may be
    common from young stars and traceable through CO lines.

The high spectral resolution and sensitivity of CRIRES allow a more
comprehensive overview of emission lines from weaker
isotopologues. $^{13}$CO lines suffer less confusion from optical
depth effects than $^{12}$CO lines and thus provide better constraints
on the gas properties. The $^{13}$CO lines are detected with much
greater frequency than in previous studies and have lower excitation
temperatures, indicating an origin further out radially or deeper
within the vertical structure. The $^{13}$CO line widths are generally
narrower than those of $^{12}$CO, indicating dominance by slower
velocity gas arising from larger radii. The $^{13}$CO line strengths
for the objects with the coldest gas imply large column
densities/emitting areas if only collisional excitation is assumed.
IR resonant scattering may contribute to the line fluxes, as has been
proposed for Herbig Ae/Be stars, so that the inferred column densities
are upper limits. The derived temperatures are roughly consistent with
current thermo/chemical models of the inner disk.

Non-thermal excitation is clearly reflected in the CO lines, hinting
at the complex photoprocesses at work in the molecular layer.
Vibrational emission from thermally inaccessible energy levels is
detected in about 50\% of the sources. This emission is correlated
with accretion luminosity indicating that the UV radiation arising
from accretion drives the fluorescent pumping, particularly in
late-type stars where the stellar UV continuum is weak.

Finally, we examine the large sample for trends between different
types of sources. The change from class I embedded protostar to class
II disk produces dramatic changes in the CO spectra, as the removal of
the protostellar envelope results in the disappearance of the strong
absorption lines and ice feature characteristic of class I
spectra. However, the emission lines from class I and II sources, both
arising from close to the star, are similar in detection frequency,
excitation and line shape, indicating that disk characteristics are
established early. The beginning of disk dispersal seen in transition
disks also brings changes to the CO spectra. Transition disk CO lines
are generally Keplerian with a tendency for narrow line widths
indicating gas at larger radii. Herbig Ae/Be stars also typically have
narrower lines, but in this case, it is likely due to an increase in
the dust sublimation radius and the disk region with temperatures
suitable for detecting rovibrational CO.

We have highlighted general trends from the large sample to provide an
overview of CO gas within young systems. However, the individual
spectra are all unique, reflecting physical differences among the
stars, disks and surrounding environments. We hope that future
detailed modeling will be able to exploit fully all the information
contained in these rich CO spectra.

\acknowledgments{The authors thank Colette Salyk, Jeanette Bast,
  Wing-Fai Thi, Bill Dent and Kevin France for discussions and
  contributions to the programme. We also thank the anonymous referee for
  comments which improved this paper. This work is based on
  observations collected at the European Southern Observatory Very
  Large Telescope under programme ID 179.C-0151. J.M.B. acknowledges
  the Smithsonian Astrophysical Observatory for support from a SMA
  fellowship.  Support for K.M.P. was provided by NASA through Hubble
  Fellowship grant no. 01201.01 awarded by the Space Telescope Science
  Institute, which is operated by the Association of Universities for
  Research in Astronomy, Inc., for NASA, under contract NAS
  5-26555. Astrochemistry at Leiden is supported by a Spinoza grant
  from the Netherlands Organization for Scientific Research (NWO), by
  NWO grant 614.000.605 and by the Netherlands Research School for
  Astronomy (NOVA). }

\begin{appendix}

\section{Sources with extended emission}

\begin{figure*}[]
\begin{minipage}{0.5\linewidth}
\includegraphics[angle=90,scale=0.4]{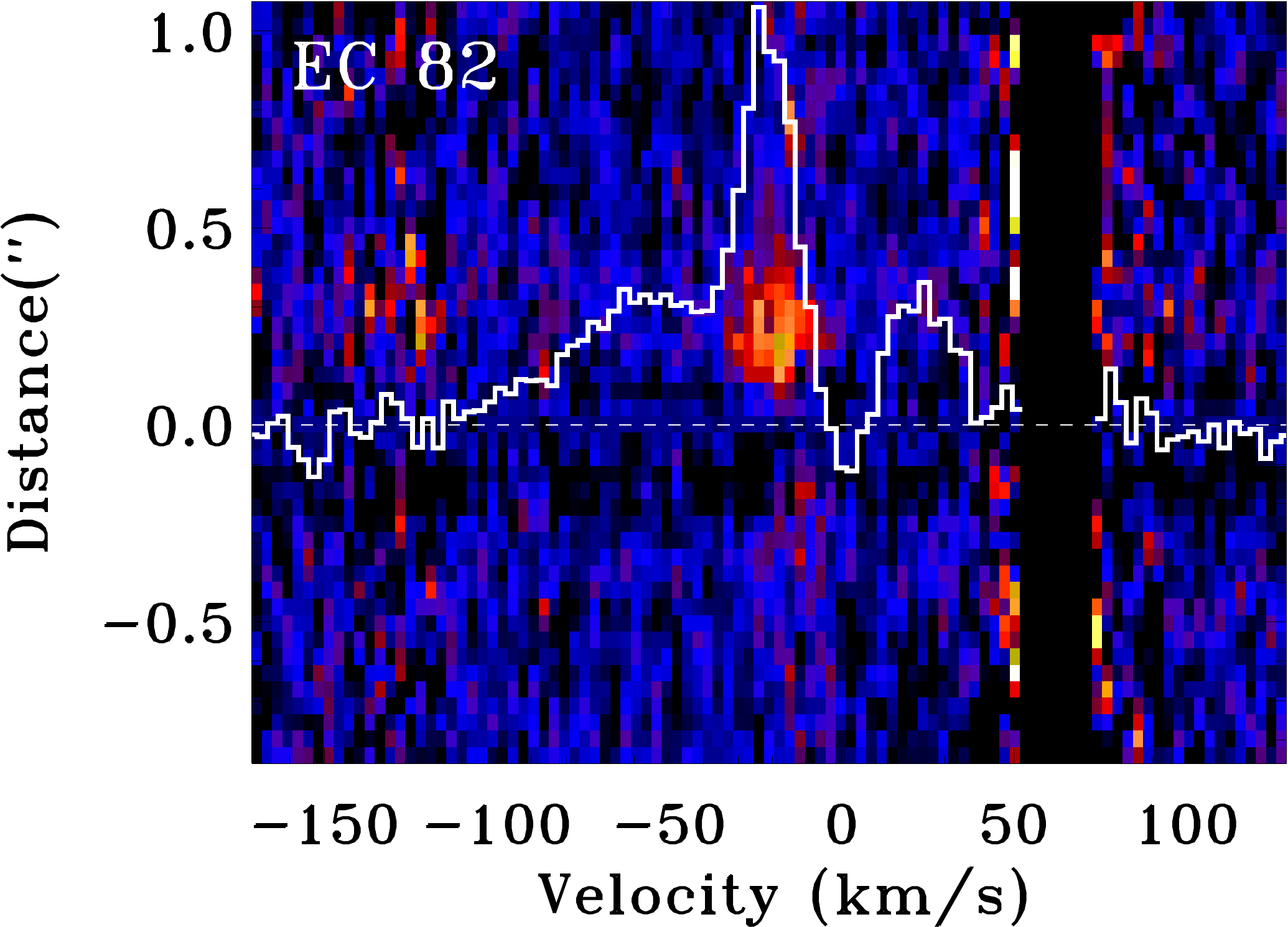} 
\end{minipage}
\begin{minipage}{0.5\linewidth}
\includegraphics[angle=90,scale=0.4]{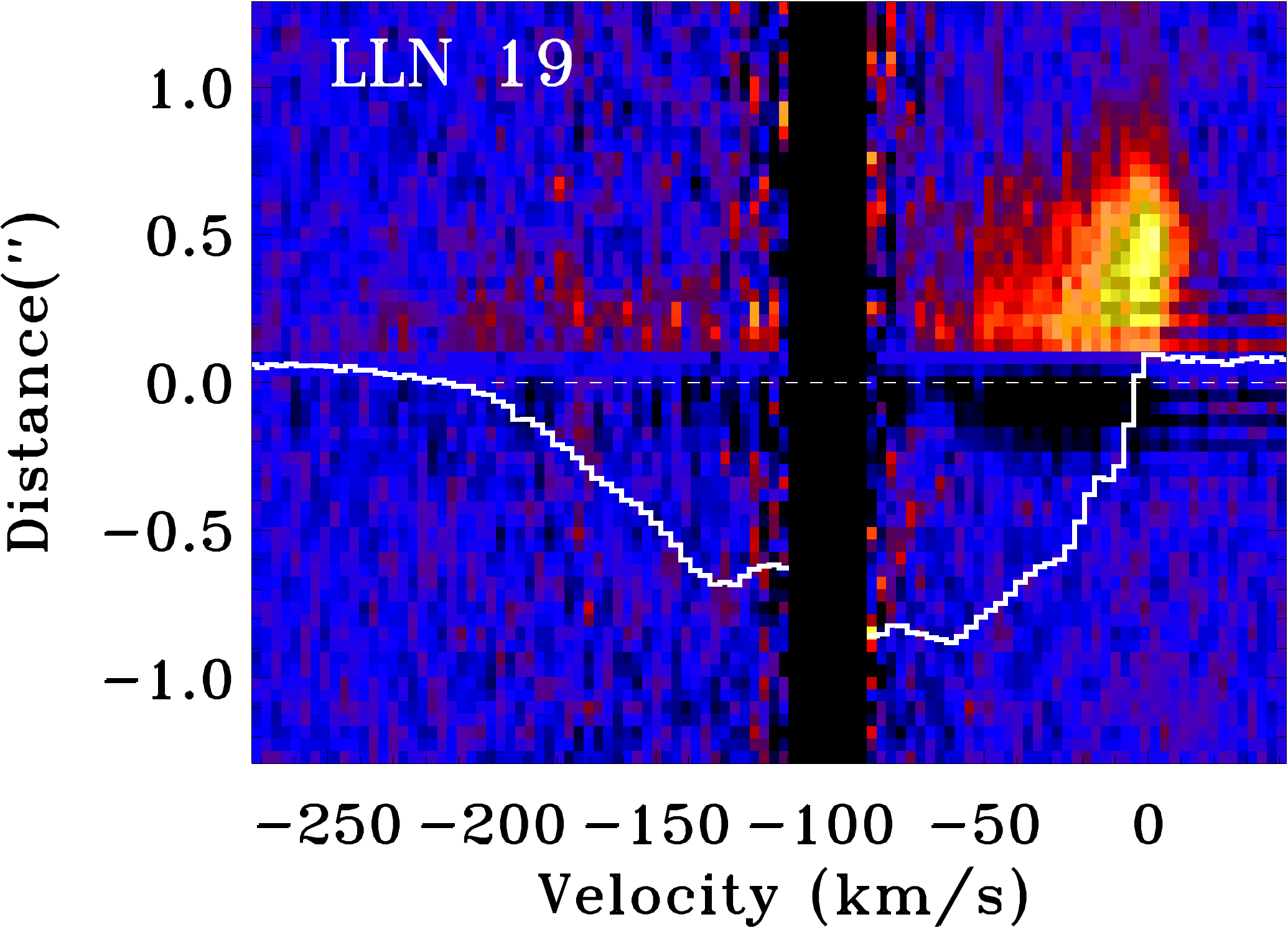} 
\end{minipage}
\begin{minipage}{0.5\linewidth}
\includegraphics[angle=90,scale=0.4]{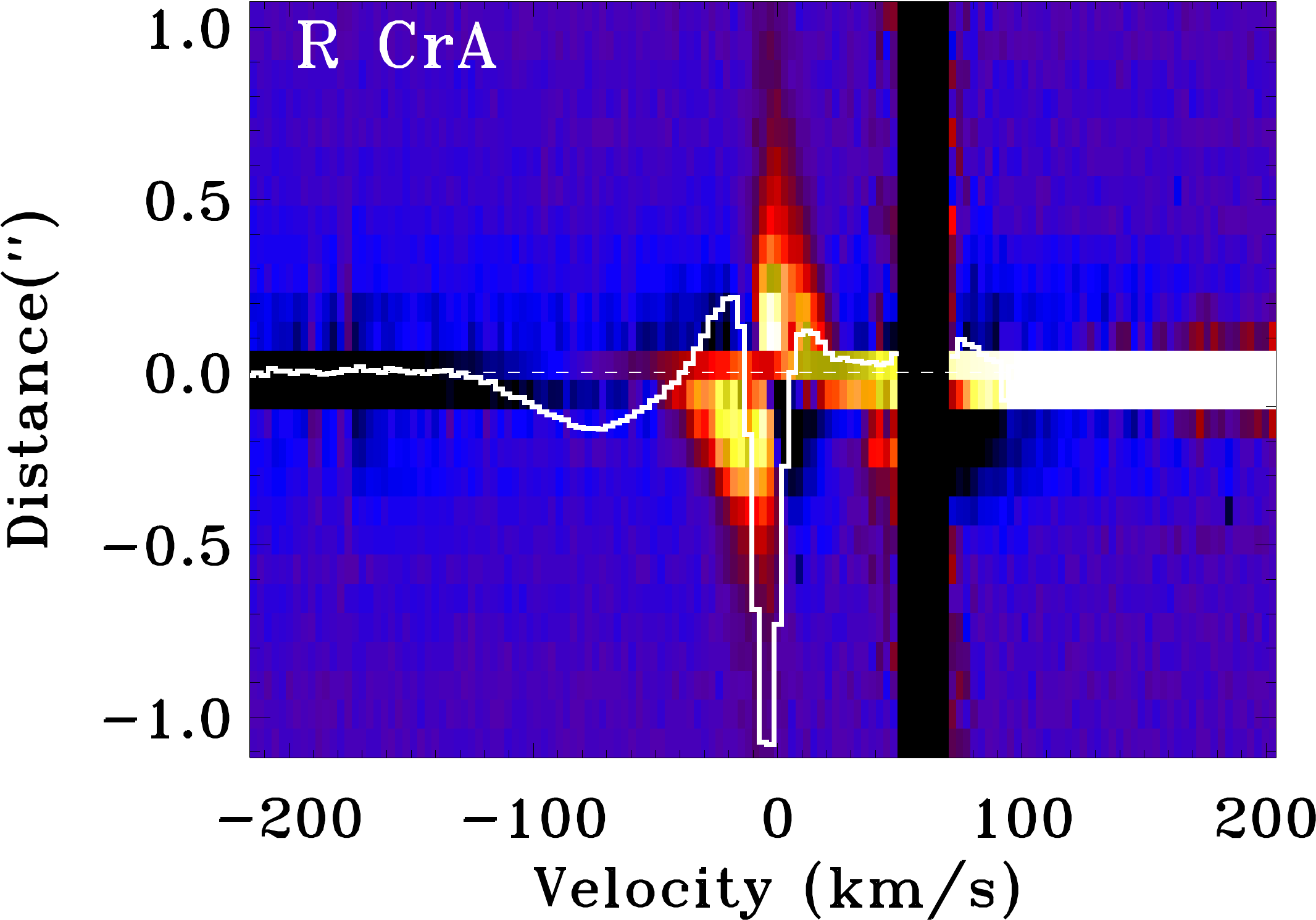} 
\end{minipage}
\begin{minipage}{0.5\linewidth}
\includegraphics[angle=90,scale=0.4]{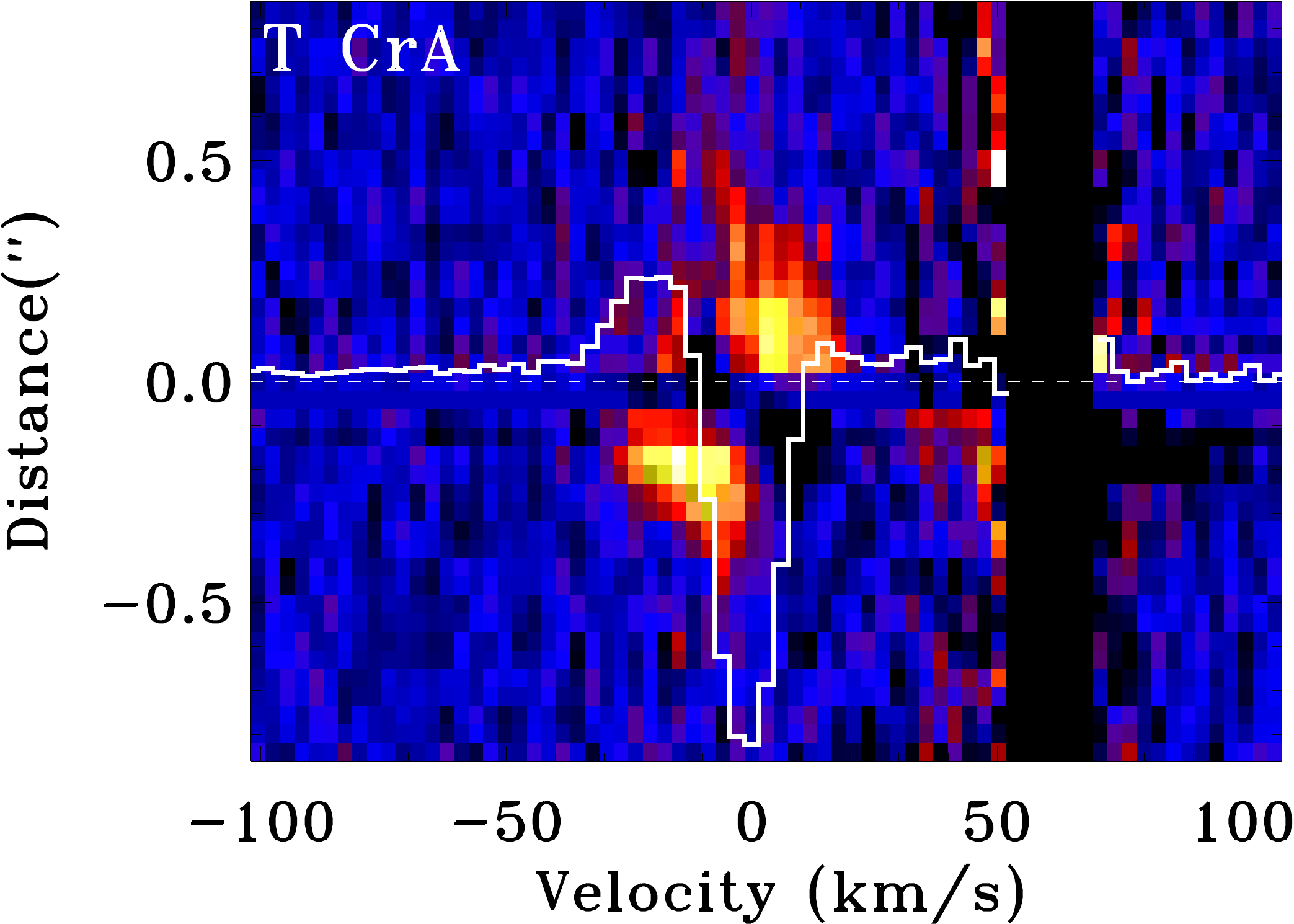} 
\end{minipage}
\caption{Continuum subtracted 2-dimension stacked spectra of sources with
spatially extended CO lines. The integrated flux spectrum is
overplotted in white to show which components of the line profile are
extended. EC 82 and T CrA are at P.A. 90, while R Cra and LLN 19 are at
P.A. 0. \label{fig:extent}}
\end{figure*}

Fig.~\ref{fig:extent} presents the emission of the three additional
sources in our disk sample for the lines can be directly imaged: EC
82, R CrA and T CrA.  LLN 19, a massive embedded object, also shows
extended emission. As found previously for the embedded
sources GSS 30 and IRS 43 by \citet{herczeg11}, most of the visually
extended sources appear to be outflows with the spatially extended
emission occurring on only one side of the spectral trace, probably in
the outflow cavity wall (see Figure \ref{fig:extent}).

Evidence for outflows is also seen in broad absorption lines for these
and other sources (Figures \ref{fig:profiles} and \ref{fig:gallery})
\citep[see also][]{thi10,herczeg11}. In contrast with the slow
molecular winds analyzed through spectro-astrometry
\citep{pontoppidan11}, the outflows show large blue-shifts up to a few 100
km s$^{-1}$ from source velocity.

\clearpage

\section{Inner radii}
\label{rin}

The gas closest to the star rotates most rapidly producing the wings
of the line profile. Assuming that the CO lines arise from such a
disk, the innermost extent of the CO can be measured based on the
width of the line base. Two different methods of measuring the width
of the stacked line profiles were used to overcome this problem (Table
\ref{table:innervel}). In the first, the width was measured between
the points where the line rose above 3 times the continuum noise. In
the second, the width was measured at 5\% of the peak flux
values. These two measurements are often similar for our typical noise
levels. When the continuum noise is low the first method provides a
better grasp of the actual extent of the gas. Inner radii were
determined by stacking clean lines to decrease continuum noise.

The derived number is actually in terms of $v\, {\rm sin}\, i$ and
thus the resulting radius is dependent on the inclination and stellar
mass. The corresponding $R {\rm sin}^2 i$ was calculated assuming
Keplerian rotation and a typical stellar mass for the spectral type
(O, B - 5 M$_\odot$, A - 3 M$_\odot$, F, G - 1 M$_\odot$, and K, M -
0.5 M$_\odot$). The resulting numbers are highly dependent on
inclination particularly for low inclination sources. For example, TW
Hya, which has a known inclination of 4-7$^{\rm o}$
\citep{pontoppidan08, qi04}, has an $R {\rm sin}^2 i$ of 9.9 AU but an
actual inner radius of less than 0.048-0.15 AU. Also, any disk
turbulence will broaden the line resulting in a smaller estimated
inner radius than is actually the case. The disk turbulence is likely
to be small compared to the line widths of $>$50 km s$^{-1}$.

The distribution of $v\, {\rm sin}\, i$ for the T Tauri stars is flat within
the errors as expected if inclination is the dominant effect. The
range of values is consistent with Keplerian emission from the inner
disk close to the dust sublimation radii as measured by infrared
interferometry (e.g. \citealt{akeson05}). The smallest $v\, {\rm sin}\, i$
values belong to the single peaked Keplerian sources, which in our
sample are mainly transition disks with known low inclinations. The
largest values tend to belong to the double peaked Keplerian sources.

\section{Gallery of spectra}

\begin{figure*}[]
\includegraphics[scale=0.8]{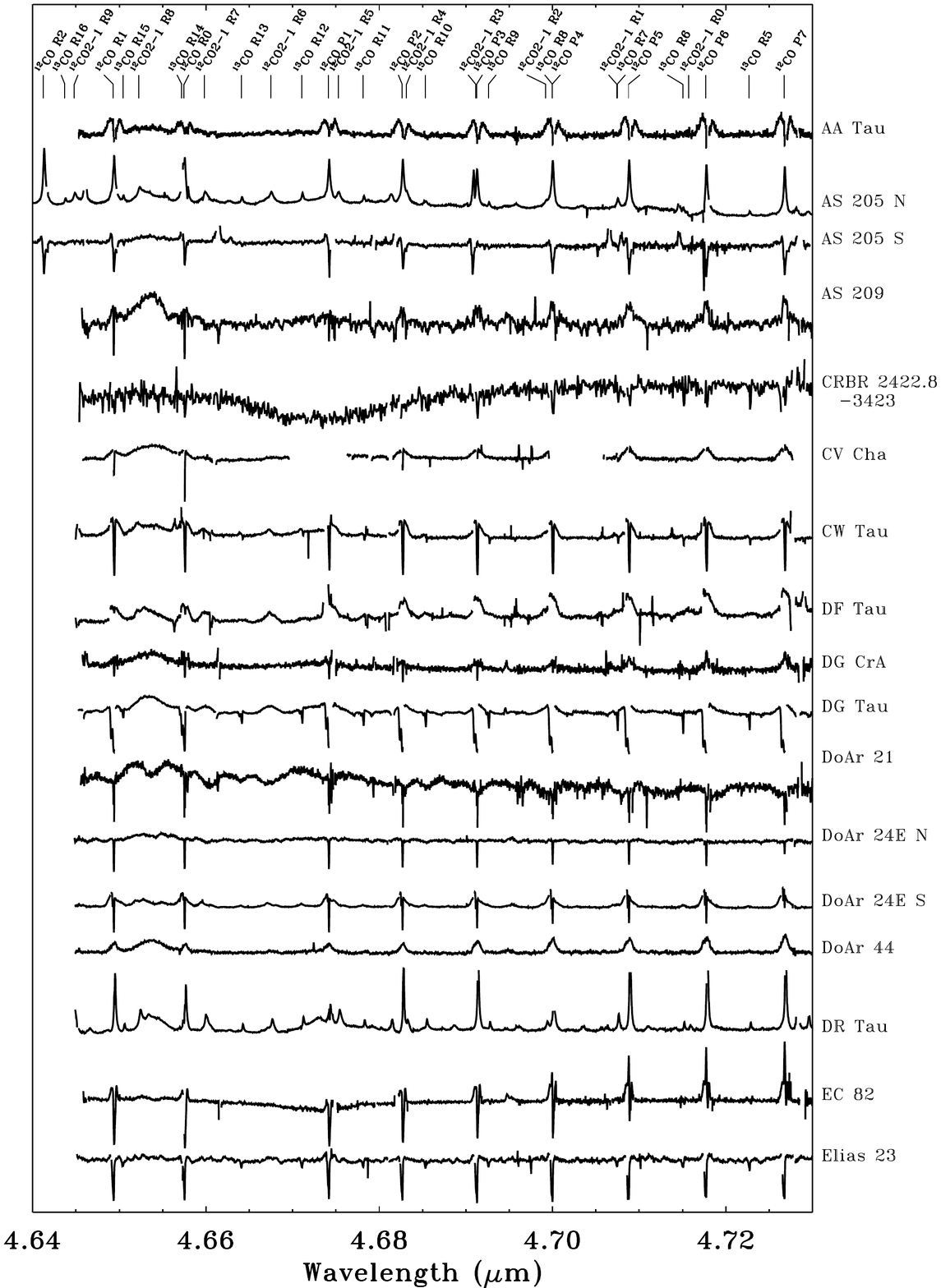} 
\caption{Spectra of the entire CRIRES disk sample in alphabetical order. \label{fig:overview1}}
\end{figure*}

\begin{figure*}[]
\includegraphics[scale=0.8]{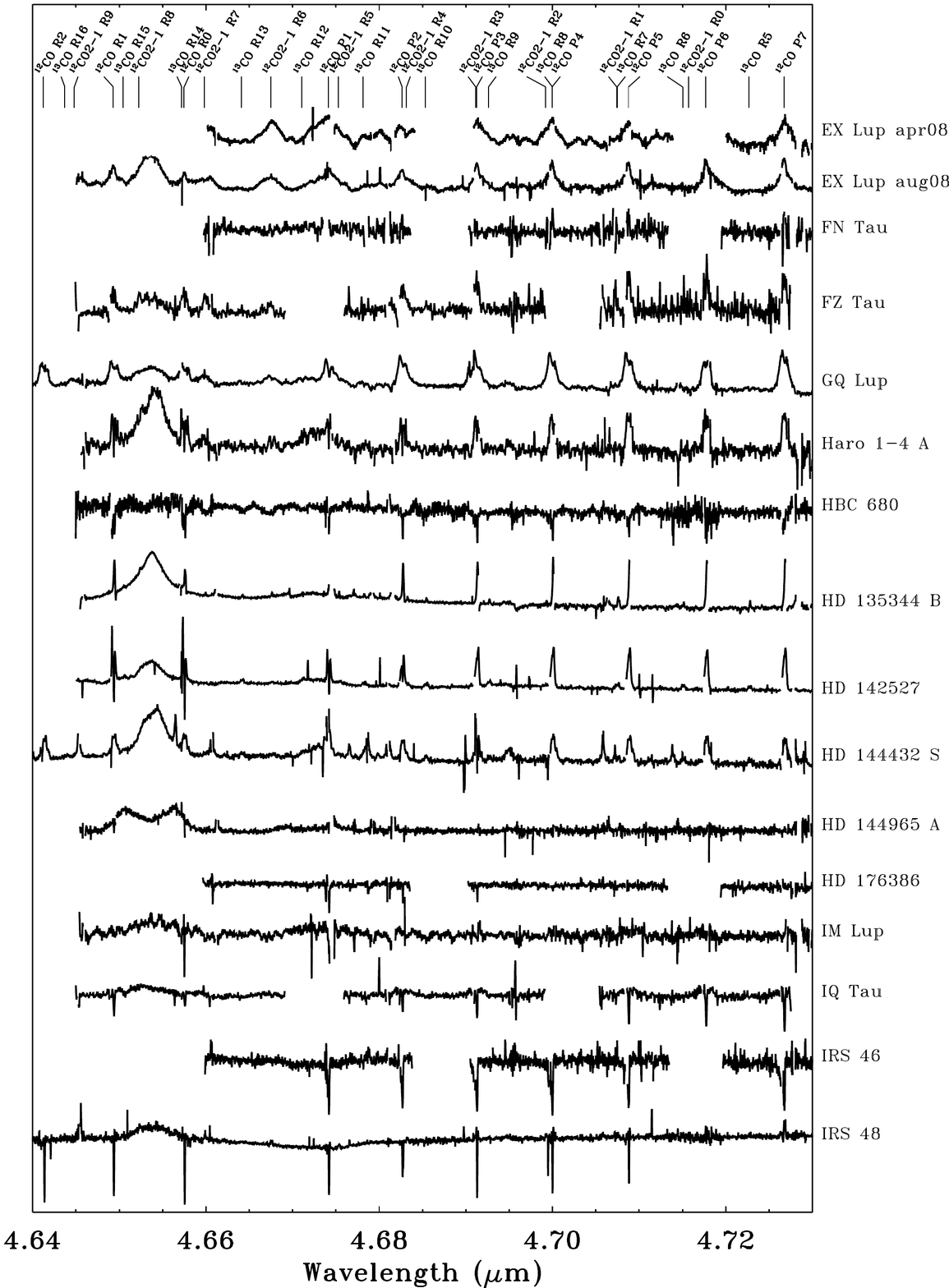} 
\caption{\label{fig:overview2}}
\end{figure*}

\begin{figure*}[]
\includegraphics[scale=0.8]{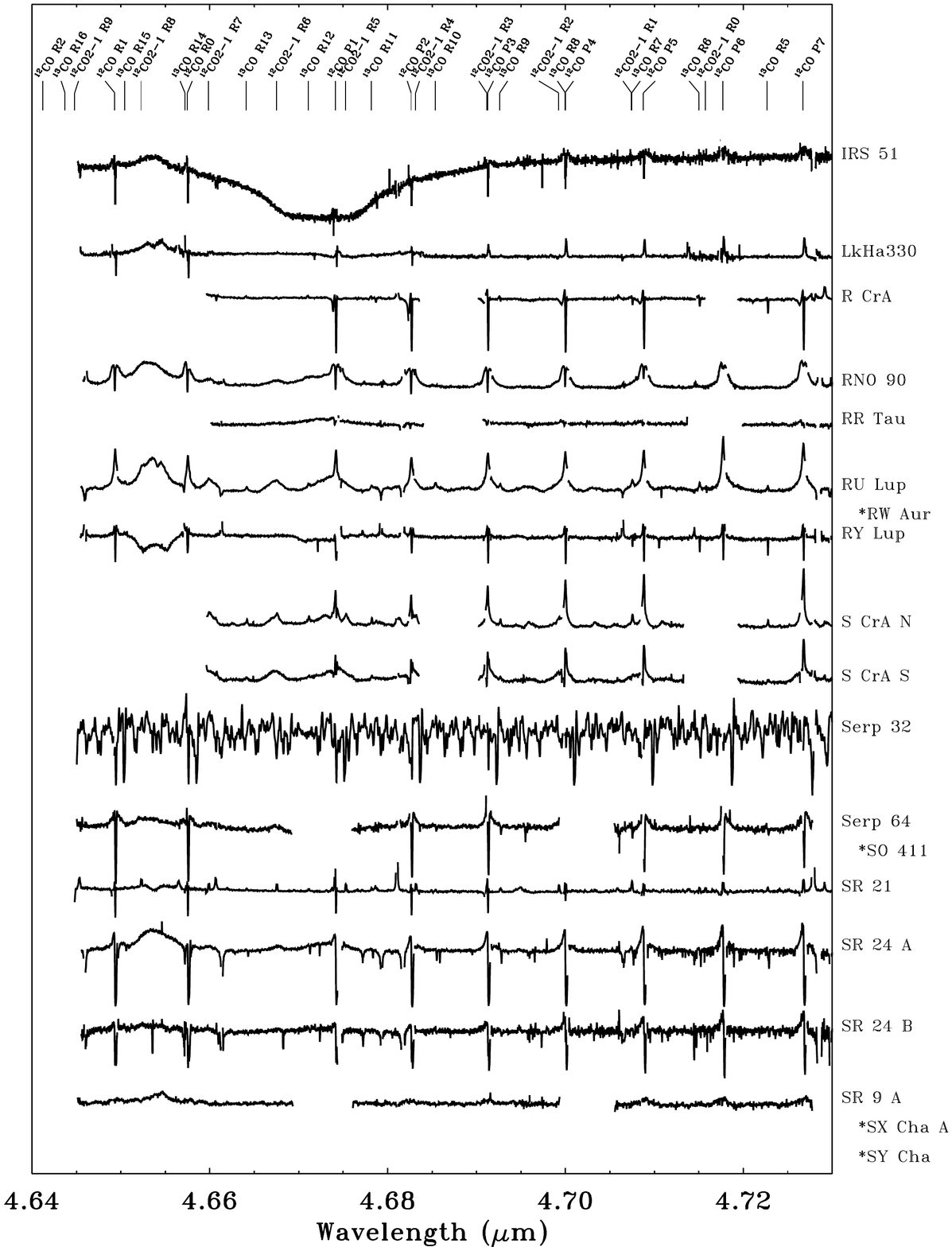} 
\caption{\label{fig:overview3}}
\end{figure*}

\begin{figure*}[]
\includegraphics[scale=0.8]{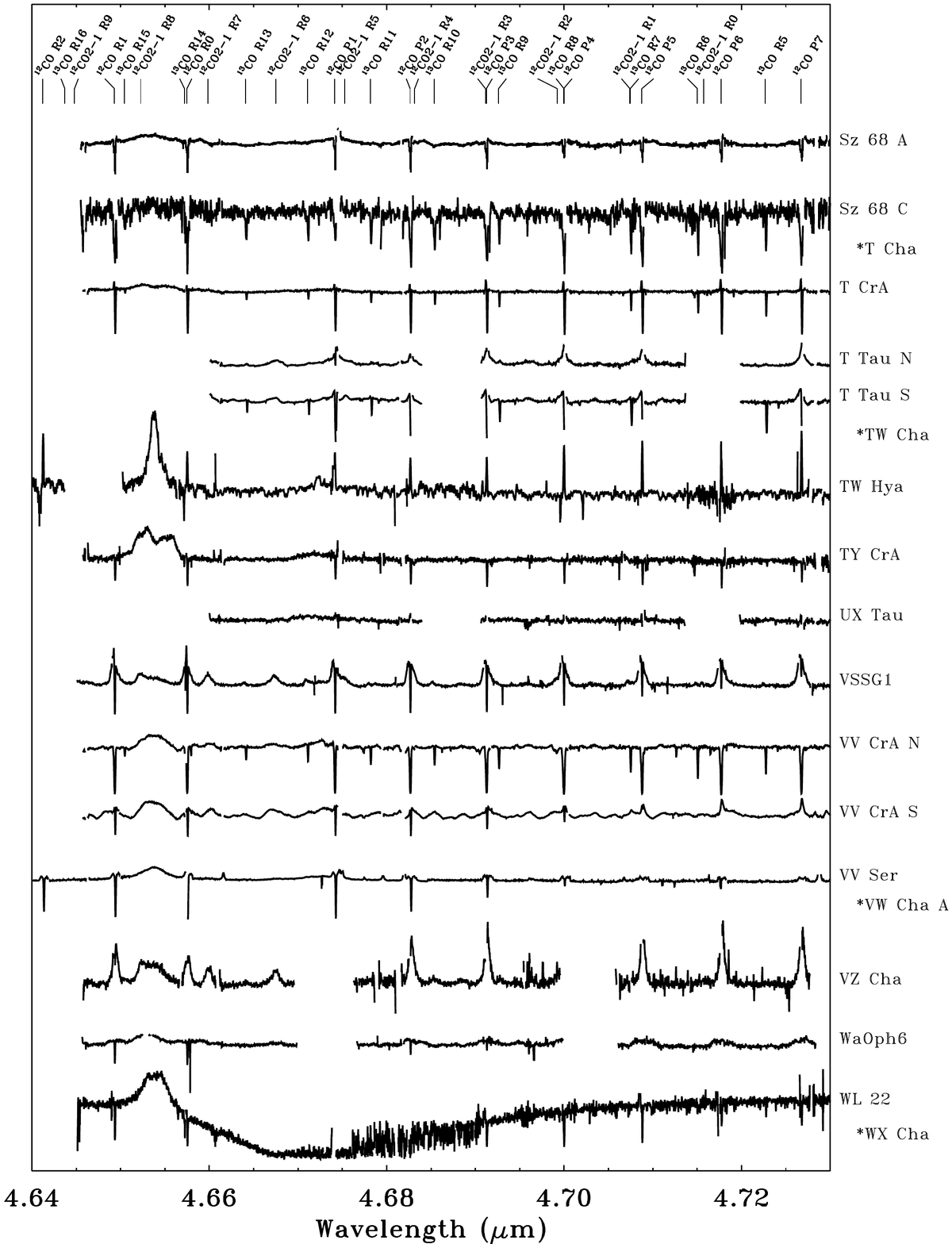} 
\caption{\label{fig:overview4}}
\end{figure*}

\begin{figure*}[]
\vspace{-10cm}
\includegraphics[scale=0.8]{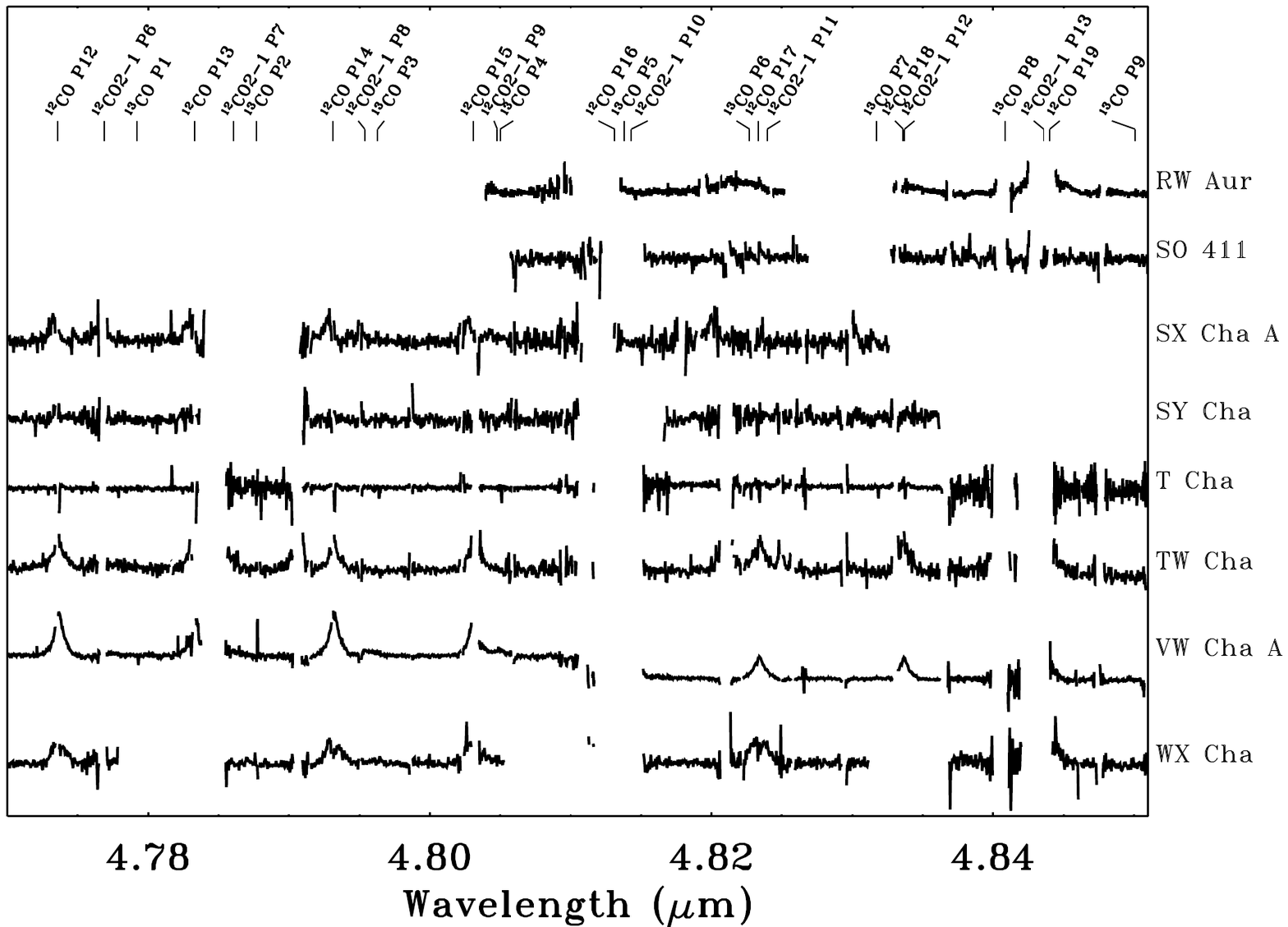} 
\caption{\label{fig:overviewhi}}
\end{figure*}

\begin{figure*}[]
\includegraphics[scale=0.8]{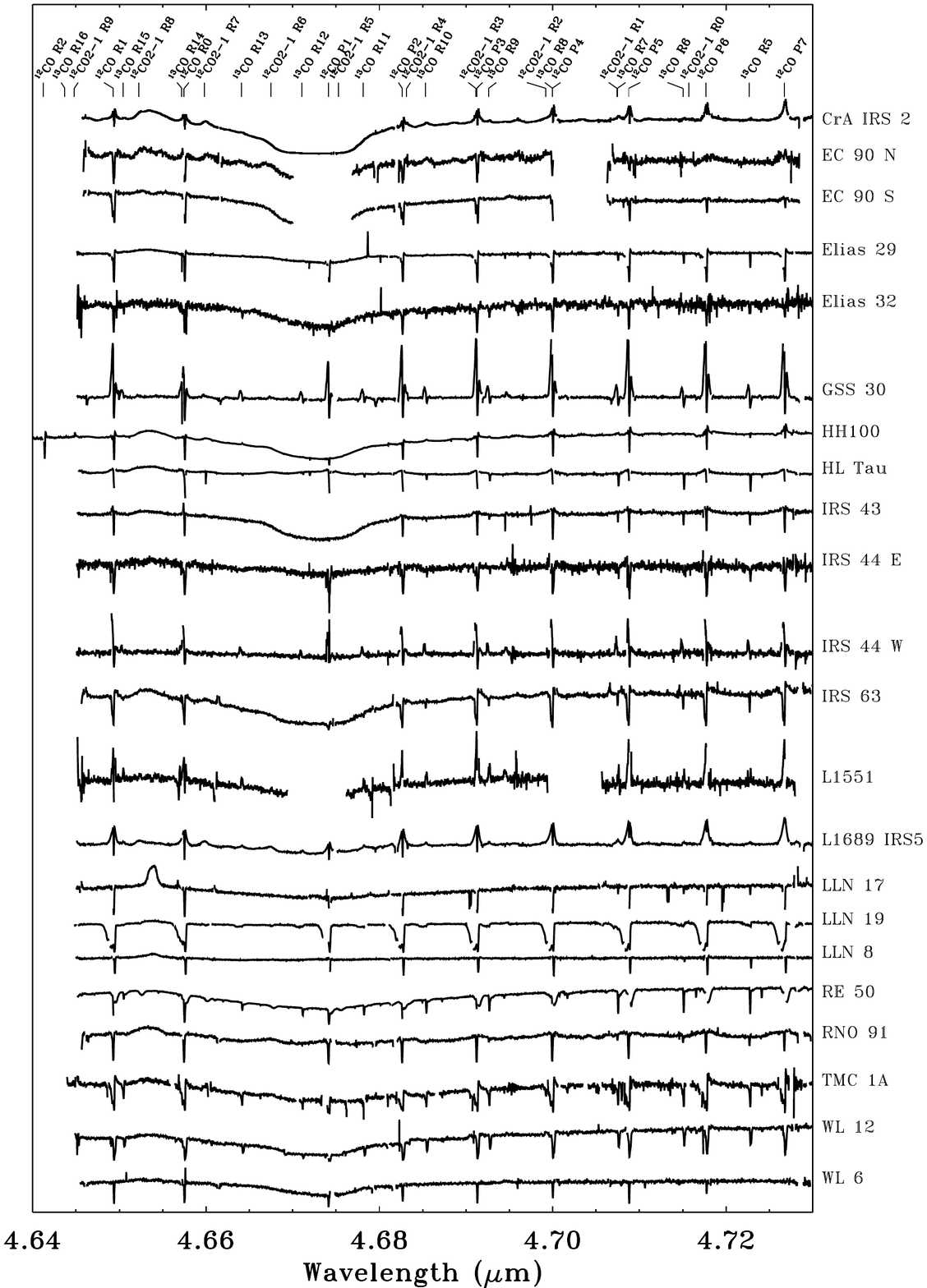} 
\caption{\label{fig:overviewembed}}
\end{figure*}

Spectra of the entire CRIRES sample are presented in Figures
\ref{fig:overview1}-\ref{fig:overviewembed}. Figures
\ref{fig:overview1} - \ref{fig:overview4} show 4.64-4.63 $\mu$m
[$^{12}$CO R(0)-P(7)] for all the disks in alphabetical order
(Fig. \ref{fig:overview1}: A-EL, Fig. \ref{fig:overview2}: EX-IR,
Fig. \ref{fig:overview2}: IR-SY, Fig. \ref{fig:overview2}:
SY-Z). Eight disk sources, mainly in Chamaeleon, were observed only in
the longer wavelength settings. These sources are denoted with $^*$ in
Figs. \ref{fig:overview1}-\ref{fig:overview4} and the spectra are
shown in Figure \ref{fig:overviewhi} covering 4.77-4.85 $\mu$
[$^{12}$CO P(12)-P(18)]. The 22 embedded sources, including higher
mass sources not covered in \citet{herczeg11}, are shown in
Fig. \ref{fig:overviewembed}.

\end{appendix}

\clearpage

\clearpage

\clearpage

\clearpage

\clearpage

\clearpage 

\clearpage

\begin{deluxetable}{lll}
\tablecaption{\label{table:obs}Summary of CRIRES disk observations}
\tablehead{\colhead{Source} & \colhead{Wavelength\tablenotemark{1}} &
\colhead{Date of observation\tablenotemark{1} }  \\ 
\colhead{} & \colhead{(nm)} & \colhead{} }
\tabletypesize{\footnotesize}
\startdata
AA Tau              & 4616,4730,4868 & 2007 Oct 12 \\
AS 205 N (A)        & 4662.1,4676.1,4760.8,4773.6; 4730 & 2007 Apr 21; 2007 Sep 1, 2008 May 2, 2007 Aug 29 \\
AS 205 S            & 4662.1,4676.1,4760.8,4773.6 & 2007 Apr 21 \\
AS 209              & 4716,4730 & 2008 May 1 \\ 
CRBR 2422.8-3423    & 4716,4730;4868 & 2008 Aug 3; 2008 Aug 4 \\
CV Cha              & 4716,4800,4946.2 & 2009 Jan 3 \\
CW Tau              & 4716,4868,4946.2;4730 & 2008 Dec 30; 2009 Jan 1 \\
DF Tau              & 4716,4730 & 2009 Jan 1 \\
DG CrA              & 4716,4730,4833;4946.2 & 2007 Apr 25; 2008 Aug 9\\
DG Tau              & 4716,4730,4868 & 2007 Oct 15\\
DoAr 21             & 4716,4730 & 2008 Apr 30\\
DoAr 24E S (B)      & 4716,4730;4730 & 2007 Sep 2; 2007 Sep 3\\
DoAr 24E N (A)      & 4716,4730 & 2007 Sep 2\\
DoAr 44 (Haro 1-16) & 4716,4730;4833 & 2007 Sep 4,2008 Apr 27;2008 Apr 29\\
DR Tau              & 4716,4730,4833;4716,4946.2 & 2007 Oct 10;2008 Dec 30\\
EC 82               & 4716,4730 & 2007 Apr 22\\
Elias 23	    & 4716,4730,4868;4946 & 2008 Aug 06; 2010 Mar 4 \\
EX Lup              & 4716,4730;4716,4730,4868 & 2008 Apr 27; 2008 Aug 6\\
FN Tau              & 4868;4730 & 2007 Oct 16; 2008 Dec 29\\
FZ Tau              & 4716 & 2009 Jan 1\\
GQ Lup              & 4662.1,4676.1,4760.8,4773.6;4730 & 2007 Apr 21; 2008 May 2, 2008 Aug 4\\
HBC 680             & 4716,4730 & 2007 Sep 1\\
HD 135344 B         & 4716,4730,4929.3,4946.2 & 2007 Apr 22\\
HD 142527           & 4716,4730,4868;4710;4076,4101 & 2008 Aug 5; 2008 Aug 7; Aug 9\\
HD 144432 S \& N$^*$& 4710,4730 & 2008 Aug 2\\
HD 144965 A (N) \& B$^*$ & 4716,4730 & 2008 Apr 29\\
HD 176386           & 4730 & 2007 Sep 1\\
Haro 1-4 A (S) \& B$^*$  & 4716,4730,4833 & 2008 Apr 29\\
IM Lup              & 4716,4730,4833 & 2008 Apr 26\\
IQ Tau              & 4716,4946.2 & 2008 Dec 30\\
IRS 46              & 4730 & 2008 Aug 7\\
IRS 48              & 4730;4730,4833;4710,4730;4868 & 2007 Sep 5; 2008 May3; Aug 2; Aug 5\\
IRS 51              & 4716,4730 & 2008 Aug 3\\
LkHa 330            & 4730;4868;4716,4730 & 2007 Oct 10; Oct 12; 2008 Dec 29\\
R CrA               & 4760.8;4730;4946.2 & 2007 Apr 21; Sep 1; 2008 Aug 9\\
RNO 90              & 4716,4730,4833,4929.3;4730 & 2007 Apr 25; Apr 26\\
RR Tau              & 4730,4868 & 2007 Oct 15\\
RU Lup              & 4716,4730,4833,4929.3;4730 & 2007 Apr 26; 2008 Apr 27\\
RW Aur              & 4868 & 2007 Oct 15\\
RY Lup              & 4716,4730,4840,4929.3;4716,4730,4833 & 2007 Apr 24; 2008 Apr 26\\
S CrA N \& S        & 4716,4730;4730;4946.2 & 2007 Apr 22; 2007 Sep 3; 2008 Aug 9\\
Serp 32             & 4716,4730 & 2007 Sep 2\\
Serp 64             & 4716 & 2007 Aug 30\\
SO 411              & 4868 & 2007 Oct 14, Oct 16\\
SR 9 A (S) \& B$^*$ & 4716 & 2008 Aug 8\\
SR 21               & 4716,4730;4730,4833 & 2007 Aug 30; Aug 31\\
SR 24 A (S) \& B/C(N)& 4716,4730,4833 & 2008 May 3\\
SX Cha A \& B$^*$   & 4800,4946.2 & 2008 Dec 30\\
SY Cha              & 4800,4946.2 & 2009 Jan 3\\
Sz 68 A,B \& C      & 4716,4730,4833,4929.3 & 2007 Apr 26\\
T Cha               & 4800,4820;4946.2 & 2008 Dec 29; Dec 30\\
T CrA               & 4716,4730;4730,4833,4929.3 & 2007 Apr 23; Apr 26\\
T Tau N \& S (A)    & 4730 & 2007 Oct 10\\
TW Cha              & 4800,4820,4946.2 & 2009 Jan 2\\
TW Hya              & 4662.1,4760.8,4773.6;4730 & 2007 Apr 21; Apr 26\\
TY CrA              & 4716,4730 & 2007 Apr 24\\
UX Tau              & 4730,4868 & 2007 Oct 12\\
VSSG 1              & 4716,4730,4868 & 2008 Aug 6\\
VV CrA N \& S       & 4716,4730,4840;4770,4779.5;4946.2 & 2007 Apr 24; Aug 31; 2008 Aug 9\\
VV Ser              & 4662.1,4676.1,4760.8,4773.6;4730 & 2007 Apr 21; 2008 May 1\\
VW Cha A \& B/C$^*$ & 4800,4820;4946.2 & 2008 Dec 29; Dec 30\\
VZ Cha              & 4800,4820;4946.2;4716 & 2008 Dec 29; Dec 30; 2009 Jan 3\\
Wa Oph 6            & 4716,4730,4833 & 2008 May 1\\
WL 22               & 4716,4730 & 2008 Aug 3\\
WX Cha              & 4946.2;4830 & 2008 Dec 30; 2009 Jan 2
\enddata
\tablenotetext{a}{This table summarizes the disk sources. See appendix table \ref{table:embedobs} for the complete embedded source list (also \citealt{herczeg11} for low mass embedded sources).}
\tablenotetext{1}{Semicolons (;) divide different groups of observations (e.g. for VW Cha A, the 4800 and 4820 settings were both observed on Dec 29 while the 4946.2 setting was observed on Dec 30).}
\tablenotetext{2}{Serp 32 is a Spitzer selected source at 18:28:45.6 -00:07:21.6, and is probably a background giant/AGB star based on the high radial velocity.The source is not counted in the disk statistics.}
\tablenotetext{3}{Serp 64 is a Spitzer selected source at 18:29:01.7 +00:29:38.7.}
\tablenotetext{*}{Detected companion was too faint for further analysis}
\end{deluxetable}

\begin{deluxetable}{lllll}
\tablecaption{\label{table:cat}Categorization of line profiles of disk sources}
\tablehead{\colhead{}& \colhead{}& \colhead{}& \colhead{}}
\tabletypesize{\footnotesize}
\startdata
1. Narrow      & 2. Broad single peaked    &  3. Broad & 4. Double & No lines         \\    
\hline  \hline                              	      \hline     
FN Tau       & AS 205 N  & AS 209         &  AA Tau       & HD 144965 A           \\
HD 135344 B  & DR Tau    & CV Cha         &  GQ Lup       & HD 176386             \\  
LkHa 330     & S CrA N   & DF Tau         &  IRS 48       & IM Lup         \\ 
SR 21	     & S CrA S   & DG CrA         &  RNO90        & SO 411        \\ 
TW Hya       & RU Lup    & DoAr 24 E S    &  VV Ser       & SY Cha       \\ 
             & VV CrA S  & DoAr 44        &  WaOph 6      & UX Tau  \\
             & VW Cha A  & EX Lup$^a$     &  WX Cha       &         \\ 
             & VZ Cha    & FZ Tau          &     &  \\   
             &           & HD 142527       &       &\\     
             &           & HD 144432 S      &  &  \\     
	     &           & Haro 1-4 A        &  & \\
	     &           & IRS 51	     &  & \\
	     &           & RR Tau	     &  & \\
	     &           & RW Aur	   &  & \\
	     &           & SR 9 A	   &  & \\
             &           & SX Cha A	      &        &  \\
             &           &  TW Cha                     &        &  \\
             &           &  VSSG 1        &        &  \\
\hline
5. Narrow absorption  & 6. Emission with broad & 7. Blue  & 8. Blue emission/ & Unclassifiable  \\ 
\,\,\, only  & \,\,\, central absorption & \,\,\, absorption & \,\,\, red absorption  &  \\ 
		     \hline   \hline        
CRBR 2422.8-3423 & CW Tau                   &  DG Tau     &  AS 205 S & Serp 32       \\           
DoAr 21          & Elias 23$^b$             &  HBC 680    &  EC 82  & T Tau N \\         
DoAr 24 E N      & IQ Tau                   &  IRS 46     &  R CrA            \\          
T Cha            & RY Lup                   &  Sz 68 A/B  &  SR 24 A                \\          
TY Cra           & Serp 64                  &  Sz 68 C    &  SR 24 B/C   \\           
WL 22            & T Tau A  		    &  VV CrA N   &  T CrA     \\
 \\        

\enddata
\tablenotetext{a}{EX Lup was undergoing outburst during this time which affected the line profiles. A detailed examination of the CO emission based on this data can be found in \citet{goto11}.}
\tablenotetext{b}{The CO v=2-1 line profiles are double peaked so the underlying v=1-0 emission line profile could fall in category 4.}
\end{deluxetable}

\begin{deluxetable}{llllll}
\tablecaption{\label{table:radvel}Radial velocities}
\tablehead{\colhead{Source} & \colhead{$V_*(\rm Helio)$} & 
\colhead{$V_*(\rm LSR)$} & \colhead{$V_{\rm CO}(\rm LSR)$} & 
\colhead{Ref.} \\
\colhead{} & \colhead{(km s$^{-1}$)} &  
\colhead{(km s$^{-1}$)} & \colhead{(km s$^{-1}$)} & 
\colhead{} }
\startdata
AA Tau	   &     16.1  $\pm$ 2	  & 5.6   &  9.0 & 1 \\
AS 205 N   &     \,-9.4  $\pm$ 1.5  & 2.0   &  5.0 & 2 \\
AS 209	   &     \,-8.5  $\pm$ 1    & 5.0   &  5.7 & 2 \\
CV Cha	   &     16.1  $\pm$ 1.3  & 5.5   &  3.0 & 3 \\
CW Tau	   &     14.5  $\pm$ 2    & 5.6   & -8.7 & 1  \\
DF Tau     &     \,\,\,\,12    $\pm$ 5    & 2.1   &  8.7 & 1  \\
DG CrA	   &     \,-1.8  $\pm$ 0.4  & 5.4   &  0.7 & 3 \\
DG Tau     &     19.3  $\pm$  2.7 & 9.5   & -3.4 & 4 \\
DoAr 24 E N&     \,-8.0  $\pm$ 2.5  & 2.3   &  3.5 & 2 \\
DoAr 44    &     \,-5.9  $\pm$ 0.4  & 4.7   & 11.4 & 2 \\
EX Lup	   &     \,-1.5  $\pm$ 2.3  & 3.6   &  3.3 & 2 \\
FN Tau     &     14.9  $\pm$ 0.4  & 6.1   &  4.6 & 5 \\
GQ Lup 	   &     \,-3.2  $\pm$ 0.6  & 2.8   &  6.0 & 2 \\
Haro 1-4 A &     \,-7.6  $\pm$ 0.1  & 2.8   &  5.4 & 2 \\
HD 135344 B&     \,\,\,\, -3   $\pm$ 3    & 1.3   &  7.4 & 6 \\
RU Lup	   &     \,-0.9  $\pm$ 1.1  & 4.6   &  2.0 & 2 \\
RW Aur     &      16.0   $\pm$  1.7   & 6.5   & -50  & 4 \\
RY Lup	   &     \,-2.0  $\pm$ 0.8  & 2.9   & -2.0 & 2 \\
S CrA 	   &     \,-0.3  $\pm$ 1.7  & 7.1   &  3.3 & 2 \\
Sz 68      &	\,-2.3   $\pm$  0.6 &  3.9  &  4.5 & 2 \\
T Cha	   &     14.0  $\pm$  1.3 &  4.1  &  6.0 & 3 \\
T Tau N    &     19.1  $\pm$  2   &  7.6  &  8.7 & 1 \\
TW Cha	   &     17.8  $\pm$  1.  &  7.0  &  5.0 & 3 \\
TW Hya     &    12.4   $\pm$  0.2 &  3.0  &  3.3 & 7 \\
VW Cha A   &     17.2  $\pm$  0.5 &  6.6  &  6.0 & 3 \\
VZ Cha	   &	 18.0  $\pm$  1.1 &  7.3  & 12.7 & 2 \\
\enddata
\tablenotetext{1}{\citet{hb88}; $^2$\citet{Melo03}; $^3$\citet{Gunther07}; $^4$ \citet{White04}; $^5$ \citet{Muzerolle03}; $^6$ \citet{Dunkin97}; $^{7}$ \citet{Setiawan08}}
\end{deluxetable}

\begin{deluxetable}{lllll}
\tablecaption{\label{table:rotpar}Summary of $^{13}$CO rotation diagrams}
\tablehead{\colhead{Source} & \colhead{$T$($^{13}$CO)} & \colhead{Opt. thin} & 
\colhead{N$_{\rm CO,100 AU^2}$} & \colhead{Gas mass} \\ 
\colhead{} & \colhead{(K)} & \colhead{radius (AU)$^*$} &
\colhead{(10$^{17}$ cm$^{-2}$)} & \colhead{(10$^{-4}$ M$_\odot$)}}
\startdata
AS 205 N        & 480 & 0.8   & 1.2   & 0.0058  \\
DoAr 24E S      & 490 & 0.35  & 0.089 & 0.00045 \\
DR Tau          & 570 & 1.3   & 0.95  & 0.0048  \\
HD 135344 B     & 250 & 5.0   & 41.   & 0.21    \\
HD 142527	& 420 & 3.1   & 0.26 & 0.088        \\
IRS 48          & 270 & 5.0   & 40.   & 0.20    \\
LkH$\alpha$ 330 & 240 & 13    & 460.  & 2.3     \\
RU Lup          & 310 & 4.4   & 25.   & 0.12    \\
S CrA N         & 380 & 5.6   & 17.   & 0.086   \\
S CrA S         & 280 & 8.9   & 120.  & 0.60    \\
SR 21           & 220 & 18    & 770.  & 3.9     \\
T Tau N         & 640 & 0.8   & 0.54  & 0.0027  
\enddata
\tablenotetext{*}{Where $A_{\rm emit} = \pi R_{\rm emit}^2$}
\end{deluxetable}

\begin{deluxetable}{ll}
\tablecolumns{2}
\setlength{\tabcolsep}{0.08in}
\tablewidth{0pt}
\tabletypesize{\footnotesize}
\tablecaption{\label{table:vib}Vibrational flux ratios}
\tablehead{\colhead{Source} & \colhead{Vibrational flux ratio} \\
\colhead{} & \colhead{(CO v=2-1/v=1-0)}}
\startdata
AA Tau		&  0.1   \\
AS 205 N    	&  0.2   \\
AS 209		&  $<$0.2 \\
CV Cha		&  0.25  \\
CW Tau		&  0.26  \\
DF Tau 	    	&  0.37  \\
DoAr 24E S	&  0.12  \\
DoAr 44		&  $<$0.068 \\
DR Tau		&  0.20  \\
FZ Tau		&  $<$0.15 \\
GQ Lup		&  0.19  \\
HD 135344 B	&  0.067 \\
HD 144432 S 	&  $<$0.12  \\
IRS 48 		&  0.5   \\
LkH$\alpha$ 330 &  0.11  \\
RNO 90		&  0.18  \\
RU Lup		&  0.29  \\
S CrA N		&  0.42  \\
S CrA S		&  0.2   \\
TW Hya		&  0.053 \\
WaOph 6  	&  0.25  \\
VSSG 1		&  0.27  \\
VV CrA S 	&  0.4   \\
VV Ser		&  $<$0.2 \\
VW Cha A	&  0.13  \\
VZ Cha		&  0.31  \\
\enddata
\end{deluxetable}

\begin{deluxetable}{lll}
\footnotesize
\tablecaption{\label{table:embedobs}CRIRES observations of embedded young stars}
\tablehead{\colhead{Source} & \colhead{Wavelength\tablenotemark{1}} &
\colhead{Date of observation\tablenotemark{1} }  \\ 
\colhead{} & \colhead{(nm)} &
\colhead{} }
\tabletypesize{\footnotesize}
\startdata
CrA IRS 2	& 4716,4730,4833,4929.3 & 2007 Apr 25 \\
EC 90 N	 \& S	& 4716,4730,4929.3,4946.2 & 2007 Apr 23 \\
Elias 29	& 4716,4730,4868  & 2008 Aug 8 \\
Elias 32	& 4716,4730,4833;4716;4730 & 2008 May 2; 2008 Aug 8; 9 \\ 
GSS 30		& 4716,4730;5115 & 2010 Mar 4; 2008 Aug 4 \\
HH 100		& 4710,4730,4868,4946.2;4716,4730 & 2008 Aug 2; 2007 Sep 2 \\
IRS 43		& 4716,4730,4868;4831 & 2008 Aug5; 2008 Aug 6 \\
IRS 44 E \& W	& 4716;4730,4833;4946.2;4868;4716,4730 & 2008 Apr 26; 29; 30; 2008 Aug 5; 6 \\
IRS 63		& 4716,4730,4833,4929.3;4716,4730,4868 & 2007 Apr 25; 2008 Aug 5\\
L1551		& 4716,4730 & 2007 Oct 16 \\
L1689 IRS 5	& 4716,4730;4716,4730,4800,5115 & 2010 Mar 10; 2010 Mar 15\\
LLN 8		& 4716,4730 & 2007 Oct 11\\
LLN 17		& 4716,4730;4716,4730,4868 & 2007 Oct 11; 2007 Oct 13\\
LLN 19		& 4716,4730,4833 & 2007 Apr 25\\
RE 50		& 4716,4730,4868;4770 & 2007 Oct 11,2007 Oct 17 \\
RNO 91		& 4716,4730,4929.3,4946.2;4770,4779.5; & 2007 Apr 23; Apr 24;\\
		& 4716,4730;4800,4820 & 2010 Mar 19; 2010 Mar 9\\
TMC 1A		& 4716,4730;4800,4946.2 & 2010 Feb 10; 2010 Feb 3\\
WL 12		& 4716,4730;4833 & 2007 Sep 1; 2007 Sep 3 \\
WL 6		& 4716,4730,4833 & 2008 Apr 26
\enddata
\tablenotetext{1}{Equivalent to Table \ref{table:obs} for the embedded sample}
\end{deluxetable}

\begin{deluxetable}{llllll} 
\tablecolumns{6}
\setlength{\tabcolsep}{0.08in} 
\tablewidth{0pt}
\tabletypesize{\footnotesize} 
\tablecaption{\label{table:feat}Summary
of observed line morphology$^1$} 
\tablehead{\colhead{Source} &
\colhead{M flux} & \colhead{P(8) flux} & \colhead{$^{12}$CO} &
\colhead{$^{13}$CO} & \colhead{$^{12}$CO 2-1} \\ \colhead{} &
\colhead{(Jy)} & \colhead{(Jy km s$^{-1}$)} & \colhead{} & \colhead{}
} 
\startdata 
AA Tau           & 0.3$^a$      & 11.0          & E/A   & E & E \\ 
AS 205 N 	 & 4.3$^b$      & 112 		& E     & E & E \\ 
AS 205 S 	 & 5.4$^b$      & 16.8/-59.7 	& E/A   & A & \\ 
AS 209 		 & 1.4$^c$      & 15.2 		& E/FA  & & \\ 
CRBR 2422.8-3423 & 0.4$^c$      & -4.6 		& A 	& A & \\ 
CV Cha 		 & 1.4$^c$      & 32.9 		& E/FA 	& E/FA & E \\ 
CW Tau 		 & 1.6$^d$      & 44.9/-14.3 	& E/A 	& E/A & E \\ 
DF Tau 		 & 1.5$^d$      & 43.2 		& E/FA 	& E & E \\ 
DG CrA 		 & 0.5$^c$      & 7.0 		& E/FA 	& & \\ 
DG Tau 		 & 2.4$^{d,e}$  & 25.8/-74.3 	& E/A 	& E/A & E \\ 
DoAr 21 	 & 0.88$^{f}$   & -0.26 	& A 	& A & \\ 
DoAr 24E N 	 & 0.15$^{f,c}$ & -0.04 	& A 	& A & \\ 
DoAr 24E S 	 & 0.07$^{f,c}$ & 1.7/-0.1 	& E/FA 	& E/FA & E \\ 
DoAr 44 	 & 0.54$^f$ 	& 12.2 		& E 	& & \\ 
DR Tau 	 	 & 1.3$^{g}$ 	& 43.8 		& E 	& E & E \\ 
Elias 23	 & 1.4$^c$      & 15.1/-22.3    & E/A	& E/A & E \\
EC 82 		 & 0.6$^c$ 	& 17$^*$ 	& E/A 	& E/A & E \\ 
EX Lup 		 & 1.3$^h$ 	& 20.8 		& E 	& E & E \\ 
FN Tau 		 & 0.26$^i$ 	& 1.3 		& E 	& & \\ 
FZ Tau 		 & 1.0$^a$ 	& 51.1 		& E 	& & E \\ 
GQ Lup 		 & 1.0$^c$ 	& 26.6 		& E 	& & E \\ 
Haro 1-4 S 	 & 0.6$^c$ 	& 10.0 		& E/FA 	& & E \\ 
HBC 680 	 & 0.5$^c$ 	& -4.3 		& A 	& A & \\ 
HD 135344 B 	 & 2.3$^j$ 	& 6.4 		& E 	& E & E \\ 
HD 142527 	 & 6.8$^j$ 	& 40.9 		& E 	& E & E \\ 
HD 144432 S 	 & 1.9$^e$ 	& 10.1 		& E 	& E & \\ 
HD 144965 A 	 & 0.5$^c$ 	& $<$0.7$^3$ 	& FA 	& & \\ 
HD 176386 	 & 0.4$^k$ 	& $<$0.3$^3$ 	& FA 	& & \\ 
IM Lup 		 & 0.6$^c$ 	& $<$0.9$^3$ 	& FA 	& & \\ 
IQ Tau 		 & 0.5$^a$ 	& 3.3/-2.9 	& E/A 	& & E \\ 
IRS 46 		 & 0.6$^c$ 	& -16.1 	& A 	& A & \\ 
IRS 48 		 & 2.0$^j$ 	& 4.2 		& E/FA 	& E/FA & E \\ 
IRS 51 		 & 0.92$^f$ 	& 12.1 		& E/FA 	& FA & \\ 
LkHa 330 	 & 0.9$^i$ 	& 4.1           & E/FA 	& E & E \\ 
R CrA 		 & 69$^m$ 	& -183/159/-375 & E/A 	& E/A & \\ 
RNO 90 		 & 1.3$^c$ 	& 44.7 		& E/FA 	& E/FA & E \\ 
RR Tau 		 & 1.1$^m$ 	& 1.7 		& E/FA 	& & \\ 
RU Lup 		 & 1.1$^{n}$ 	& 32.1 		& E 	& E & E \\
RW Aur 		 & 0.9$^e$ 	& -- 		& E 	& & \\ 
RY Lup 		 & 1.2$^c$ 	& 8.6/-3.0 	& E/A 	& E/A & \\ 
S CrA N 	 & 1.1$^b$ 	& 42.5 		& E/FA 	& E & E \\ 
S CrA S 	 & 3.3$^b$ 	& 74.3 		& E/FA 	& E & E \\ 
Serp 32 	 & 2.8$^f$ 	& 53.2/-70.4 	& E/A 	& & \\
Serp 64 	 & 0.3$^f$ 	& 4.0/-3.0 	& E/A 	& A & E \\ 
SO 411 		 & 0.4$^c$ 	& -- 		& 	& & \\ 
SR 9 A 		 & 0.5$^{f,b}$ 	& 5.3 		& E 	& & \\ 
SR 21 		 & 1.2$^o$ 	& 2.8 		& E/FA 	& E/FA & E \\ 
SR 24 A 	 & 1.4$^{f,c}$  & 42.3/-28.9 	& E/A 	& A & E \\
SR 24 B/C 	 & 0.6$^{f,c}$  & 8.3/-6.9$^*$  & E/A 	& A & \\ 
SX Cha A 	 & 0.6$^c$ 	& 8.9  		& E 	& & \\ 
SY Cha 		 & 0.5$^c$ 	& $<$3.8$^3$    & 	& & \\ 
Sz 68 A/B 	 & 1.3$^{p,c}$  & 5.8/-6.3 	& E/A 	& A & \\ 
Sz 68 C 	 & 0.1$^{p,c}$ 	& 8.3/-6.9$^*$ 	& A 	& A & \\ 
T Cha 		 & 1.2$^q$ 	& --$^*$ 	& A 	& A & \\ 
T CrA 		 & 3.7$^m$ 	& 7.0/-18.7 	& E/A 	& A & \\ 
T Tau N 	 & 11$^d$ 	& --$^*$ 	& E 	& E & E \\ 
T Tau S 	 & 15$^d$ 	& --$^*$ 	& E/A 	& E/A & E \\ 
TW Cha 		 & 0.5$^c$ 	& 15.3 		& E 	& & \\ 
TW Hya 		 & 0.3$^e$ 	& 2.7 		& E 	& & E \\ 
TY CrA 		 & 0.81$^m$ 	& -0.3 		& A 	& A & \\ 
UX Tau 		 & 3.9$^c$ 	& --$^*$ 	& 	& & \\ 
VSSG 1 		 & 1.0$^f$ 	& 30.5/-1.2 	& E/FA 	& E/FA & E \\ 
VV CrA N 	 & 0.7$^{k,c}$  & -7.8 		& A 	& A & E \\ 
VV CrA S 	 & 2.2$^{k,c}$  & 26 		& E/FA 	& E/FA & E \\ 
VV Ser 		 & 3.0$^m$ 	& 15.2 		& E/FA 	& FA & \\ 
VW Cha A 	 & 0.4$^r$ 	& 21.1 		& E 	& & E \\ 
VZ Cha 		 & 0.8$^r$ 	& 35.5 		& E 	& & E \\ 
Wa Oph 6 	 & 0.9$^s$ 	& 13.1 		& E/FA 	& & E\\ 
WL 22 		 & 0.6$^c$ 	& -1.0 		& A 	& A & \\ 
WX Cha 		 & 0.6$^c$ 	& -- 		& E 	& & E 
\enddata 
\tablenotetext{1}{E - emission, A - absorption, FA-
foreground absorption, only present for $J<$8. Positive fluxes are in
emission while negative fluxes denote absorption.}
\tablenotetext{*}{Strongly affected by telluric features}
\tablenotetext{2}{a.) \citet{luhman06}; b.) \citet{mccabe06}; c.)
Derived from the CRIRES spectra; d.) \citet{kenyon95}; e.)
\citet{hartmann05}; f.) \citet{evans09}; g.) \citet{salyk08}; h.)
\citet{goto11}; i.) \citet{luhman10}; j.) \citet{malfait98}; k.)
\citet{peterson11}; l.) \citet{brown12}; m.)\citet{hillenbrand92}; n.)
\citet{kessler-silacci06}; o.) \citet{brown07}; p.) \citet{merin08};
q.) \citet{alcala93}; r.) \citet{luhman08}; s.) \citet{padgett06}}
\tablenotetext{3}{3 $\sigma$ upper limit on integrated line flux
assuming a line width of 100 km s$^{-1}$.}

\end{deluxetable}

\footnotesize
\begin{deluxetable}{lrrrr}
\tablecolumns{5}
\setlength{\tabcolsep}{0.08in}
\tablewidth{0pt}
\tabletypesize{\footnotesize}
\tablecaption{\label{table:innervel}Line widths}
\tablehead{\colhead{Source} & \colhead{width 3x noise} & 
\colhead{width 5\% peak} & \colhead{$R_{\rm in} sin^2 i$}  & \colhead{Inclin-} \\
\colhead{} & \colhead{(km s$^{-1}$)} & 
\colhead{(km s$^{-1}$)} & \colhead{(AU)} & \colhead{ation($^o$)}}
\startdata
AA Tau           &  172  & 172   & 0.060 & 75 \\
AS 205 N         &  103  & 78    & 0.29 & 25\\
AS 209           &  165  & 90    & 0.22 & \\
CV Cha           &  151  & 154   & 0.15 &\\
CW Tau           &  142  & 142   & 0.088 &\\
DF Tau           &  102  & 118   & 0.13 & \\
DG CrA           &  200  & 200   & 0.089 & \\
DG Tau           &  142  & 155   & 0.073 & \\
DoAr 24E S       &  139  & 135   & 0.097 &\\
DoAr 44          &  90   & 102   & 0.17 & \\
DR Tau           &  62   & 54    & 0.62 & 37 \\
EC 82            &  74   & 76    & 0.30 & \\
EX Lup apr       &  173  & 250   & 0.028 & \\
EX Lup aug       &  134  & 150   & 0.079 & \\
FZ Tau           &  50   & 105   & 0.16 &\\
GQ Lup           &  139  & 139   & 0.091 &\\
Haro 1-4 A       &  64   & 68    & 0.38 & \\
HD 135344 B      &  27   & 32    & 3.43 & 10\\
HD 142527        &  35   & 40    & 2.20 & \\
HD 144432 S      &  51   & 64    & 2.57 & 48\\
IRS 48           &  17   & 17    & 37.4 & \\
IRS 51           &  66   & 109   & 0.15 & \\
LkHa 330         &  34   & 35    & 2.92 & 10 \\
RNO 90           &  186  & 181   & 0.11 & \\
RU Lup           &  135  & 122   & 0.12 & \\
RY Lup           &  51   & 146   & 0.083 & 85 \\
S CrA N          &  101  & 99    & 0.36 & \\
S CrA S          &  166  & 133   & 0.10 & \\
Serp 64          &  99   & 123   & 0.11 &\\
SR 9 A           &  44   & 141   & 0.090 &\\
SR 21            &  13   & 16    & 13.7 & 20\\
SR 24 A          &  122  & 145   & 0.085 &\\
SR 24 B/C        &  54   & 83    & 0.26 & \\
T CrA            &  39   & 71    & 0.70 & \\
T Tau N          &  83   & 146   & 0.17 & \\
T Tau S          &  62   & 95    & 0.39 & \\
TW Cha           &  67   & 132   & 0.10 &\\
TW Hya           &  12   & 13    & 9.9  & 4\\
VSSG 1           &  118  & 117   & 1.31 & 53 \\
VV CrA S         &  60   & 102   & 0.17 & \\
VV Ser           &  74   & 79    & 2.84 & \\
VW Cha A         &  98   & 139   & 0.091 &\\
VZ Cha           &  70   & 99    & 0.18 & \\
Wa Oph 6         &  114  & 197   & 0.046 & 39 \\
WX Cha           &  127  & 332   & 0.016 & \\
\enddata
\end{deluxetable}


\begin{thebibliography}{0}
\expandafter\ifx\csname natexlab\endcsname\relax\def\natexlab#1{#1}\fi
\expandafter\ifx\csname bibnamefont\endcsname\relax
  \def\bibnamefont#1{#1}\fi
\expandafter\ifx\csname bibfnamefont\endcsname\relax
  \def\bibfnamefont#1{#1}\fi
\expandafter\ifx\csname citenamefont\endcsname\relax
  \def\citenamefont#1{#1}\fi
\expandafter\ifx\csname url\endcsname\relax
  \def\url#1{\texttt{#1}}\fi
\expandafter\ifx\csname urlprefix\endcsname\relax\def\urlprefix{URL }\fi
\providecommand{\bibinfo}[2]{#2}
\providecommand{\eprint}[2][]{\url{#2}}

\end{thebibliography}


\begin{thebibliography}{87}
\expandafter\ifx\csname natexlab\endcsname\relax\def\natexlab#1{#1}\fi

\bibitem[{{Aikawa} {et~al.}(2002){Aikawa}, {van Zadelhoff}, {van Dishoeck}, \&
  {Herbst}}]{aikawa02}
{Aikawa}, Y., {van Zadelhoff}, G.~J., {van Dishoeck}, E.~F., \& {Herbst}, E.
  2002, \aap, 386, 622

\bibitem[{{Akeson} {et~al.}(2005){Akeson}, {Boden}, {Monnier}, {Millan-Gabet},
  {Beichman}, {Beletic}, {Calvet}, {Hartmann}, {Hillenbrand}, {Koresko},
  {Sargent}, \& {Tannirkulam}}]{akeson05}
{Akeson}, R.~L., {Boden}, A.~F., {Monnier}, J.~D., {Millan-Gabet}, R.,
  {Beichman}, C., {Beletic}, J., {Calvet}, N., {Hartmann}, L., {Hillenbrand},
  L., {Koresko}, C., {Sargent}, A., \& {Tannirkulam}, A. 2005, \apj, 635, 1173

\bibitem[{{Alcala} {et~al.}(1993){Alcala}, {Covino}, {Franchini}, {Krautter},
  {Terranegra}, \& {Wichmann}}]{alcala93}
{Alcala}, J.~M., {Covino}, E., {Franchini}, M., {Krautter}, J., {Terranegra},
  L., \& {Wichmann}, R. 1993, \aap, 272, 225

\bibitem[{{Armitage}(2010)}]{armitage10}
{Armitage}, P.~J. 2010, {Astrophysics of Planet Formation}, ed. {Armitage,
  P.~J.} (Cambridge University Press)

\bibitem[{{Armitage}(2011)}]{armitage11}
---. 2011, \araa, 49, 195

\bibitem[{{Armitage} {et~al.}(2003){Armitage}, {Clarke}, \&
  {Palla}}]{armitage03}
{Armitage}, P.~J., {Clarke}, C.~J., \& {Palla}, F. 2003, \mnras, 342, 1139

\bibitem[{{Bast} {et~al.}(2011){Bast}, {Brown}, {Herczeg}, {van Dishoeck}, \&
  {Pontoppidan}}]{bast11}
{Bast}, J.~E., {Brown}, J.~M., {Herczeg}, G.~J., {van Dishoeck}, E.~F., \&
  {Pontoppidan}, K.~M. 2011, \aap, 527, A119+

\bibitem[{{Blake} \& {Boogert}(2004)}]{blake04}
{Blake}, G.~A. \& {Boogert}, A.~C.~A. 2004, \apjl, 606, L73

\bibitem[{{Boogert} {et~al.}(2002{\natexlab{a}}){Boogert}, {Hogerheijde}, \&
  {Blake}}]{boogert02b}
{Boogert}, A.~C.~A., {Hogerheijde}, M.~R., \& {Blake}, G.~A.
  2002{\natexlab{a}}, \apj, 568, 761

\bibitem[{{Boogert} {et~al.}(2002{\natexlab{b}}){Boogert}, {Hogerheijde},
  {Ceccarelli}, {Tielens}, {van Dishoeck}, {Blake}, {Latter}, \&
  {Motte}}]{boogert02a}
{Boogert}, A.~C.~A., {Hogerheijde}, M.~R., {Ceccarelli}, C., {Tielens},
  A.~G.~G.~M., {van Dishoeck}, E.~F., {Blake}, G.~A., {Latter}, W.~B., \&
  {Motte}, F. 2002{\natexlab{b}}, \apj, 570, 708

\bibitem[{{Brittain} {et~al.}(2009){Brittain}, {Najita}, \&
  {Carr}}]{brittain09}
{Brittain}, S.~D., {Najita}, J.~R., \& {Carr}, J.~S. 2009, \apj, 702, 85

\bibitem[{{Brittain} {et~al.}(2007){Brittain}, {Simon}, {Najita}, \&
  {Rettig}}]{brittain07}
{Brittain}, S.~D., {Simon}, T., {Najita}, J.~R., \& {Rettig}, T.~W. 2007, \apj,
  659, 685

\bibitem[{{Brown}(2008)}]{brown08t}
{Brown}, J.~M. 2008, PhD thesis, California Institute of Technology

\bibitem[{{Brown} {et~al.}(2007){Brown}, {Blake}, {Dullemond}, {Mer{\'{\i}}n},
  {Augereau}, {Boogert}, {Evans}, {Geers}, {Lahuis}, {Kessler-Silacci},
  {Pontoppidan}, \& {van Dishoeck}}]{brown07}
{Brown}, J.~M., {Blake}, G.~A., {Dullemond}, C.~P., {Mer{\'{\i}}n}, B.,
  {Augereau}, J.~C., {Boogert}, A.~C.~A., {Evans}, II, N.~J., {Geers}, V.~C.,
  {Lahuis}, F., {Kessler-Silacci}, J.~E., {Pontoppidan}, K.~M., \& {van
  Dishoeck}, E.~F. 2007, \apjl, 664, L107

\bibitem[{{Brown} {et~al.}(2012){Brown}, {Herczeg}, {Pontoppidan}, \& {van
  Dishoeck}}]{brown12}
{Brown}, J.~M., {Herczeg}, G.~J., {Pontoppidan}, K.~M., \& {van Dishoeck},
  E.~F. 2012, \apj, 744, 116

\bibitem[{{Bruderer} {et~al.}(2012){Bruderer}, {van Dishoeck}, {Doty}, \&
  {Herczeg}}]{bruderer12}
{Bruderer}, S., {van Dishoeck}, E.~F., {Doty}, S.~D., \& {Herczeg}, G.~J. 2012,
  \aap, 541, A91

\bibitem[{{Correia} {et~al.}(2006){Correia}, {Zinnecker}, {Ratzka}, \&
  {Sterzik}}]{correia06}
{Correia}, S., {Zinnecker}, H., {Ratzka}, T., \& {Sterzik}, M.~F. 2006, \aap,
  459, 909

\bibitem[{{Dunkin} {et~al.}(1997){Dunkin}, {Barlow}, \& {Ryan}}]{Dunkin97}
{Dunkin}, S.~K., {Barlow}, M.~J., \& {Ryan}, S.~G. 1997, \mnras, 290, 165

\bibitem[{{Evans} {et~al.}(2003)}]{evans03}
{Evans}, N.~J. {et~al.} 2003, \pasp, 115, 965

\bibitem[{{Evans} {et~al.}(2009)}]{evans09}
---. 2009, \apjs, 181, 321

\bibitem[{{Field} {et~al.}(1972){Field}, {Tilford}, {Howard}, \&
  {Simmons}}]{field72}
{Field}, R.~W., {Tilford}, S.~G., {Howard}, R.~A., \& {Simmons}, J.~D. 1972, J.
  Mol. Spectrosc., 44, 347

\bibitem[{{Flaherty} {et~al.}(2012){Flaherty}, {Muzerolle}, {Rieke},
  {Gutermuth}, {Balog}, {Herbst}, {Megeath}, \& {Kun}}]{flaherty12}
{Flaherty}, K., {Muzerolle}, J., {Rieke}, G., {Gutermuth}, R., {Balog}, Z.,
  {Herbst}, W., {Megeath}, S.~T., \& {Kun}, M. 2012, ArXiv e-prints

\bibitem[{{France} {et~al.}(2011){France}, {Schindhelm}, {Burgh}, {Herczeg},
  {Harper}, {Brown}, {Green}, {Linsky}, {Yang}, {Abgrall}, {Ardila}, {Bergin},
  {Bethell}, {Brown}, {Calvet}, {Espaillat}, {Gregory}, {Hillenbrand},
  {Hussain}, {Ingleby}, {Johns-Krull}, {Roueff}, {Valenti}, \&
  {Walter}}]{france11}
{France}, K., {Schindhelm}, E., {Burgh}, E.~B., {Herczeg}, G.~J., {Harper},
  G.~M., {Brown}, A., {Green}, J.~C., {Linsky}, J.~L., {Yang}, H., {Abgrall},
  H., {Ardila}, D.~R., {Bergin}, E., {Bethell}, T., {Brown}, J.~M., {Calvet},
  N., {Espaillat}, C., {Gregory}, S.~G., {Hillenbrand}, L.~A., {Hussain}, G.,
  {Ingleby}, L., {Johns-Krull}, C.~M., {Roueff}, E., {Valenti}, J.~A., \&
  {Walter}, F.~M. 2011, \apj, 734, 31

\bibitem[{{Furlan} {et~al.}(2009){Furlan}, {Watson}, {McClure}, {Manoj},
  {Espaillat}, {D'Alessio}, {Calvet}, {Kim}, {Sargent}, {Forrest}, \&
  {Hartmann}}]{furlan09}
{Furlan}, E., {Watson}, D.~M., {McClure}, M.~K., {Manoj}, P., {Espaillat}, C.,
  {D'Alessio}, P., {Calvet}, N., {Kim}, K.~H., {Sargent}, B.~A., {Forrest},
  W.~J., \& {Hartmann}, L. 2009, \apj, 703, 1964

\bibitem[{{Gammie}(1996)}]{gammie96}
{Gammie}, C.~F. 1996, \apj, 457, 355

\bibitem[{{Gilijamse} {et~al.}(2007){Gilijamse}, {Hoekstra}, {Meek},
  {Mets\"al\"a}, {van de Meerakker}, {Meijer}, \& {Groenenboom}}]{gilijamse07}
{Gilijamse}, J.~J., {Hoekstra}, S., {Meek}, S.~A., {Mets\"al\"a}, M., {van de
  Meerakker}, S.~Y.~T., {Meijer}, G., \& {Groenenboom}, G.~C. 2007, J. Chem.
  Phys., 127, 221102

\bibitem[{{Glassgold} {et~al.}(2009){Glassgold}, {Meijerink}, \&
  {Najita}}]{glassgold09}
{Glassgold}, A.~E., {Meijerink}, R., \& {Najita}, J.~R. 2009, \apj, 701, 142

\bibitem[{{Gorti} \& {Hollenbach}(2008)}]{gorti08}
{Gorti}, U. \& {Hollenbach}, D. 2008, \apj, 683, 287

\bibitem[{{Gorti} {et~al.}(2011){Gorti}, {Hollenbach}, {Najita}, \&
  {Pascucci}}]{gorti11}
{Gorti}, U., {Hollenbach}, D., {Najita}, J., \& {Pascucci}, I. 2011, \apj, 735,
  90

\bibitem[{{Goto} {et~al.}(2011){Goto}, {Reg{\'a}ly}, {Dullemond}, {van den
  Ancker}, {Brown}, {Carmona}, {Pontoppidan}, {{\'A}brah{\'a}m}, {Blake},
  {Fedele}, {Henning}, {Juh{\'a}sz}, {K{\'o}sp{\'a}l}, {Mosoni},
  {Sicilia-Aguilar}, {Terada}, {van Boekel}, {van Dishoeck}, \&
  {Usuda}}]{goto11}
{Goto}, M., {Reg{\'a}ly}, Z., {Dullemond}, C.~P., {van den Ancker}, M.,
  {Brown}, J.~M., {Carmona}, A., {Pontoppidan}, K., {{\'A}brah{\'a}m}, P.,
  {Blake}, G.~A., {Fedele}, D., {Henning}, T., {Juh{\'a}sz}, A.,
  {K{\'o}sp{\'a}l}, {\'A}., {Mosoni}, L., {Sicilia-Aguilar}, A., {Terada}, H.,
  {van Boekel}, R., {van Dishoeck}, E.~F., \& {Usuda}, T. 2011, \apj, 728, 5

\bibitem[{{Goto} {et~al.}(2006){Goto}, {Usuda}, {Dullemond}, {Henning}, {Linz},
  {Stecklum}, \& {Suto}}]{goto06}
{Goto}, M., {Usuda}, T., {Dullemond}, C.~P., {Henning}, T., {Linz}, H.,
  {Stecklum}, B., \& {Suto}, H. 2006, \apj, 652, 758

\bibitem[{{Greene} \& {Meyer}(1995)}]{greene95}
{Greene}, T.~P. \& {Meyer}, M.~R. 1995, \apj, 450, 233

\bibitem[{{Guenther} {et~al.}(2007){Guenther}, {Esposito}, {Mundt}, {Covino},
  {Alcal{\'a}}, {Cusano}, \& {Stecklum}}]{Gunther07}
{Guenther}, E.~W., {Esposito}, M., {Mundt}, R., {Covino}, E., {Alcal{\'a}},
  J.~M., {Cusano}, F., \& {Stecklum}, B. 2007, \aap, 467, 1147

\bibitem[{{Hartmann} {et~al.}(2005){Hartmann}, {Megeath}, {Allen}, {Luhman},
  {Calvet}, {D'Alessio}, {Franco-Hernandez}, \& {Fazio}}]{hartmann05}
{Hartmann}, L., {Megeath}, S.~T., {Allen}, L., {Luhman}, K., {Calvet}, N.,
  {D'Alessio}, P., {Franco-Hernandez}, R., \& {Fazio}, G. 2005, \apj, 629, 881

\bibitem[{{Herbig} \& {Bell}(1988)}]{hb88}
{Herbig}, G.~H. \& {Bell}, K.~R. 1988, {Third Catalog of Emission-Line Stars of
  the Orion Population : 3 : 1988}

\bibitem[{{Herczeg} {et~al.}(2011){Herczeg}, {Brown}, {van Dishoeck}, \&
  {Pontoppidan}}]{herczeg11}
{Herczeg}, G.~J., {Brown}, J.~M., {van Dishoeck}, E.~F., \& {Pontoppidan},
  K.~M. 2011, \aap, 533, A112

\bibitem[{{Herczeg} {et~al.}(2002){Herczeg}, {Linsky}, {Valenti},
  {Johns-Krull}, \& {Wood}}]{herczeg02}
{Herczeg}, G.~J., {Linsky}, J.~L., {Valenti}, J.~A., {Johns-Krull}, C.~M., \&
  {Wood}, B.~E. 2002, \apj, 572, 310

\bibitem[{{Hillenbrand} {et~al.}(1992){Hillenbrand}, {Strom}, {Vrba}, \&
  {Keene}}]{hillenbrand92}
{Hillenbrand}, L.~A., {Strom}, S.~E., {Vrba}, F.~J., \& {Keene}, J. 1992, \apj,
  397, 613

\bibitem[{{Hollenbach} {et~al.}(1994){Hollenbach}, {Johnstone}, {Lizano}, \&
  {Shu}}]{hollenbach94}
{Hollenbach}, D., {Johnstone}, D., {Lizano}, S., \& {Shu}, F. 1994, \apj, 428,
  654

\bibitem[{{Kaeufl} {et~al.}(2004)}]{kaeufl04}
{Kaeufl}, H. {et~al.} 2004, in Presented at the Society of Photo-Optical
  Instrumentation Engineers (SPIE) Conference, Vol. 5492, Society of
  Photo-Optical Instrumentation Engineers (SPIE) Conference Series, ed.
  {A.~F.~M.~Moorwood \& M.~Iye}, 1218--1227

\bibitem[{{Kenyon} \& {Hartmann}(1995)}]{kenyon95}
{Kenyon}, S.~J. \& {Hartmann}, L. 1995, \apjs, 101, 117

\bibitem[{{Kessler-Silacci} {et~al.}(2006){Kessler-Silacci}, {Augereau},
  {Dullemond}, {Geers}, {Lahuis}, {Evans}, {van Dishoeck}, {Blake}, {Boogert},
  {Brown}, {J{\o}rgensen}, {Knez}, \& {Pontoppidan}}]{kessler-silacci06}
{Kessler-Silacci}, J., {Augereau}, J.-C., {Dullemond}, C.~P., {Geers}, V.,
  {Lahuis}, F., {Evans}, II, N.~J., {van Dishoeck}, E.~F., {Blake}, G.~A.,
  {Boogert}, A.~C.~A., {Brown}, J., {J{\o}rgensen}, J.~K., {Knez}, C., \&
  {Pontoppidan}, K.~M. 2006, \apj, 639, 275

\bibitem[{{Kley} \& {Nelson}(2012)}]{kley12}
{Kley}, W. \& {Nelson}, R.~P. 2012, ArXiv e-prints

\bibitem[{{K{\"o}hler} {et~al.}(2008){K{\"o}hler}, {Neuh{\"a}user},
  {Kr{\"a}mer}, {Leinert}, {Ott}, \& {Eckart}}]{koehler08}
{K{\"o}hler}, R., {Neuh{\"a}user}, R., {Kr{\"a}mer}, S., {Leinert}, C., {Ott},
  T., \& {Eckart}, A. 2008, \aap, 488, 997

\bibitem[{{Le Floch}(1992)}]{lefloch91}
{Le Floch}, A. 1992, Mol. Phys., 72, 133

\bibitem[{{Le Roy}(2004)}]{leroy04}
{Le Roy}, R.~J. 2004, Chemical Physics Research, Report No. CP-657R

\bibitem[{{Luhman} {et~al.}(2008){Luhman}, {Allen}, {Allen}, {Gutermuth},
  {Hartmann}, {Mamajek}, {Megeath}, {Myers}, \& {Fazio}}]{luhman08}
{Luhman}, K.~L., {Allen}, L.~E., {Allen}, P.~R., {Gutermuth}, R.~A.,
  {Hartmann}, L., {Mamajek}, E.~E., {Megeath}, S.~T., {Myers}, P.~C., \&
  {Fazio}, G.~G. 2008, \apj, 675, 1375

\bibitem[{{Luhman} {et~al.}(2010){Luhman}, {Allen}, {Espaillat}, {Hartmann}, \&
  {Calvet}}]{luhman10}
{Luhman}, K.~L., {Allen}, P.~R., {Espaillat}, C., {Hartmann}, L., \& {Calvet},
  N. 2010, \apjs, 186, 111

\bibitem[{{Luhman} {et~al.}(2006){Luhman}, {Whitney}, {Meade}, {Babler},
  {Indebetouw}, {Bracker}, \& {Churchwell}}]{luhman06}
{Luhman}, K.~L., {Whitney}, B.~A., {Meade}, M.~R., {Babler}, B.~L.,
  {Indebetouw}, R., {Bracker}, S., \& {Churchwell}, E.~B. 2006, \apj, 647, 1180

\bibitem[{{Malfait} {et~al.}(1998){Malfait}, {Bogaert}, \&
  {Waelkens}}]{malfait98}
{Malfait}, K., {Bogaert}, E., \& {Waelkens}, C. 1998, \aap, 331, 211

\bibitem[{{Mandell} {et~al.}(2012){Mandell}, {Bast}, {van Dishoeck}, {Blake},
  {Salyk}, {Mumma}, \& {Villanueva}}]{mandell12}
{Mandell}, A.~M., {Bast}, J., {van Dishoeck}, E.~F., {Blake}, G.~A., {Salyk},
  C., {Mumma}, M.~J., \& {Villanueva}, G. 2012, \apj, 747, 92

\bibitem[{{Markwick} {et~al.}(2002){Markwick}, {Ilgner}, {Millar}, \&
  {Henning}}]{markwick02}
{Markwick}, A.~J., {Ilgner}, M., {Millar}, T.~J., \& {Henning}, T. 2002, \aap,
  385, 632

\bibitem[{{McCabe} {et~al.}(2006){McCabe}, {Ghez}, {Prato}, {Duch{\^e}ne},
  {Fisher}, \& {Telesco}}]{mccabe06}
{McCabe}, C., {Ghez}, A.~M., {Prato}, L., {Duch{\^e}ne}, G., {Fisher}, R.~S.,
  \& {Telesco}, C. 2006, \apj, 636, 932

\bibitem[{{Melo}(2003)}]{Melo03}
{Melo}, C.~H.~F. 2003, \aap, 410, 269

\bibitem[{{Mer{\'{\i}}n} {et~al.}(2008){Mer{\'{\i}}n}, {J{\o}rgensen},
  {Spezzi}, {Alcal{\'a}}, {Evans}, {Harvey}, {Prusti}, {Chapman}, {Huard}, {van
  Dishoeck}, \& {Comer{\'o}n}}]{merin08}
{Mer{\'{\i}}n}, B., {J{\o}rgensen}, J., {Spezzi}, L., {Alcal{\'a}}, J.~M.,
  {Evans}, II, N.~J., {Harvey}, P.~M., {Prusti}, T., {Chapman}, N., {Huard},
  T., {van Dishoeck}, E.~F., \& {Comer{\'o}n}, F. 2008, \apjs, 177, 551

\bibitem[{{Muzerolle} {et~al.}(2009){Muzerolle}, {Flaherty}, {Balog}, {Furlan},
  {Smith}, {Allen}, {Calvet}, {D'Alessio}, {Megeath}, {Muench}, {Rieke}, \&
  {Sherry}}]{muzerolle09}
{Muzerolle}, J., {Flaherty}, K., {Balog}, Z., {Furlan}, E., {Smith}, P.~S.,
  {Allen}, L., {Calvet}, N., {D'Alessio}, P., {Megeath}, S.~T., {Muench}, A.,
  {Rieke}, G.~H., \& {Sherry}, W.~H. 2009, \apjl, 704, L15

\bibitem[{{Muzerolle} {et~al.}(2003){Muzerolle}, {Hillenbrand}, {Calvet},
  {Brice{\~n}o}, \& {Hartmann}}]{Muzerolle03}
{Muzerolle}, J., {Hillenbrand}, L., {Calvet}, N., {Brice{\~n}o}, C., \&
  {Hartmann}, L. 2003, \apj, 592, 266

\bibitem[{{Nagasawa} {et~al.}(2007){Nagasawa}, {Thommes}, {Kenyon}, {Bromley},
  \& {Lin}}]{nagasawa07}
{Nagasawa}, M., {Thommes}, E.~W., {Kenyon}, S.~J., {Bromley}, B.~C., \& {Lin},
  D.~N.~C. 2007, Protostars and Planets V, 639

\bibitem[{{Najita} {et~al.}(2003){Najita}, {Carr}, \& {Mathieu}}]{najita03}
{Najita}, J., {Carr}, J.~S., \& {Mathieu}, R.~D. 2003, \apj, 589, 931

\bibitem[{{Najita} {et~al.}(2011){Najita}, {{\'A}d{\'a}mkovics}, \&
  {Glassgold}}]{najita11}
{Najita}, J.~R., {{\'A}d{\'a}mkovics}, M., \& {Glassgold}, A.~E. 2011, \apj,
  743, 147

\bibitem[{{Oliveira} {et~al.}(2009){Oliveira}, {Mer{\'{\i}}n}, {Pontoppidan},
  {van Dishoeck}, {Overzier}, {Hern{\'a}ndez}, {Sicilia-Aguilar}, {Eiroa}, \&
  {Montesinos}}]{oliveira09}
{Oliveira}, I., {Mer{\'{\i}}n}, B., {Pontoppidan}, K.~M., {van Dishoeck},
  E.~F., {Overzier}, R.~A., {Hern{\'a}ndez}, J., {Sicilia-Aguilar}, A.,
  {Eiroa}, C., \& {Montesinos}, B. 2009, \apj, 691, 672

\bibitem[{{Owen} {et~al.}(2010){Owen}, {Ercolano}, {Clarke}, \&
  {Alexander}}]{owen10}
{Owen}, J.~E., {Ercolano}, B., {Clarke}, C.~J., \& {Alexander}, R.~D. 2010,
  \mnras, 401, 1415

\bibitem[{{Padgett} {et~al.}(2006){Padgett}, {Cieza}, {Stapelfeldt}, {Evans},
  {Koerner}, {Sargent}, {Fukagawa}, {van Dishoeck}, {Augereau}, {Allen},
  {Blake}, {Brooke}, {Chapman}, {Harvey}, {Porras}, {Lai}, {Mundy}, {Myers},
  {Spiesman}, \& {Wahhaj}}]{padgett06}
{Padgett}, D.~L., {Cieza}, L., {Stapelfeldt}, K.~R., {Evans}, II, N.~J.,
  {Koerner}, D., {Sargent}, A., {Fukagawa}, M., {van Dishoeck}, E.~F.,
  {Augereau}, J.-C., {Allen}, L., {Blake}, G., {Brooke}, T., {Chapman}, N.,
  {Harvey}, P., {Porras}, A., {Lai}, S.-P., {Mundy}, L., {Myers}, P.~C.,
  {Spiesman}, W., \& {Wahhaj}, Z. 2006, \apj, 645, 1283

\bibitem[{{Pani{\'c}} {et~al.}(2009){Pani{\'c}}, {Hogerheijde}, {Wilner}, \&
  {Qi}}]{panic09}
{Pani{\'c}}, O., {Hogerheijde}, M.~R., {Wilner}, D., \& {Qi}, C. 2009, \aap,
  501, 269

\bibitem[{{Perez-Becker} \& {Chiang}(2011)}]{perez11}
{Perez-Becker}, D. \& {Chiang}, E. 2011, \apj, 735, 8

\bibitem[{{Peterson} {et~al.}(2011){Peterson}, {Caratti o Garatti}, {Bourke},
  {Forbrich}, {Gutermuth}, {J{\o}rgensen}, {Allen}, {Patten}, {Dunham},
  {Harvey}, {Mer{\'{\i}}n}, {Chapman}, {Cieza}, {Huard}, {Knez}, {Prager}, \&
  {Evans}}]{peterson11}
{Peterson}, D.~E., {Caratti o Garatti}, A., {Bourke}, T.~L., {Forbrich}, J.,
  {Gutermuth}, R.~A., {J{\o}rgensen}, J.~K., {Allen}, L.~E., {Patten}, B.~M.,
  {Dunham}, M.~M., {Harvey}, P.~M., {Mer{\'{\i}}n}, B., {Chapman}, N.~L.,
  {Cieza}, L.~A., {Huard}, T.~L., {Knez}, C., {Prager}, B., \& {Evans}, N.~J.
  2011, \apjs, 194, 43

\bibitem[{{Pontoppidan} {et~al.}(2011{\natexlab{a}}){Pontoppidan}, {Blake}, \&
  {Smette}}]{pontoppidan11}
{Pontoppidan}, K.~M., {Blake}, G.~A., \& {Smette}, A. 2011{\natexlab{a}}, \apj,
  733, 84

\bibitem[{{Pontoppidan} {et~al.}(2008){Pontoppidan}, {Blake}, {van Dishoeck},
  {Smette}, {Ireland}, \& {Brown}}]{pontoppidan08}
{Pontoppidan}, K.~M., {Blake}, G.~A., {van Dishoeck}, E.~F., {Smette}, A.,
  {Ireland}, M.~J., \& {Brown}, J. 2008, \apj, 684, 1323

\bibitem[{{Pontoppidan} {et~al.}(2003){Pontoppidan}, {Fraser}, {Dartois},
  {Thi}, {van Dishoeck}, {Boogert}, {d'Hendecourt}, {Tielens}, \&
  {Bisschop}}]{pontoppidan03}
{Pontoppidan}, K.~M., {Fraser}, H.~J., {Dartois}, E., {Thi}, W., {van
  Dishoeck}, E.~F., {Boogert}, A.~C.~A., {d'Hendecourt}, L., {Tielens},
  A.~G.~G.~M., \& {Bisschop}, S.~E. 2003, \aap, 408, 981

\bibitem[{{Pontoppidan} {et~al.}(2011{\natexlab{b}}){Pontoppidan}, {van
  Dishoeck}, {Blake}, {Smith}, {Brown}, {Herczeg}, {Bast}, {Mandell}, {Smette},
  {Thi}, {Young}, {Morris}, {Dent}, \& {K{\"a}ufl}}]{pontoppidan11_mess}
{Pontoppidan}, K.~M., {van Dishoeck}, E., {Blake}, G.~A., {Smith}, R., {Brown},
  J., {Herczeg}, G.~J., {Bast}, J., {Mandell}, A., {Smette}, A., {Thi}, W.-F.,
  {Young}, E.~D., {Morris}, M.~R., {Dent}, W., \& {K{\"a}ufl}, H.~U.
  2011{\natexlab{b}}, The Messenger, 143, 32

\bibitem[{{Qi} {et~al.}(2004){Qi}, {Ho}, {Wilner}, {Takakuwa}, {Hirano},
  {Ohashi}, {Bourke}, {Zhang}, {Blake}, {Hogerheijde}, {Saito}, {Choi}, \&
  {Yang}}]{qi04}
{Qi}, C., {Ho}, P.~T.~P., {Wilner}, D.~J., {Takakuwa}, S., {Hirano}, N.,
  {Ohashi}, N., {Bourke}, T.~L., {Zhang}, Q., {Blake}, G.~A., {Hogerheijde},
  M., {Saito}, M., {Choi}, M., \& {Yang}, J. 2004, \apjl, 616, L11

\bibitem[{{Raymond} {et~al.}(2004){Raymond}, {Quinn}, \& {Lunine}}]{raymond04}
{Raymond}, S.~N., {Quinn}, T., \& {Lunine}, J.~I. 2004, Icarus, 168, 1

\bibitem[{{Reg{\'a}ly} {et~al.}(2010){Reg{\'a}ly}, {S{\'a}ndor}, {Dullemond},
  \& {van Boekel}}]{regaly10}
{Reg{\'a}ly}, Z., {S{\'a}ndor}, Z., {Dullemond}, C.~P., \& {van Boekel}, R.
  2010, \aap, 523, A69

\bibitem[{{Rettig} {et~al.}(2006){Rettig}, {Brittain}, {Simon}, {Gibb},
  {Balsara}, {Tilley}, \& {Kulesa}}]{rettig06}
{Rettig}, T., {Brittain}, S., {Simon}, T., {Gibb}, E., {Balsara}, D.~S.,
  {Tilley}, D.~A., \& {Kulesa}, C. 2006, \apj, 646, 342

\bibitem[{{Salyk} {et~al.}(2009){Salyk}, {Blake}, {Boogert}, \&
  {Brown}}]{salyk09}
{Salyk}, C., {Blake}, G.~A., {Boogert}, A.~C.~A., \& {Brown}, J.~M. 2009, \apj,
  699, 330

\bibitem[{{Salyk} {et~al.}(2011){Salyk}, {Blake}, {Boogert}, \&
  {Brown}}]{salyk11}
---. 2011, \apj, 743, 112

\bibitem[{{Salyk} {et~al.}(2013){Salyk}, {Herczeg}, {Brown}, {Blake},
  {Pontoppidan}, \& {van Dishoeck}}]{salyk12}
{Salyk}, C., {Herczeg}, G.~J., {Brown}, J.~M., {Blake}, G.~A., {Pontoppidan},
  K.~M., \& {van Dishoeck}, E.~F. 2012, \apj, in press

\bibitem[{{Salyk} {et~al.}(2008){Salyk}, {Pontoppidan}, {Blake}, {Lahuis}, {van
  Dishoeck}, \& {Evans}}]{salyk08}
{Salyk}, C., {Pontoppidan}, K.~M., {Blake}, G.~A., {Lahuis}, F., {van
  Dishoeck}, E.~F., \& {Evans}, II, N.~J. 2008, \apjl, 676, L49

\bibitem[{{Schindhelm} {et~al.}(2012){Schindhelm}, {France}, {Herczeg},
  {Bergin}, {Yang}, {Brown}, {Brown}, {Linsky}, \& {Valenti}}]{schindhelm12}
{Schindhelm}, E., {France}, K., {Herczeg}, G.~J., {Bergin}, E., {Yang}, H.,
  {Brown}, A., {Brown}, J.~M., {Linsky}, J.~L., \& {Valenti}, J. 2012, \apjl,
  756, L23

\bibitem[{{Setiawan} {et~al.}(2008){Setiawan}, {Henning}, {Launhardt},
  {M{\"u}ller}, {Weise}, \& {K{\"u}rster}}]{Setiawan08}
{Setiawan}, J., {Henning}, T., {Launhardt}, R., {M{\"u}ller}, A., {Weise}, P.,
  \& {K{\"u}rster}, M. 2008, \nat, 451, 38

\bibitem[{{Smith} {et~al.}(2009){Smith}, {Pontoppidan}, {Young}, {Morris}, \&
  {van Dishoeck}}]{smith09}
{Smith}, R.~L., {Pontoppidan}, K.~M., {Young}, E.~D., {Morris}, M.~R., \& {van
  Dishoeck}, E.~F. 2009, \apj, 701, 163

\bibitem[{{Thi} {et~al.}(2010){Thi}, {van Dishoeck}, {Pontoppidan}, \&
  {Dartois}}]{thi10}
{Thi}, W., {van Dishoeck}, E.~F., {Pontoppidan}, K.~M., \& {Dartois}, E. 2010,
  \mnras, 406, 1409

\bibitem[{{Torres} {et~al.}(2006){Torres}, {Quast}, {da Silva}, {de La Reza},
  {Melo}, \& {Sterzik}}]{torres06}
{Torres}, C.~A.~O., {Quast}, G.~R., {da Silva}, L., {de La Reza}, R., {Melo},
  C.~H.~F., \& {Sterzik}, M. 2006, \aap, 460, 695

\bibitem[{{van der Plas} {et~al.}(2009){van der Plas}, {van den Ancker},
  {Acke}, {Carmona}, {Dominik}, {Fedele}, \& {Waters}}]{vanderplas09}
{van der Plas}, G., {van den Ancker}, M.~E., {Acke}, B., {Carmona}, A.,
  {Dominik}, C., {Fedele}, D., \& {Waters}, L.~B.~F.~M. 2009, \aap, 500, 1137

\bibitem[{{van Dishoeck} {et~al.}(2003){van Dishoeck}, {Dartois},
  {Pontoppidan}, {Thi}, {D'Hendecourt}, {Boogert}, {Fraser}, {Schutte}, \&
  {Tielens}}]{vanDishoeck03}
{van Dishoeck}, E.~F., {Dartois}, E., {Pontoppidan}, K.~M., {Thi}, W.~F.,
  {D'Hendecourt}, L., {Boogert}, A.~C.~A., {Fraser}, H.~J., {Schutte}, W.~A.,
  \& {Tielens}, A.~G.~G.~M. 2003, The Messenger, 113, 49

\bibitem[{{van Kempen} {et~al.}(2007){van Kempen}, {van Dishoeck}, {Brinch}, \&
  {Hogerheijde}}]{vankempen07}
{van Kempen}, T.~A., {van Dishoeck}, E.~F., {Brinch}, C., \& {Hogerheijde},
  M.~R. 2007, \aap, 461, 983

\bibitem[{{Walsh} {et~al.}(2012){Walsh}, {Nomura}, {Millar}, \&
  {Aikawa}}]{walsh12}
{Walsh}, C., {Nomura}, H., {Millar}, T.~J., \& {Aikawa}, Y. 2012, \apj, 747,
  114

\bibitem[{{Walter} {et~al.}(2003){Walter}, {Herczeg}, {Brown}, {Ardila},
  {Gahm}, {Johns-Krull}, {Lissauer}, {Simon}, \& {Valenti}}]{walter03}
{Walter}, F.~M., {Herczeg}, G., {Brown}, A., {Ardila}, D.~R., {Gahm}, G.~F.,
  {Johns-Krull}, C.~M., {Lissauer}, J.~J., {Simon}, M., \& {Valenti}, J.~A.
  2003, \aj, 126, 3076

\bibitem[{{Wang} {et~al.}(2004){Wang}, {Mundt}, {Henning}, \& {Apai}}]{wang04}
{Wang}, H., {Mundt}, R., {Henning}, T., \& {Apai}, D. 2004, \apj, 617, 1191

\bibitem[{{White} \& {Hillenbrand}(2004)}]{White04}
{White}, R.~J. \& {Hillenbrand}, L.~A. 2004, \apj, 616, 998

\bibitem[{{Woitke} {et~al.}(2009){Woitke}, {Kamp}, \& {Thi}}]{woitke09}
{Woitke}, P., {Kamp}, I., \& {Thi}, W.-F. 2009, \aap, 501, 383

\bibitem[{{Woods} \& {Willacy}(2009)}]{woods09}
{Woods}, P.~M. \& {Willacy}, K. 2009, \apj, 693, 1360

\bibitem[{{Yang} {et~al.}(2012){Yang}, {Herczeg}, {Linsky}, {Brown},
  {Johns-Krull}, {Ingleby}, {Calvet}, {Bergin}, \& {Valenti}}]{yang12}
{Yang}, H., {Herczeg}, G.~J., {Linsky}, J.~L., {Brown}, A., {Johns-Krull},
  C.~M., {Ingleby}, L., {Calvet}, N., {Bergin}, E., \& {Valenti}, J.~A. 2012,
  \apj, 744, 121

\end{thebibliography}
\end{document}